\documentclass[11pt]{article}
\usepackage{graphicx}
\usepackage{amssymb}
\usepackage{amsmath}
\usepackage{graphicx}
\usepackage{amstext}
\usepackage{leftidx}
\usepackage{esint}
\usepackage[utf8]{inputenc}
\usepackage{color}
\usepackage{cite}
\usepackage{hyperref}
\usepackage{multido}

\newcommand{\rL}{\rho_\Lambda}

\newcommand{\CC}{\Lambda}

\newcommand{\rv}{\rho_{\rm vac}}
\newcommand{\Pv}{P_{\rm vac}}
\newcommand{\rvo}{\rho^0_{\rm vac}}
\newcommand{\drv}{\dot{\rho}_{\rm vac}}

\newcommand{\OMo}{\Omega_{m}^0}

\newcommand{\Ovo}{\Omega^0_{\rm vac}}

\newcommand{\rco}{\rho^0_{c}}
\newcommand{\rmo}{\rho_{m}^0}

\newcommand{\rM}{\rho_m}
\newcommand{\rmr}{\rho_m}
\newcommand{\pmr}{p_m}
\newcommand{\rMo}{\rho_{m}^0}

\newcommand{\wm}{\omega_m}

\newcommand{\nueff}{\nu_{\rm eff}}

\newcommand{\bk}{{\bf k}}
\newcommand{\mpl}{m_{\rm Pl}}
\newcommand{\MPl}{{\cal M}_{\rm Pl}}

\newcommand{\be}{\begin{equation}}
\newcommand{\ee}{\end{equation}}

\newcommand{\LQCD}{\Lambda_{\rm QCD}}

\newcommand{\tnu}{\tilde{\nu}}
\newcommand{\tmu}{\mu}

\newcommand{\cH}{\mathcal{H}}

\newcommand{\ha}{\hat{a}}
\newcommand{\astar}{a_{*}}

\newcommand{\wv}{w_{\rm vac}}

\newcommand{\calH}{\mathcal{H}}
\begin{document}



 \hyphenation{nu-cleo-syn-the-sis u-sing si-mu-la-te ma-king par-ti-cu-lar-ly
cos-mo-lo-gy know-led-ge e-vi-den-ce stu-dies be-ha-vi-or streng-thens
res-pec-ti-ve-ly appro-xi-ma-te-ly gra-vi-ty sca-ling ha-ving
ge-ne-ra-li-zed re-gu-la-ri-za-tion mo-del mo-dels po-wers ex-cee-din-gly ho-we-ver me-tric pa-ra-me-ter  vacu-um per-for-ming appro-xi-ma-tion li-te-ra-tu-re pro-pa-ga-tor}




\begin{center}
{\bf \Large  Running Vacuum in the expanding Universe: a unified QFT paradigm for Inflation and Dark Energy} \vskip 2mm

 \vskip 8mm

\textbf{Joan Sol\`a Peracaula\footnote{Invited review contribution to the the 40th Anniversary Special Issue of IJMPA and MPLA.}}

\vskip 0.5cm
Departament de F\'isica Qu\`antica i Astrof\'isica, \\
and \\ Institute of Cosmos Sciences,\\ Universitat de Barcelona, \\
Av. Diagonal 647, E-08028 Barcelona, Catalonia, Spain

\vskip0.5cm

E-mail: sola@fqa.ub.edu

 \vskip2mm

\end{center}
\vskip 15mm

\begin{quotation}
\noindent {\large\it \underline{Abstract}}.
\begin{small}
The concordance $\CC$CDM model, based on a rigid $\CC$-term for the entire cosmic history, has been in crisis for a long time. In our expanding Universe, an evolving $\CC$ with the expansion is intuitively much more reasonable, especially if a fundamental theoretical reason can be given for it.  In the running vacuum model (RVM) framework, based on quantum field theory (QFT) in curved spacetime,  quantum effects induce a  vacuum energy density (VED) $\rho_{\rm vac}=\Lambda/(8\pi G)$  which is a function of the Hubble rate $H$ and its time derivatives,  $\rho_{\rm vac}=\rho_{\rm vac}(H, \dot{H},\ddot{H},\dots)$. Currently, $\rho_{\rm vac}$ evolves very slowly with the expansion,
$\delta\rho_{\rm vac}\sim {\cal O}(\mpl ^2 H^2)$, and this fact provides a possible fundamental origin of dark energy (DE), conceived as dynamical vacuum energy. The RVM framework leads to a ``renormalized $\CC$CDM'', an upgraded version of the concordance model  which  can help cure the cosmological tensions. In it, Newton's  $G$ is also evolving, but much more slowly (logarithmically with $H$): $G=G(\ln H)$.  In the very early universe, on the other hand,  the vacuum fluctuations of the quantized matter fields induce higher (even) powers, e.g.  $\sim H^4$, capable of triggering fast inflation in a very short period, in which $H$ is very large and approximately constant.  This is the mechanism of `RVM-inflation'. It does not require an `inflaton' field since inflation is brought about by pure QFT  effects on the dynamical background. It differs from Starobinsky's inflation, where $H$ is never constant.
Furthermore, the dynamics of $\rho_{\rm vac}(H)$ and $G(H)$ can also have implications on the frequently discussed possibility that the fundamental `constants' of Nature can be mildly evolving with the cosmic expansion. Putting things together,  a unified QFT framework of dark energy  and inflation ensues as a realistic theory for the description of the universe as a whole on fundamental grounds. In it,  dynamical VED is predicted and is much welcomed, since it fits in with current DESI measurements, preferring dynamical DE over a rigid $\Lambda$ term.
\end{small}
\end{quotation}
\vskip 5mm

\newpage

\tableofcontents

\newpage

\section{Introduction}\label{sec:introduction}
 An essential component of Einstein's equations is  the cosmological term, $\CC$, usually assumed to be constant, which is  why it is still called the cosmological constant (CC)\cite{Einstein1917}. The reasons adduced by Einstein in 1917 for introducing this term are now only of mere historical interest and, in fact, scientifically flawed with regard to the presumed static model of the universe at that time\footnote{What is not flawed at all is that $\Lambda g_{\mu\nu}$ is a valid term in the field equations, since is allowed by general covariance.  Once this fundamental fact  is recognized, it cannot be withdrawn by fiat. So, as the saying goes,  every cloud has a silver lining. Still,  we are left with $\CC$: but what is it?...}.  Ironically, his reasons for abandoning $\CC$ in 1931\,\cite{Einstein1931},  14 years after its introduction,  were no less flawed, since $\CC$ is no obstacle at all to the expansion of the universe, discovered by Hubble two years before\cite{Hubble:1929ig}.   In addition, as we now know for almost three decades, $\CC$ may provide a timely candidate for what we call the dark energy (DE) of the universe, which is purportedly responsible for the observed speeding up of the cosmological expansion, provided $\CC>0$\,\cite{SupernovaSearchTeam:1998fmf,Planck:2018vyg}.
 The DE in this case would be associated with the `vacuum energy density' (VED), $\rv=\CC/(8\pi G)$, where $G$ is Newton's gravitational coupling. The parameter $\CC$, assumed to be constant throughout cosmic history,  is a fundamental ingredient of the standard or `concordance' $\CC$CDM model of cosmology \cite{peebles:1993,Turner:2022gvw}. Although $\CC$ is nominally the “simplest” candidate, it is anything but simple.  In fact, we know very little about its  nature and physical interpretation. Even though we tend to interpret it in terms of the density of vacuum energy,  we do not really know what that nonmaterial density is. It was  G. Lema\^\i tre in 1934 (when Einstein had already rejected $\CC$)  who first discussed the notion of `primeval atom' and vacuum energy in cosmology \cite{Lemaitre1934} through the mentioned association  $\rv=\CC/(8\pi G)$, but he did not make any connection of this concept with quantum theory and the vacuum fluctuations, despite the many consolidated developments of quantum mechanics existing at that time. This conceptual transition came 34 years later with Y. Zeldovich\cite{Zeldovich:1968}, who realized that any quantum field of nonvanishing mass $m$ should produce a contribution to the VED of order $\sim \hbar\, m^4$, a pure quantum effect (hence proportional to $\hbar$). With this observation, it was made clear that the idea  of $\CC$ (or, more properly, of  $\rv$) as the  ``simplest'' candidate for the DE might not be granted at all, since these quantum contributions are generally many orders of magnitude bigger than the observed value  $\rvo\sim 10^{-47}$ GeV$^4$ (in natural units). This is the so-called `old cosmological constant problem' (CCP)\cite{Weinberg:1988cp}, the biggest theoretical conundrum of all time; in fact, the most fascinating and challenging enigma of all theoretical physics ever. For some of the historical development of the problem (and more), see e.g. \cite{Sola:2013gha}.

 That there is a huge problem worth meditating on here is beyond doubt. Indeed, the typically predicted value for $\rv$ in the standard model (SM) of particle physics is given by the vacuum expectation value of the Higgs potential,  $\langle V_H\rangle \sim G_F^{-1} M_H^2\sim 10^9$GeV$^4$ (with negative sign), where $G_F\simeq1.166\times 10^{-5}$ GeV$^{-2}$ is Fermi's constant and $M_H\simeq 125$ GeV is the measured value of the Higgs boson mass. Hence $|\langle V_H\rangle|$ is  $56$ orders of magnitude larger than the observed vacuum energy density $\rvo$ in cosmology.  There are additional contributions within the SM, such as those arising from a variety of QCD vacuum condensates induced by strong interactions, which should also affect, in principle,  the vacuum energy of the SM (see, however, Refs. \cite{Brodsky:2009zd,Brodsky:2022fqy}), although they are comparatively much smaller. Therefore, paradoxically, the main offending contribution comes from the time-honored Higgs sector of the SM, which represents the ``jewel of the crown'' of the particle physics world. Whereby the bulk of the  CCP ultimately originates from the fundamental concept of spontaneous symmetry breaking (SSB), or Higgs mechanism, on which the very SM architecture is built upon! So, as it happens, the world-wide celebrated success of the Higgs boson finding at CERN in 2012 proved to be a fiasco in the deep conceptual domain of cosmology, since it instantly detonated a `cluster bomb' in the very core of theoretical physics, with myriad consequences on all sectors of particle physics and quantum field theory (QFT). Actually, on top of the fantastic SSB contribution from the Higgs sector, we still have to add the zero-point energies (ZPE's) of all the quantum fields in the SM (and beyond), which are not just crumbs underneath the table. The ZPE ensues from pure quantum effects associated with the vacuum fluctuations in QFT, and  are of order $\pm m_i^4$ for all bosons and fermions (following Zeldovich's estimate). They can be particularly large in some cases, comparable to the SSB part. This is the case with respect to the zero-point energy of  the Higgs particle itself and that of the top quark, which has a huge mass $m_t\simeq 173$ GeV. They have different signs, but there is no chance of cancellation.

This murky state of affairs between  theoretical cosmology and particle physics  defines the dramatic birth of the  modern form of the CCP, namely the experimental confirmation that the electroweak (EW) vacuum energy, and in fact all vacuum effects emerging from the ZPE and the SSB mechanism,  are extremely large compared to the total vacuum energy density that we have measured in the universe. The CCP is usually formulated in even more dramatic terms in the literature by invoking mind-boggling contributions associated with the Planck scale $\mpl\sim 10^{19}$ GeV, which generate effects of order $\mpl^4\sim 10^{76}$ GeV$^4$ on the vacuum energy, hence $\sim 123$ orders of magnitude bigger than the measured $\rvo\sim 10^{-47}$ GeV$^{4}$. However, there is no need to appeal to the Planck scale, we have already a $100\%$ certified `real' problem at `low energies' of order $|\langle V_H\rangle|^{1/4}\sim 100-200$ GeV (fully accessible to our accelerators), where the discrepancy is `only' of 56 orders of magnitude!

A first naive thought may cross our mind: performing a fine tuning among the various contributions. But it is actually impracticable, since it is further aggravated by a tower of quantum corrections from higher order perturbative effects which unavoidably emerge from QFT, including a variety of non-perturbative contributions of various sorts. The CCP is therefore a very hard beast to deal with, in fact indomitable at present\,\cite{SolaPeracaula:2022hpd}. For many, it may be thought of as a theoretical fiction. For a serious theoretical physicist, however,  it is a formidable intellectual challenge. At face value the CCP should be  as “real” as the Higgs boson itself, one cannot exist without the other, which is a very uncomfortable situation! There are virtually as many attempts at solving the CCP as stars are in the sky; for a very incomplete list, see e.g.
\cite{Weinberg:1988cp,Sahni:1999gb,Carroll:2000fy,PeeblesRatra2003,Padmanabhan2003,Nobbenhuis:2004wn,Copeland2006,Aitchison2009,Sola:2013gha,SolaPeracaula:2022hpd,Bauer:2010wj,Burgess:2013ara} and references therein.  Unless one appeals to the Anthropic principle\cite{Weinberg:1987dv},  all known attempts involve (sometimes quite surreptitiously) extreme fine tuning of the parameters. See \cite{Bauer:2010wj} for a vigorous attempt based on an automatic adjustment mechanism. Old examples are \cite{Dolgov:1982gh,Abbott:1984qf,Banks:1984cw,Ford:1987de,Sola:1988nz}; and, in particular,  the well-known `cosmon model' \cite{Peccei:1987mm}, which was further discussed by Weinberg \cite{Weinberg:1988cp} in the context of his `No-go Theorem' (NGT)\footnote{Weinberg's NGT states that no local field theory, in  particular classical gravity,  can have a flat spacetime (Minkowski) solution for generic values of the parameters. Put another way: it forbids the existence of any solution to the CCP within local field theories without fine tuning. The theorem (based on classical gravity) is claimed to be valid also in quantum gravity  (QG) upon using  BRST invariance of the effective action\,\cite{Oda:2017qzs}.}.  
For a milder and thought provoking presentation of the CCP and related matters, more accessible to general readers, see \cite{JSPCosmoverse}.

Despite the difficulties outlined above with the CCP, recent theoretical studies on renormalizing the vacuum energy in curved spacetime, specifically in the context of the running vacuum model (RVM) in Friedmann-Lemaître-Robertson-Walker (FLRW) cosmology, point to new theoretical avenues for ameliorating  the status of the CCP, see \cite{Moreno-Pulido:2020anb,Moreno-Pulido:2022phq,Moreno-Pulido:2022upl,Moreno-Pulido:2023ryo,SolaPeracaula:2022hpd}. The root of the problem lies in the renormalization procedure, which should be more physical. The RVM approach is old in origin; see e.g.\,\cite{Sola:2013gha} and references therein, but it was only after the aforementioned recent works arose that the RVM framework has been put in a solid theoretical status.  Explicit calculations within QFT in FLRW spacetime, which we shall outline in the present review,  demonstrate that the VED is a smooth function $\rv(H)$ of the Hubble rate $H$ (and its time derivatives), and hence it is predicted to be a time-evolving quantity from first principles. The RVM approach is therefore a bold proposal implying a breakthrough in the subject and certainly a change of paradigm beyond the concordance $\CC$CDM model. Moreover, the time evolution of the VED as  predicted by the RVM is as mild as the cosmic expansion itself, since it follows the (currently) slow pace defined by the expansion rate $H$. We shall review all these aspects here; see the aforementioned studies for more details.  Needless to say, a variable VED, $\rv(H)$, also leads to a `variable CC',  $\CC=\CC(H)$,  since the latter is just $8\pi G\rv(H)$. It is worth noticing that the dynamical character of the predicted vacuum energy in the RVM fits in with timely with the recent measurements of the DESI collaboration that favor in a significant way a mild dynamical evolution of the DE \,\cite{DESI:2024mwx,DESI:2024aqx,DESI:2025zgx,DESI:2025fii}.

In fact, the RVM framework was initially aimed at a QFT theory of cosmological evolution. This perspective  was pinpointed long ago using semi-qualitative arguments  based on the renormalization group (RG)\cite{Shapiro:1999zt,ShapSol,Shapiro:2004ch,Sola:2007sv,Shapiro:2009dh,Babic:2001vv,Guberina:2002wt,Babic:2004ev}.  For earlier RG work in cosmology, see \cite{Nelson:1982kt}.  The main purpose\cite{Sola:2011qr,Sola:2013gha} was to provide a dynamical evolution to the $\CC$-term without resorting to \textit{ad hoc} scalar fields (quintessence and the like\cite{PeeblesRatra2003,Padmanabhan2003,Copeland2006,Tsujikawa:2013fta,Amendola:2015ksp,Avsajanishvili:2023jcl}) and without imposing a variety of (no less \textit{ad hoc}) time dependencies on $\CC$, as it was done in the old and extensive class of `$\CC(t)$ scenarios' in the literature, still under scrutiny, see e.g. \cite{Ozer:1985wr,Freese:1986dd,Carvalho:1991ut,Overduin:1998zv,Macedo:2023zrd,Chaudhary:2026qpg,Avsajanishvili:2026fqi}.  A further step in the RVM road was to try to incorporate inflation \cite{Sola:2007sv} from the anomaly-induced effective action \cite{Shapiro:2001rh} using the framework of the  aforementioned `cosmon model'\cite{Peccei:1987mm}. Later on, an alternative unification  of the cosmic history within the RVM pipeline was vindicated on phenomenological grounds \cite{Lima:2013dmf,Perico:2013mna,Basilakos:2013xpa,Lima:2014hia,Lima:2015mca ,Sola:2015rra,Sola:2015csa,Sola:2014tta,Yu2020}.  However, an explicit QFT approach was still missing beyond the mentioned semiqualitative RG ideas. Only very recently a full-fledged QFT formulation was systematically put forward  in\cite{Moreno-Pulido:2020anb,Moreno-Pulido:2022phq,Moreno-Pulido:2022upl,Moreno-Pulido:2023ryo}  and \cite{SolaPeracaula:2025yco,SolaPeracaula:2026trz} which has definitely revamped the RVM framework and put it on solid theoretical  ground.

The practical outcome of the RVM approach\cite{SolaPeracaula:2022hpd} for the current universe is that the standard $\CC$CDM  is promoted into a  `renormalization group improved $\CC$CDM' (denoted $\bar{\CC}$CDM in \cite{Sola:2015rra}),  in which the fundamental cosmological parameters become slightly shifted with respect to the original ones, and moreover the small predicted shifts (quantum corrections) are dynamical with the expansion: $\alpha_i\to \alpha_i+\delta\alpha_i(H)$, with $\alpha_i=\Lambda, \rv, G, g_i, m_i...$. Here,  $g_i$  and $m_i$ stand respectively for the gauge couplings of strong and EW interactions and the particle masses, all of them subject to  cosmic time dependence since they become functions of the Hubble rate $H$.  All parameters are indeed sensitive to quantum effects and thereby become ``effective charges'' in the literal sense of the RG, see \cite{Sola:2013gha,SolaPeracaula:2022hpd} and references therein. As such, their physical (renormalized) values differ only very mildly from the original parameters for the entire (post-inflationary) cosmic history: $\delta\alpha_i(H)\ll \alpha_i$. Parameters closely related to cosmology, such as $\CC$ and $G$,  acquire a theoretical status similar to ordinary gauge couplings, and the small (time-evolving)  shifts $\delta\Lambda(H)$ and $\delta G(H)$ can have fruitful phenomenological implications, as these small renormalization effects may provide a better description of the cosmological data\cite{SolaPeracaula:2021gxi,SolaPeracaula:2023swx,deCruzPerez:2025dni}. In actual fact, the RVM approach is more ambitious since it points to a generalization of the RG in which the gauge couplings $g_i$ (e.g. the electromagnetic charge $e$ and the associated fine structure constant $\alpha_{\rm em}=e^2/4\pi$) now depend on \textit{two} renormalization scales\cite{Fritzsch:2012qc}:  $g_i=g_i(\mu; M)$, namely, one scale (usually denoted $\mu$)  related to the ordinary (high energy) running of the gauge theories of strong and electroweak interactions;  and another one, $M$, which describes the (low-energy) cosmic running of these couplings. The same applies to physical masses $m_i=m_i(\mu;M)$\cite{Fritzsch:2012qc,Fritzsch:2015lua,Fritzsch:2016ewd}. Cosmological parameters such as $\Lambda$ or $G$, instead,  depend only on $M$ in the RVM framework.  As we shall see, $M=M(t)$ is dynamical and is naturally associated with the Hubble rate $H(t)$, as was implicit in the above discussion.

Renormalization effects are small in the present universe, but not in the early universe. In the RVM, inflation is dominated by quartic quantum vacuum effects $\rv(H)\sim H^4$, which are huge in the primeval epoch and negligible at present. In contrast, the current universe (and for that matter the entire post-inflationary era) is dominated by quantum vacuum effects of order $\sim \mpl^2H^2$. Notice that in gravity the running can be power-law, because of the dimensional coupling constants involved. This is at variance with the usual renormalizable gauge field theories of strong and electroweak interactions, where the running is in all cases logarithmic because couplings are dimensionless\footnote{For Newton's $G$, the term in the action is $\sim R/G$ (with $R$ the curvature scalar). The natural dimension of $R$ is $+2$ and that of $G$ is $-2$, which is carried by  $G_N=1/\mpl^2$, the local value of the gravitational coupling, obtained e.g. from Cavendish-like (torsion balance) experiments. Thus the running of $G$ can be dimensionless and in fact is logarithmic (cf. Secs.\,\ref{sec:RGinvariance} and \ref{sec:RunningG}). In stark contrast, the running of $\rv$ is power-law since $\rv$ is additive in the action and has dimension $+4$.}). This is an important distinction that appears when we discuss renormalization in gravity and in particular in the RVM approach that we present here.

The implication that neither $\CC$ nor $G$ are constant but subject to renormalization group running leads to a  new paradigm that, in fact, goes beyond Einstein's theory\cite{Einstein1917} and Brans-Dicke theory\cite{Brans:1961sx} (in which $\CC$ was never considered as a part of the discussion). As noted above, the reach of this paradigm is even more ambitious since it suggests that all of the  `fundamental constants'  of Nature are actually time-evolving with the cosmological expansion\cite{Fritzsch:2012qc}, a possibility which has been (and is being) repeatedly tested by high precision lab and astrophysical experiments, see e.g.\cite{Uzan:2010pm,Chiba:2011bz,Martins:2017yxk,Uzan:2024ded} and references therein.  Not surprisingly, the golden rule in this field says the following:  when one `fundamental constant' varies,  all of them vary!

Hereafter, I will review the RVM as a unified theoretical paradigm for explaining inflation and dynamical dark energy on fundamental  QFT grounds, see also \cite{Sola:2013gha,SolaPeracaula:2022hpd} and \cite{Sola:2011qr,Sola:2014tta,Sola:2015rra}, and references therein. Now, theory without phenomenology is not keen of physics. Fortunately, the RVM has been extensively tested in the literature. The  vacuum dynamics inherent in it improves the quality fits to the cosmological data and provides a possible solution or attenuation of the famous tensions on $H_0$ and $\sigma_8$ that still affect the concordance $\CC$CDM, see \cite{DiValentino:2020zio,DiValentino:2020vvd,Perivolaropoulos:2021jda,Abdalla:2022yfr} and \cite{Vagnozzi:2023nrq,Vagnozzi:2019ezj,CosmoVerseNetwork:2025alb} (and the extensive bibliography cited in these papers). The RVM paradigm provides its own solutions to these problems \cite{SolaPeracaula:2021gxi,SolaPeracaula:2023swx,deCruzPerez:2025dni} due to the dynamical character of the vacuum in that context.  The idea that dark energy could be dynamical has a long pedigree. Already about ten years ago, the  positive impact of dynamical DE was highlighted by several devoted studies which considered  large sets of cosmological data of various kinds; some of these old works involved the RVM and provided remarkable fits to the observations, see particularly \cite{Sola:2015wwa,Sola:2016jky,SolaPeracaula:2016qlq,Sola:2016hn,Sola:2016zeg,Sola:2017znb,SolaPeracaula:2017esw,Sola:2016sbt} and \cite{Gomez-Valent:2017idt,Gomez-Valent:2018nib,SolaPeracaula:2018wwm,Rezaei:2019xwo}. For related analyses involving the RVM, see  e.g. \cite{Basilakos:2009wi,Basilakos:2010rs,Grande:2011xf,Gomez-Valent:2014rxa,Gomez-Valent:2014fda,Gomez-Valent:2015pia,Rezaei:2021qwd}.

In addition, several RVM-inspired studies (including running vacuum extensions of Brans-Dicke theories\cite{SolaPeracaula:2019zsl,SolaPeracaula:2020vpg,deCruzPerez:2023wzd,SolaPeracaula:2018dsw,deCruzPerez:2018cjx}) have produced interesting results to alleviate the cosmological tensions. Let us also mention the $\CC$XCDM \cite{Grande:2006nn}, a composite DE model that incorporates the notion of `phantom matter', a cosmic fluid which recently has proven useful to fit modern cosmological data \cite{Gomez-Valent:2024tdb,Gomez-Valent:2024ejh,Gomez-Valent:2025mfl} -- cf. also the related approach\cite{Akarsu:2023mfb,Akarsu:2019hmw,Akarsu:2022typ}.
Besides, the RVM framework admits effective (low-energy) `stringy' extensions\cite{Mavromatos:2020kzj,PhantomVacuum2021,Gomez-Valent:2023hov} (denoted StRVM), which enrich its theoretical scope and possible phenomenological implications\cite{BasMavSol,basilakos2, basilakos3,NickPhiloTrans,Dorlis2024,Dorlis2024b,Dorlis:2025zzz,Dorlis:2025amf,
Mavromatos:2026juq}. Finally, the RVM behavior can be mimicked by other gravitational frameworks, including $f(T)$ theories \cite{Cai:2015emx,Errahmani:2025vde},  entropic-force and holography\,\cite{Basilakos:2012ra,Basilakos:2014tha,Komatsu:2013qia,Rezaei:2022bkb,Gohar:2025huk,Gohar:2023lta}, finite-temperature renormalization effects\,\cite{Park:2024kfn,Hatefi:2025lqe} and lattice quantum gravity\,\cite{Dai:2024vjc}.
As indicated, it might even provide an explanation  for the time variation of the fundamental constants of Nature\,\cite{Fritzsch:2012qc,Fritzsch:2015lua,Fritzsch:2016ewd}. For an in depth exposition of the RVM in both QFT and  stringy StRVM formulations, see \cite{NickJoan_PR}.

The flow chart of this review is as follows. In Sec. \ref{sec:RVM-QFT}  the basics of the QFT approach to the RVM is presented, in particular the off-shell adiabatic renormalization procedure and the calculation of the $\beta$-function of the running VED. In Sec.\,\ref{sec:effAction} we consider the effective action approach as an alternative method.  In Sec. \ref{sec:RVMinflation} we study the different aspects of RVM-inflation and also briefly compare with de Sitter vacuum decay. In Sec.\,\ref{eq:RVMPheno} we review some phenomenological aspects for the present universe; in particular,
the equation of state (EoS) of the RVM and the running of $\rv$ and $G$. We also present a solution to the entropy problem. Sec.\,\ref{eq:GeneralRVM} shows how the RVM could explain  the possible cosmic drift of the fundamental  `constants' of Nature.  In the last section, we recapitulate.  Finally, in three appendices, we illustrate the method of calculating the off-shell adiabatic expansion and its connection with the effective action, and revisit the quantum vacuum in QED.
In this paper, natural units ($\hbar=c=1$) will be used throughout, except in special cases.


\section{Running vacuum framework: quantum field theory aspects}\label{sec:RVM-QFT}

In what follows, I will try to summarize the recent theoretical developments setting the foundations of the RVM, a framework aiming to describe the entire cosmological expansion from a  fundamental QFT point of view. Full-fledged technical details can be found in  \cite{Moreno-Pulido:2020anb,Moreno-Pulido:2022phq,Moreno-Pulido:2022upl,Moreno-Pulido:2023ryo}.  For earlier  reviews, see also \cite{Sola:2013gha,SolaPeracaula:2022hpd}. For more recent developments along the lines of seeking a unified QFT theory of dynamical dark energy and inflation inspired by the RVM framework can be found in \cite{SolaPeracaula:2025yco,SolaPeracaula:2026trz}.

\subsection{Classical versus quantum field theory and quantum gravity}\label{sec:cTQFTQG}

As can be expected, QFT calculations in cosmology are generally rather cumbersome.  The main difficulty is that we have to deal with  renormalization in curved spacetime, and this is more demanding than QFT in Minkowski spacetime.  A consequence of this fact is that apart from  the ultraviolet (UV) divergent integrals of the type we usually encounter in flat spacetime calculations, in the present context,  we meet a new type of UV-divergent integrals whose existence is inherent to the spacetime curvature, meaning that with zero curvature these contributions would not be present.

This is not a minor issue, since it can have a direct bearing on the CCP. In this kind of subtle problem,  every bit counts and we cannot afford to introduce any level of noise into the result.  For example, conventional approaches to the CCP within QFT in flat spacetime based on, say, noncovariant cutoff procedures or even the minimal subtraction scheme (MSS)\cite{Collins84,Brown:1992db,Peskin1995} are unsuitable, as they lack a direct physical meaning and end up introducing some level of arbitrariness in the results and also involving serious fine tuning among the parameters \cite{SolaPeracaula:2022hpd}.  Thus, despite the many valuable attempts in the literature discussing the computation of the vacuum energy in Minkowski spacetime and the corresponding implications in cosmology (sometimes even involving considerations on the running of the gravitational quantities) we deem this is not a realistic pathway to approach the problem. For different considerations on these matters showing the limitations of this kind of approach, see, e.g., \cite{Shapiro:1999zt,ShapSol,Shapiro:2004ch,Sola:2007sv,Shapiro:2009dh,Babic:2001vv,Guberina:2002wt,Babic:2004ev} and \cite{Brown:1992db,Akhmedov:2002ts,Ossola:2003ku,Martin:2012bt,Koksma:2011cq,BennieW2013,Visser:2016mtr,Antipin:2017pbt,Donoghue:2020hoh}.  We have to admit that discussing the CCP in QFT in flat spacetime is a bit of a cheat. It  has been over-exploited in the literature and should finally be abandoned, as the vacuum energy in QFT in flat spacetime (if there is any vacuum energy at all in this case, see Sec.\,\ref{sec:VEDMinkowski} and  Appendix \ref{sec:appendixC})  is completely unrelated to the cosmological constant. There is no mutual implication between the two concepts. The logical link usually works only one-way:  $\CC\neq 0$ implies the existence of  $\rv\neq 0$ in curved spacetime,  but $\rv\neq 0$ does not always imply the existence of $\CC$; in fact, it never implies it in flat spacetime, since $\CC$ cannot be defined in it, except if $\CC=0$, as only then Minkowski spacetime can be considered  a solution of Einstein's equations.

To properly discuss the  physical running of the cosmological quantities requires us to consider background curvature effects from the very beginning, possibly including considerations into the QG arena\,\cite{Reuter:1996cp,Bonanno:2001xi, Donoghue:2019clr,Donoghue:2024uay,Kawai:2025wkp,Christensen:1979iy}. We shall, however, not deal with QG here, not even in the context of effective field theory\cite{Donoghue:1994dn} or in the quadratic gravity framework\,\cite{Stelle:1976gc} where (it is claimed) that perturbative QFT may consistently be formulated within the strict renormalizability criterion --  see e.g. \cite{Buoninfante:2025dgy} and references therein. The bare reality is that we are still far away from achieving a proficient knowledge of QG as a ultimate theory of gravity, provided it exists at all. For, it is not even clear whether QG is necessary since gravity could be just an `emergent phenomenon'\,\cite{Jacobson:1995ab,Padmanabhan:2003gd,Padmanabhan:2007tm}. Therefore, we will not quantize the gravitational field here, only the matter fields. This means that we will work in a semiclassical formalism, i.e. in the context of QFT in curved spacetime, where the metric acts as an external field and is not subject to the quantization process, which exclusively affects the matter fields\cite{Birrell:1982ix,Fulling89,Parker:2009uva,MukhanovWinitzki07}.  We know how to apply this approach in a safe way: if  only `small' fluctuations of the `vacuum EMT', i.e. of $\langle T_{\mu\nu}\rangle\equiv \langle 0| T_{\mu\nu}|0\rangle $,  are involved, the semiclassical approach leads to a perfectly consistent and renormalizable picture of quantum matter fields in a curved classical background (see \cite{Kuo:1993if} for a precise definition of `small' in this context). For a variety of studies related to the VED in the semiclassical approach, see e.g.\cite{Antoniadis:2006wq,Bilic2011,Capozziello:2011et,Mottola2022} and references therein.

We adopt the adiabatic renormalization prescription (ARP), which is very convenient in cosmological spacetimes for which the comoving angular frequency $\Omega_k$ is slowly varying at early and late times (cf. Sec.\,\ref{sec:AdiabaticVacuum}), so that an expansion of the solution in terms of higher order adiabatic corrections is asymptotically convergent (meaning, as usual, only up to a finite order, as in the case of a WKB series). This method is particularly adequate for the removal of divergences in the vacuum expectation value (VEV) of the energy-momentum tensor (EMT) of a quantized field propagating in the FLRW background. The conventional formulation of the ARP is well known and we refer the reader once again to classic textbooks\, \cite{Birrell:1982ix,Fulling89,Parker:2009uva,MukhanovWinitzki07} (and comprehensive reports \cite{DeWitt1975,ParkerCargese1978})  on the subject,  and references therein. However, the RVM approach makes one further step in the adiabatic procedure.  It offers a genuinely new formulation of this method that relies on an off-shell renormalization scale, $M$. Such a generalization (``off-shell ARP'' \cite{Moreno-Pulido:2020anb,Moreno-Pulido:2022phq,Moreno-Pulido:2022upl,Moreno-Pulido:2023ryo})  is crucial to enable us to explore the cosmological evolution of the vacuum energy and in general of the VEV of the EMT, $\langle T_{\mu\nu}\rangle$ (a quantity which we already called the `vacuum EMT' for short)  at any cosmic epoch.  Calculations in the RVM therefore will depend on a floating scale $M$\footnote{We prefer to denote this renormalization scale as $M$ rather than $\mu$  to emphasize that our renormalization scheme is not MSS with dimensional regularization (DR)\,\cite{Collins84,Brown:1992db,Peskin1995}. Notwithstanding, DR can still be used here as an auxiliary tool for intermediate calculations related to regularization, but not with renormalization. All our final results will be independent of $\mu$; yet the renormalized VED in the off-shell ARP will still depend on $M$, as it should since it is only the full renormalized effective action (of which the vacuum effective action is only one part)  which must be overall independent of $M$; not so the renormalized vacuum action taken in isolation\,\cite{Moreno-Pulido:2020anb,Moreno-Pulido:2022phq} -- see Sec.\ref{sec:effAction}.}. This is after all customary in QFT and especially within the context of the RG. At the end we have to make contact with physics and we associate the renormalization point  $M$ with the value of the Hubble rate $H$ corresponding to the cosmic epoch under study, since $H$ is indeed the characteristic energy scale (in natural units)  of cosmological spacetime at any  given epoch; and $H^2$ provides a measure of its four-dimensional curvature in the considered epoch. Notice that such a `scale setting' is in consonance with the usual practice in particle physics based on selecting the RG scale near the typical energy of the process. In our case,  the `process'  is nothing but the cosmic expansion of the universe at a given epoch. So, for FLRW spacetime, the scale setting $M=H$ looks entirely natural.

The resulting expansion dynamics of the VED that emerges from the RVM is tantamount to saying that the physical cosmological term $\CC(H)=8\pi G(H)\rv(H)$ becomes a running quantity with the expansion, where even the Newton's coupling may also develop some small evolution. In this context, therefore, $\CC$ no longer appears as a true `cosmological constant'.  The implications of this approach are far reaching, as indeed the internal consistency of the theory (e.g. with the Bianchi identity in Einstein's equations) leads also to the cosmic evolution of the gravitational `constant', $G=G(H)$,  and in general of all the so-called fundamental `constants' of Nature (cf. Sec.\,\ref{eq:GeneralRVM}). This connection of the RVM framework with a potential variation of the fundamental constants was first pointed out in \cite{Fritzsch:2012qc}.

Let us start from the classical Einstein-Hilbert (EH) action for gravity plus matter\footnote{Conventions:  ${\rm sign} (g_{\mu\nu})=(-, +,+,+ )$; Riemann tensor
$R^\lambda_{\,\,\,\,\mu \nu \sigma} = \partial_\nu \, \Gamma^\lambda_{\,\,\sigma\mu} + \Gamma^\lambda_{\,\, \nu\rho}\Gamma^\rho_{\,\, \sigma\mu} \,  - (\nu \leftrightarrow \sigma)$; Ricci tensor $R_{\mu\nu} = R^\lambda_{\,\,\,\,\mu \lambda \nu}$; and Ricci scalar  $R = g^{\mu\nu} R_{\mu\nu}$. Overall, these conventions correspond to  $(+, +, +)$ in the classification by Misner-Thorn-Wheeler\,\cite{MTW}.}:
\begin{equation}\label{eq:EH}
\begin{split}
S_{\rm EH+m} &= \frac{1}{16\pi G}\int d^4 x \sqrt{-g}\, (R  - 2\CC) + S_{\rm m}\\
&=\frac{1}{16\pi G}\int d^4 x \sqrt{-g}\, R  -  \int d^4 x \sqrt{-g}\, \rL+ S_{\rm m}\,,
\end{split}
\end{equation}
where the quantity
\begin{equation}\label{eq:rLCC}
\rL=\frac{\CC}{8\pi G}
\end{equation}
has dimension of energy density.
At the moment,  for us the term $\rL$  is  just a bare parameter of the EH action, as $G$ itself. In a QFT context, they cannot be called at this point vacuum energy and Newton's constant, respectively. In fact, in the presence of quantum fluctuations the physical values of the various parameters and quantities cannot be identified at the level of the classical action;  we need to go to the quantum level (i.e. we need to quantize the matter fields) and renormalize the theory. Only at this point we will be in position to speak of physical quantities. However, once renormalization is implemented, an analogous relation to \eqref{eq:rLCC}  holds for the renormalized quantities.
\begin{equation}\label{eq:rvCC}
\rL(M)=\frac{\CC(M)}{8\pi G(M)}\,,
\end{equation}
where $M$ is the renormalization scale, which, as indicated, will eventually be identified with $H$ in cosmological spacetime. However, not even the renormalized $\rL(M)$ is to be naively  associated with the  physical VED since the vacuum fluctuations contribute also to the vacuum energy, of course, and therefore we have to carefully distinguish between $\rL(M)$ and the renormalized VED value $\rv(M)$, which we will define later on.

For illustrative purposes it will suffice to consider  the semiclassical calculation of the energy density of the vacuum fluctuations for a single quantized, neutral,  scalar field $\phi$ non-minimally coupled to the scalar curvature of FLRW spacetime. This field plays the role of our matter component and will be used to define the matter action $S_{\rm m}$.  Therefore, the matter action associated with $\phi$ takes on the form
\begin{equation}\label{eq:Sphi}
  S[\phi]=-\int d^4x \sqrt{-g}\left(\frac{1}{2}g^{\mu \nu}\partial_{\mu} \phi \partial_{\nu} \phi+\frac{1}{2}(m^2+\xi R)\phi^2 \right)\,.
\end{equation}
For $\xi=0$ the matter field $\phi$ is said to be only minimally coupled to gravity, but for $\xi\neq 0$ it is non-minimally coupled. The particular value $\xi=1/6$ together with the massless case ($m=0$) makes the action conformally invariant. We are not interested in this case and we will keep the non-minimal coupling $\xi$ general. We shall see that it is phenomenologically  fruitful to proceed in this way.

Since we are targeting pure quantum effects, viz. the zero-point energy (ZPE) in curved spacetime,  we shall not contemplate  a  contribution to the VED  from an effective potential for $\phi$, which could also be there, of course. The calculation is already sufficiently cumbersome without including the potential and its quantum effects, so we shall not address  this part here. However, later on we will briefly comment on the possible impact of the effective potential in our discussion and argue that nothing fundamental can change in our main qualitative conclusions.

At the moment, we remain in the classical domain. On vaying the action  \eqref{eq:EH},  first with respect to the metric,  it yields Einstein's field equations:
\begin{equation} \label{EinsteinEqs}
\frac{1}{8\pi G}G_{\mu \nu}=-\rho_\Lambda g_{\mu \nu}+T_{\mu \nu}^{\phi}\,,
\end{equation}
where  $G_{\mu\nu}=R_{\mu\nu}-(1/2) g_{\mu\nu} R$  is the usual Einstein tensor and $T_{\mu \nu}^{\phi}$ is the  energy-momentum tensor (EMT) corresponding to $\phi$.  The above are classical field equations and hence do not incorporate quantum effects yet. This will be done upon quantization in the next section.

The classical  EMT for $\phi$ is obtained from the action \eqref{eq:Sphi} in the usual way:
\begin{equation}\label{EMTScalarField}
\begin{split}
T_{\mu \nu}^{\phi}=&-\frac{2}{\sqrt{-g}}\frac{\delta S[\phi]}{\delta g^{\mu\nu}}= (1-2\xi) \partial_\mu \phi \partial_\nu\phi+\left(2\xi-\frac{1}{2} \right)g_{\mu \nu}\partial^\sigma \phi \partial_\sigma\phi\\
& -2\xi \phi \nabla_\mu \nabla_\nu \phi+2\xi g_{\mu \nu }\phi \Box \phi +\xi G_{\mu \nu}\phi^2-\frac{1}{2}m^2 g_{\mu \nu} \phi^2.
\end{split}
\end{equation}
On the other hand, varying the action \eqref{eq:Sphi} with respect to $\phi$ yields the  Klein-Gordon (KG) equation in curved spacetime:
$(\Box-m^2-\xi R)\,\phi=0\,,$
where $\Box\phi=g^{\mu\nu}\nabla_\mu\nabla_\nu\phi=(-g)^{-1/2}\partial_\mu\left(\sqrt{-g}\, g^{\mu\nu}\partial_\nu\phi\right)$.

As announced, we shall perform our QFT calculations in the background spacetime defined by the FLRW line element. We choose  spatially flat three-dimensional hypersurfaces since this is the canonical option. Furthermore, we use the conformally flat form of the metric:
\begin{equation}\label{eq:FLRWflatmetric}
ds^2=a^2(\tau)\eta_{\mu\nu}dx^\mu dx^\nu\,,
\end{equation}
where  $\eta_{\mu\nu}={\rm diag} (-1, +1, +1, +1)$ is the Minkowski  metric in our conventions,  and $a(\tau)$ is the scale factor as a function of the conformal time $\tau$. Recall that $dt= a(\tau) d\tau$.  Primes will denote derivatives with respect to $\tau$   ($^\prime\equiv d/d\tau$). The  Hubble rate in conformal time  is $\mathcal{H}(\tau)\equiv a^\prime /a$. Even though our calculations will be performed using the conformally flat metric, the final results will be delivered in terms of the usual Hubble function  $H(t)=\dot{a}/a$ ( $\dot{}\equiv d/dt$) in cosmic time.  Useful relations for the conversion are $\mathcal{H}=a H$, $\mathcal{H}^\prime=a^2(H^2+\dot{H})$,
$\mathcal{H}^{\prime\prime}=a^3\left(2H^3+4 H\dot{H}+\ddot{H}\right)$, etc., which will be used frequently in our calculations.

\subsection{Quantum effects and mode functions}\label{sec:AdiabaticVacuum}

Obviously, to further advance in the study of the vacuum effects, we must move from the above classical field theory treatment to quantum field theory, which is based on the vacuum action as a most fundamental field theory concept. Thanks to it we know that the vacuum is sort of ``boiling'' all the time with quantum fluctuations, which must be taken into account by means of an appropriate renormalization program. In our semiclassical treatment, we consider only the quantum fluctuations of the matter fields, in this case  the scalar field $\phi$,  around a classical background value $\phi_b$:
\begin{equation}
\phi(\tau,x)=\phi_b(\tau)+\delta\phi (\tau,x)\,. \label{ExpansionField}
\end{equation}
The background field $\phi_b (\tau)$ is assumed to be spatially homogeneous. Not so  the fluctuating part, which depends both on time and the spatial coordinates, $\delta\phi (\tau,x)$. It obviously satisfies the  curved spacetime KG equation  and hence it can be decomposed in Fourier frequency modes. Changing the field variable from $\phi$  to $\varphi=a\phi$ and denoting $h_k(\tau)$ the frequency modes of the fluctuating part, we can write its Fourier decomposition\cite{Moreno-Pulido:2020anb,Moreno-Pulido:2022phq}:
\begin{equation}\label{FourierModesFluc}
\delta \varphi(\tau,{\bf x})=\int \frac{d^3{k}}{(2\pi)^{3/2}} \left[ A_\bk e^{i{\bf k\cdot x}} h_k(\tau)+A_\bk^\dagger e^{-i{\bf k\cdot x}} h_k^*(\tau) \right]\,.
\end{equation}
In this expression,  $A_\bk^\dagger $ and  $A_\bk$  are  the (time-independent) Fourier coefficients.  Upon quantization they become promoted, respectively,  to creation and annihilation
operators and. satisfy the usual commutation relations:
\begin{equation}
[A_\bk, A_\bk'^\dagger]=\delta({\bf k}-{\bf k'}), \qquad [A_\bk,A_ \bk']=0\,. \label{CommutationRelation}
\end{equation}
The above  Fourier expansion then provides  the solution of the field equation in the Heisenberg representation.
Inserting such an expansion into the KG equation, we can see that the frequency modes of the fluctuations, $h_k(\tau)$, satisfy the  differential equation
\begin{equation}\label{eq:ODEmodefunctions}
h_k^{\prime \prime}+\Omega_k^2(\tau) h_k=0\,, \ \ \ \ \ \ \ \ \ \ \Omega_k^2(\tau) \equiv\omega_k^2(m)+a^2\, (\xi-1/6)R\,,
\end{equation}
where $\Omega_k(\tau)$ are the comoving angular frequencies of these modes.
Except in simple cases, finding a solution of the above (linear) equation requires an approximation method, typically a recursive self-consistent iteration procedure such as the  WKB expansion.  The procedure is conveniently initialized with a change of variable defined by  the phase integral
\begin{equation}\label{eq:phaseIntegral}
h_k(\tau)=\frac{1}{\sqrt{2W_k(\tau)}}\exp\left(i\int^\tau W_k(\tilde{\tau})d\tilde{\tau} \right)\,.
\end{equation}
The normalization factor ensures that it satisfies  the so-called  Wronskian condition
$ h_k^\prime h_k^* -  h_k^{} h_k^{*\prime}=i\,,$ 
which is necessary for the quantum field to satisfy the canonical commutation relations, given the corresponding relations \eqref{CommutationRelation} for the creation and annihilation operators.
The functions $W_k$ (which play the role of effective frequencies)  obey  a non-linear differential equation, which is nothing more than the standard WKB equation:
\begin{equation} \label{WKBIteration}
W_k^2(\tau)=\Omega_k^2(\tau) -\frac{1}{2}\frac{W_k^{\prime \prime}}{W_k}+\frac{3}{4}\left( \frac{W_k^\prime}{W_k}\right)^2\,.
\end{equation}
Since the exact vacuum cannot be obtained in this approximation,  a convenient vacuum state that makes sense out of these calculations is the so-called adiabatic vacuum\cite{Birrell:1982ix,Fulling89,Parker:2009uva}. In a way it is an analog of the geometrical optical limit\cite{Moreno-Pulido:2022phq} in that it involves only short wave lengths  and weak gravitational fields. As noted, the adiabatic approach proves useful for cosmological spacetimes for which the comoving angular frequency $\Omega_k$, defined in Eq.\,\eqref{eq:ODEmodefunctions}, slowly varies at early and late times. In this way  an expansion of the solution in adiabatic corrections (carrying a growing number of derivatives with respect to cosmic or conformal time) is asymptotically convergent; as usual, this means convergent only up to a finite order, which is characteristic of a WKB series. If the adiabaticity condition is not minimally met,  a particle physics interpretation becomes hard in quantum field theory in curved spacetime. But if it holds, the Fourier modes of the field  inside the horizon admit a particle description at asymptotic times with the help of the adiabatic vacuum. Worth noticing, the adiabatic behavior can be  useful, e.g. to estimate particle production in a time-dependent background.  In particular, it helps to estimate the gravitational generation of cosmological relics in the very early universe, see e.g.\cite{Ford:2021syk,Kolb:2023ydq}.

By adopting the adiabatic vacuum as an approximation to the true vacuum, the WKB solution of our field equation can be written as
\begin{equation}\label{WKB}
W_k=\omega_k^{(0)}+\omega_k^{(2)}+\omega_k^{(4)}+\omega_k^{(6)}\cdots
\end{equation}
in which the superscript denotes the adiabatic order. If the expansion is performed up to the $N$th adiabatic order, the vacuum state annihilated by all ladder operators $A_\bk$ satisfying \eqref{CommutationRelation} is known as the $N$th-order adiabatic vacuum.
Because we have to preserve the adiabaticity condition, the above expansion  is meaningful only if the gravitational field is not strong and if changes of the physical quantities are slow (adiabatic).

As could be expected, the counting of adiabatic orders in the WKB expansion follows the number of time derivatives.
Notice that general covariance requires the presence of only even adiabatic orders.  The  $\omega_k^{(j)}$ can be expressed in terms of $\Omega_k(\tau)$ and its time derivatives.
Following \cite{FerreiroNavarroSalas2019,Moreno-Pulido:2020anb,Moreno-Pulido:2022phq} we consider an off-shell procedure  in which the frequency $\omega_k$ of a given mode  is defined not at the mass shell, $m$, of the particle but at an arbitrary mass scale $M$:
\be\label{eq:omegaM}
\omega_k\equiv\omega_k(\tau, M)\equiv \sqrt{k^2+a^2(\tau) M^2}\,.
\ee
At zeroth order, $\omega_k^{(0)}= \omega_k$ in Eq.\,\eqref{WKB}. Although the particle mass is $m$ (on-shell value), the arbitrary mass scale $M$ (off-shell value) plays the role of floating RG scale (as previously noticed). This is characteristic of renormalization theory, but the floating mass scale can be introduced in different ways, and in some cases its physical interpretation can be more direct. As we shall see, our off-shell adiabatic renormalization prescription (ARP) enables to describe the cosmological expansion history with the help of $M$, which is identified with the Hubble rate (at the end of the calculation). In this way we can track the running of the VED with $H$ and hence also of  the cosmological `constant' $\CC$. This interpretation was first put forward on firm grounds in  \cite{Moreno-Pulido:2020anb,Moreno-Pulido:2022phq} and was used to systematically establish the off-shell ARP procedure to renormalize the EMT in cosmological spacetime (cf. Sec. \ref{sec:RenormEMT}).  The difference between the on-shell squared value $m^2$ and the square of the RG scale $M^2$ (both of them being of zero adiabatic order) generates a quantity $\Delta^2\equiv m^2-M^2$ which must be counted as being of adiabatic order 2 since $\Delta^2$ appears in the WKB expansion along with other terms of the same adiabatic order.
This can be seen by working out the second and  fourth order terms of \eqref{WKB}\footnote{See Appendix \ref{sec:appendixA}  for a summary of the calculation method of the off-shall adiabatic expansion terms \eqref{WKBexpansions} up to fourth order. One can provide a more formal approach in the context of the effective action/heat-kernel computation of the VED, as we shall see in Sec. \ref{sec:effAction}.}.
\begin{equation}
\begin{split}
\omega_k^{(0)}&= \omega_k\,,\\
\omega_k^{(2)}&= \frac{a^2 \Delta^2}{2\omega_k}+\frac{a^2 R}{2\omega_k}(\xi-1/6)-\frac{\omega_k^{\prime \prime}}{4\omega_k^2}+\frac{3\omega_k^{\prime 2}}{8\omega_k^3}\,,\\
\omega_k^{(4)}&=-\frac{1}{2\omega_k}\left(\omega_k^{(2)}\right)^2+\frac{\omega_k^{(2)}\omega_k^{\prime \prime}}{4\omega_k^3}-\frac{\omega_k^{(2)\prime\prime}}{4\omega_k^2}-\frac{3\omega_k^{(2)}\omega_k^{\prime 2}}{4\omega_k^4}+\frac{3\omega_k^\prime \omega_k^{(2)\prime}}{4\omega_k^3}\,.
\end{split}\label{WKBexpansions}
\end{equation}
As it is transparent, $\Delta ^2$ in the first term of $\omega_k^{(2)}$ appears along with other quantities  which are manifestly of adiabatic order $2$, such as $R$, $\omega_k^{\prime \prime}$ and $\omega_k^{\prime 2}$ (obviously $\omega_k$ is of order zero). The on-shell result is recovered for $M=m$, for which  $\Delta = 0$ and corresponds to the usual ARP procedure\,\cite{Birrell:1982ix,Parker:2009uva}.    It is easy to see that the adiabatic expansion becomes an expansion in powers of $\mathcal{H}$ and its time derivatives, e.g.
\begin{equation}\label{omegak0}
\omega_k^\prime=a^2\mathcal{H}\frac{M^2}{\omega_k}, \qquad\omega_k^{\prime \prime}=2a^2\mathcal{H}^2\frac{M^2}{\omega_k}+a^2\mathcal{H}^\prime \frac{M^2}{\omega_k}-a^4\mathcal{H}^2\frac{M^4}{\omega_k^3}\,.
\end{equation}
These expressions indeed confirm that the adiabatic series is an expansion in the number of time derivatives of the scale factor, and amounts to an expansion in powers of the Hubble rate and its time derivatives.  The terms with more and more time derivatives should contribute less and less. The final result appears as an expansion in even powers of $H$ and also odd derivatives of it, all these terms being of even adiabatic order. In fact, as warned before,  this is what we should expect since the result must be consistent with the covariance of the effective action, which can only admit terms with an even number of time derivatives with respect to the scale factor.

Let us now proceed to the identification of the vacuum expectation value (VEV) of the EMT, i.e. what we  call the `vacuum EMT'  $\langle T_{\mu \nu}^{\delta \phi}\rangle$,  which involves the effect of the fluctuations $\delta\phi$ of the matter fields. To start with, the VEV of the field $\phi$  is connected with the background value,  $\langle 0 | \phi (\tau, x) | 0\rangle=\phi_b (\tau)$, which can be zero or not (e.g. in the presence of SSB). The VEV of the fluctuation is, however,  always zero:  $\langle  \delta\phi  \rangle\equiv \langle 0 | \delta\phi | 0\rangle =0$, since $A_\bk |0\rangle=0\ (\forall {\bf k})$. In contrast, the VEV of the bilinear products of fluctuations is generally non-zero, $\langle (\delta\phi)^2 \rangle\neq0$,  and it leads to UV-divergences since it entails the product of distributions at the same spacetime point. This is ultimately the need for renormalization in QFT, of course,  Now, it is natural to  decompose  $\langle T_{\mu \nu}^\phi \rangle=\langle T_{\mu \nu}^{\phi_b} \rangle+\langle T_{\mu \nu}^{\delta\phi}\rangle$, where
$\langle T_{\mu \nu}^{\phi_{b}} \rangle =T_{\mu \nu}^{\phi_{b}} $
is the effect from the classical background part and $\langle T_{\mu \nu}^{\delta\phi}\rangle\equiv \langle 0 | T_{\mu \nu}^{\delta\phi}| 0\rangle$ is the sought-for vacuum contribution from the field fluctuations. The latter is precisely what we have just called the vacuum EMT, which is a pure quantum effect associated with the vacuum fluctuations of the matter fields. Its $00$th component $\langle T_{00}^{\delta\phi}\rangle$ defines the `zero-point energy' (ZPE) of the quantum field $\phi$.  However, in the absence of these pure quantum effects, we would still have the term  $\rho_\Lambda$ in the EH action \eqref{eq:EH}, which defines the naive classical vacuum term (technically, a bare term in the QFT contex). Therefore,  the `full vacuum EMT' is  the sum of these two terms:
\begin{equation}\label{EMTvacuum}
\langle T_{\mu \nu}^{\rm vac} \rangle=-\rho_\Lambda g_{\mu \nu}+\langle T_{\mu \nu}^{\delta \phi}\rangle\,.
\end{equation}
We note here that after we define $\rv$  in Sec.\,\ref{sec:RenormEMT},  the above expression generally does \textit{not} imply the relation $\langle T_{\mu\nu}^{\rm vac}\rangle=-g_{\mu\nu} \rv$, which is quite often presumed in the literature as if it were self-evident. However, it is not, and we shall see why in Sec.\,\ref{sec:RenVacuumPressure}.  The  formula above, however,  shows in a manifest way that the vacuum (and hence eventually the physical CC) receives contributions not only from  the `cosmological term' in the EH action but also from the quantum fluctuations of the EMT. Nevertheless, these quantities are both formally UV-divergent since $\rL$ is a bare parameter at this point, and,  as previously noted, the vacuum EMT $\langle T_{\mu \nu}^{\delta \phi}\rangle$  involves products of distributions (see next section), which are not generally distributions themselves, so the physical (finite) vacuum energy density $\rv$ cannot be identified at this instance, but only a posteriori upon suitable renormalization.

\subsection{Zero-point energy in curved spacetime versus flat spacetime}\label{eq:RegZPE}

As noted previously, the ZPE associated to the quantum vacuum is given by the  $00$th-component of the fluctuating part of the EMT.  In curved spacetime with FLRW metric, this follows from $\phi\to\phi+\delta\phi$ in \eqref{EMTScalarField} and selecting the quadratic terms of the corresponding VEV. The unrenormalized ZPE is then given by
\begin{equation}\label{EMTInTermsOfDeltaPhi}
\begin{split}
\langle T_{00}^{\delta \phi}\rangle =&\left\langle \frac{1}{2}\left(\delta\phi^{\prime}\right)^2+\left(\frac{1}{2}-2\xi\right)\left(\nabla\delta \phi\right)^2+6\xi\mathcal{H}\delta \phi \delta \phi^\prime\right.\\
&\left.-2\xi\delta\phi\,\nabla^2\delta\phi+3\xi\mathcal{H}^2\delta\phi^2+\frac{a^2m^2}{2}(\delta\phi)^2 \right\rangle\,.
\end{split}
\end{equation}
Here   $\left(\delta\phi^{\prime}\right)^2\equiv\left(\delta\partial_0\phi\right)^2= \left(\partial_0\delta\phi\right)^2$ stands for the fluctuation of the field derivative with respect to conformal time.  Substituting the Fourier expansion of $\delta\phi=\delta\varphi/a$  as given in \eqref{FourierModesFluc}   into Eq.\,\eqref{EMTInTermsOfDeltaPhi}, using the commutation relations and then symmetrizing  the operator field products $\delta\phi \delta\phi^\prime$ with respect to the creation and annihilation operators,  one obtains the final result in Fourier space\cite{Moreno-Pulido:2020anb,Moreno-Pulido:2022phq}.
We may recast it in terms of the field modes $h_k$:
\begin{equation}\label{eq:vacuumEMT}
        \langle T_{00}^{\delta\phi}\rangle=\frac{1}{4\pi^2 a^2} \int dk k^2 \left[ |h_k'|^2
        +(\omega_k^2+a^2\Delta^2) |h_k|^2\right.
        \left.+\left( \xi-\frac{1}{6} \right)\left( -6\calH^2 |h_k|^2+6\calH(h_k'h_k^*+{h_k^*}'h_k)\right) \right]\,.
\end{equation}
The modes $h_k$ are determined from  Eq.\,\eqref{eq:phaseIntegral} using the adiabatic expansion defined in the previous section, which up to 4th order is given by \eqref{WKBexpansions}. Upon inserting the result in Eq.\eqref{eq:vacuumEMT} we find the full form of the unrenormalized ZPE:
\begin{equation*}
\begin{split}
\langle T_{00}^{\delta \phi} \rangle^{ (0-4)} & =\frac{1}{8\pi^2 a^2}\int dk k^2 \left[ 2\omega_k+\frac{a^4M^4 \mathcal{H}^2}{4\omega_k^5}-\frac{a^4 M^4}{16 \omega_k^7}(2\mathcal{H}^{\prime\prime}\mathcal{H}-\mathcal{H}^{\prime 2}+8 \mathcal{H}^\prime \mathcal{H}^2+4\mathcal{H}^4)\right.\\
&+\frac{7a^6 M^6}{8 \omega_k^9}(\mathcal{H}^\prime \mathcal{H}^2+2\mathcal{H}^4) -\frac{105 a^8 M^8 \mathcal{H}^4}{64 \omega_k^{11}}\\
&+\left(\xi-\frac{1}{6}\right)\left(-\frac{6\mathcal{H}^2}{\omega_k}-\frac{6 a^2 M^2\mathcal{H}^2}{\omega_k^3}+\frac{a^2 M^2}{2\omega_k^5}(6\mathcal{H}^{\prime \prime}\mathcal{H}-3\mathcal{H}^{\prime 2}+12\mathcal{H}^\prime \mathcal{H}^2)\right. \\
& \left. -\frac{a^4 M^4}{8\omega_k^7}(120 \mathcal{H}^\prime \mathcal{H}^2 +210 \mathcal{H}^4)+\frac{105a^6 M^6 \mathcal{H}^4}{4\omega_k^9}\right)\\
\end{split}
\end{equation*}
\begin{equation}\label{EMTFluctuations}
\begin{split}
&+\left. \left(\xi-\frac{1}{6}\right)^2\left(-\frac{1}{4\omega_k^3}(72\mathcal{H}^{\prime\prime}\mathcal{H}-36\mathcal{H}^{\prime 2}-108\mathcal{H}^4)+\frac{54a^2M^2}{\omega_k^5}(\mathcal{H}^\prime \mathcal{H}^2+\mathcal{H}^4) \right)
\right]\\
&+\frac{1}{8\pi^2 a^2} \int dk k^2 \left[  \frac{a^2\Delta^2}{\omega_k} -\frac{a^4 \Delta^4}{4\omega_k^3}+\frac{a^4 \mathcal{H}^2 M^2 \Delta^2}{2\omega_k^5}-\frac{5}{8}\frac{a^6\mathcal{H}^2 M^4\Delta^2}{\omega_k^7} \right.\\
& \left. +\left( \xi-\frac{1}{6} \right) \left(-\frac{3a^2\Delta^2 \mathcal{H}^2}{\omega_k^3}+\frac{9a^4 M^2 \Delta^2 \mathcal{H}^2}{\omega_k^5}\right)\right]\,,
\end{split}
\end{equation}
where we recall that ${\cal H}=a'/a$ is the Hubble rate in conformal time. The expression above is manifestly UV-divergent in some of the terms, a fact that was expected for the reasons mentioned in the previous section.
Note that the dependence on the off-shell scale $M$ appears in all powers of $\omega_k\equiv\omega_k(\tau, M)$ involved in the denominators of the various terms -- cf. \eqref{eq:omegaM} -- but also in the specific contributions that carry the factor $\Delta^2$ of adiabatic order $2$.

On a simple inspection of  Eq.\,\eqref{EMTFluctuations}, we confirm that only even adiabatic orders (made out of powers of $\cal H$ and/or derivatives of it)  remain in the final result.
In \cite{Maggiore2011} a simplified form of this calculation was considered in the on-shell case ($M=m$) and minimal coupling ($\xi=0$). It proves to be a particular case of the above formula, as can be easily checked.
A much more drastic simplification occurs for flat spacetime (Minkowski) corresponding to $a=1$ (hence $\mathcal{H}=0)$ since in that case the above result shrinks to just the following:
\begin{equation}\label{eq:Minkoski}
\begin{split}
  \left.\langle T_{00}^{\delta \phi}\rangle\right|_{\rm Minkowski}&=\frac{1}{4\pi^2}\int dk k^2 \omega_k =
  \int\frac{d^3k}{(2\pi)^3}\,\left(\frac12\,\hbar\,\omega_k\right)\\
  &=\int\frac{d^3k}{(2\pi)^3}\,\left(\frac12\hbar \sqrt{{\bf k}^2+m^2}\right)\Rightarrow\frac12\,\tmu^{4-n}\, \int\frac{d^{n-1} k}{(2\pi)^{n-1}}\,
   \hbar\,\sqrt{{\bf k}^2+m^2}\\
   &= \frac12\,\beta_{\rL}^{(1)}\,\left(-\frac{2}{4-n}
-\ln\frac{4\pi\tmu^2}{m^2}+\gamma_E-\frac32\right)\,,
  \end{split}
\end{equation}
where $\hbar$ has been temporally  restored for convenience in some of the terms. In the second line, we have moved to $n$-dimensional spacetime, as usually done in a DR approach\cite{Collins84,Brown:1992db}.  Moreover,  $\gamma_E$ is Euler's constant and
\begin{equation}\label{beta4}
\beta_{\rL}^{(1)}=\frac{\hbar \,m^4}{2\,(4\pi)^2}
\end{equation}
is the one-loop coefficient of the $\beta$-function for the parameter $\rL$ (cf. Sec.\,\ref{sec:RVMbetafunction}). Finally,  $\tmu$ is the usual t'Hooft mass unit in DR\cite{Collins84}. The ZPE in Minkowski space is seen to be quartically UV-divergent (expressed as a simple pole in DR at $n=4$); so it is in curved spacetime, of course. However, in a curved background there are additional types of divergences, since they depend on the value of the Hubble rate and derivatives. This is evident from Eq.\,\eqref{EMTFluctuations}. For example, the first two terms on its \textit{r.h.s.} that are proportional to $\xi-1/6$ times ${\cal H}^2$ are quadratically and logarithmically divergent, respectively. In flat spacetime these terms are absent.

Consider now the usual procedure followed in most of the literature to renormalize the quartic divergences of the VED in the Minkowski case  through the use of the MSS\cite{Brown:1992db,Akhmedov:2002ts,Ossola:2003ku}.  Let us define the VED in a symbolic shorthand way as $\rv=\rL+$ZPE, where the ZPE is the (unrenormalized) zero-point energy given by Eq.\,\eqref{eq:Minkoski} in flat spacetime, and $\rL$ is the bare term  in the action \eqref{eq:EH}, both quantities being UV-divergent. Next, we split the bare term as  $\rL=\rL(\tmu)+\delta\rL$, where  $\rL(\tmu)$ is the renormalized value of $\rL$, and the counterterm $\delta\rL$ is then chosen to cancel the pole at $n=4$ of the result \eqref{eq:Minkoski} in DR. This renders the well-known result:
\begin{equation}\label{VEDMink}
\rv=\rL(\tmu)+\frac{m^4}{64\pi^2}\,\left(\ln\frac{m^2}{\tmu^2}+C_{\rm
vac} \right)\,.
\end{equation}
The second term on the \textit{r.h.s.} of \eqref{VEDMink} is the renormalized ZPE at one-loop in flat spacetime.  It carries an arbitrary constant  $C_{\rm vac}$  since the counterterm itself is not uniquely specified except for cancelling the pole of the unrenormalized ZPE  in $n=4$ spacetime dimensions. In the modified MSS, the counterterm absorbs additive constants such as $\gamma_E$ and $\ln 4\pi$ in \eqref{eq:Minkoski} (which can be hidden in a redefinition of the unphysical scale $\tmu$, if desired)  but still leaves $C_{\rm vac}=-3/2$. This is, of course, completely arbitrary, and such arbitrariness does not disappear in a curved spacetime calculation with DR\cite{Bunch:1979mq}, nor does this procedure allow for a proper treatment of the running of the VED and $G$ in cosmology \cite{Gorbar:2002pw}. In the case of the VED, first and foremost an appropriate identification of $\rv(M)$ is required not limited to just the renormalization of the parameter $\rL$.  In our approach,  the full vacuum EMT is correctly identified right from the start (cf. Eq.\,\eqref{EMTvacuum}) and the arbitrary additive constants cancel along with the UV-divergences in the  ARP renormalization procedure (cf. Sec. \ref{sec:RenormEMT}).  At the same time the running of $G$ is also obtained in a way fully  consistent with the running of the VED (cf. Secs.\,\ref{sec:RGinvariance} and \ref{sec:RunningG})).

In a flat spacetime approach this programme is impossible. One obvious aspect of Eq.\,\eqref{VEDMink} is that it has no connection whatsoever with the expanding universe, since it does not depend on its expansion rate $H$ or any other cosmological quantity.  This point reminds us of our discussion in Sec.\,\ref{sec:cTQFTQG}, which we wish to emphasize anew here: the existence of VED does not always lead to a CC term.  The former may exist, in principle, in a perfectly renormalized form in Minkowski spacetime, as given e.g. in Eq.\,\eqref{VEDMink}, while the latter can only exist in the context of Einstein's equations in curved spacetime. However, only in a curved spacetime we are entitled to establish a physically meaningful liaison between the two concepts. The Minkowskian results \eqref{eq:Minkoski}-\eqref{VEDMink} know nothing about the spacetime curvature;  not surprisingly, a simplified approach of this sort misses many contributions (including UV-divergent ones) that are tied to the spacetime curvature itself, as noted before. Taking the aforementioned formulas and setting $\tmu=H$, as some authors do in the literature (in an attempt  to artificially inject some cosmological physics in the last minute)  is completely artificial in a flat spacetime context.
Only in the presence of gravity the CC is  physically meaningful  and only then its value becomes naturally linked with the VED through Einstein's equations (in proper renormalized form): $\rv=\CC/(8\pi G)$.   The above Minkowskian spacetime result  illustrates  that a calculation of the VED in flat spacetime, even after renormalization,  has no impact whatsoever on the value of the cosmological constant  since the latter cannot be defined in that spacetime!  In short, but reiterating the message:  one cannot properly discuss such matters as the CCP and running of $\CC$ in Minkowski space, since having $\CC\neq 0$ is inconsistent with the solution of Einstein's equations in that spacetime. Neither can these issues be fruitfully discussed in curved spacetime with inappropriate renormalization schemes, e.g. in those leaving behind arbitrary additive constants. As said, every bit counts a lot in matters such as the CCP.

The moral is clear: in order to make contact with the physical CC in cosmology we cannot limit ourselves to compute  the value of  Eq.\,\eqref{eq:Minkoski} and adopt its renormalized form Eq.\,\eqref{VEDMink}, as done too often in the literature. In Minkowski space, the scale $\tmu$ is a pure artifact that has no connection whatsoever to cosmological physics. Furthermore, even if too obvious to say it,  the term $\rL$ in the EH action is \textit{not} the VED associated with the measured cosmological constant, and hence its renormalization group equation (RGE) is \textit{not} at all the one for the physical cosmological term or VED (cf. Sec.\,\ref{sec:RVMbetafunction}).  In order to provide a physical meaning to the scale setting $\tmu=H$ and find the correct RGE for the VED,  we must face  the QFT calculation  \textit{ab initio}  in the context of cosmological spacetime, then correctly identify the VED and, finally,  use an appropriate renormalization scheme, namely one  that has more physical meaning than just the formal renormalization performed within the minimal subtraction scheme as sketched above, which is devoid of physical interpretation.

Quite often, vacuum considerations present additional complications.  For example, if we focus on such an important theory as standard quantum electrodynamics (QED) in Minkowski spacetime, the  vacuum energy associated with the fluctuations (in this case  of the quantized electromagnetic field) must be zero to make the theory compatible with GR. But this is not the end of the story, as it raises important nontrivial issues that must be carefully  addressed, which still make the presence of the vacuum component necessary as an internal consistency condition for QED (see Appendix \ref{sec:appendixC} for an expanded reflection on these matters).

How much of the renormalization program has been covered in our approach thus far?  Up to this point, we have identified the vacuum EMT, given by Eq.\,\eqref{EMTvacuum}, and we have computed the unrenormalized ZPE in FLRW spacetime, given by Eq.\,\eqref{EMTFluctuations} rather than the naive Minkowskian result \eqref{eq:Minkoski}. What remains to be done right next is to properly renormalize the curved spacetime formula \eqref{EMTFluctuations} and extract a physically meaningful VED. So there is still quite some work to do before we can make contact with physics.

\subsection{An analogy with the Casimir effect}\label{sec:CasimirAnalogy}

Let us elaborate on a simple analogy, which despite being useful, nonetheless has its own limitations. The (dynamical) vacuum energy in our expanding universe may be thought of as being the energy stored in a giant  Casimir device in which the parallel plates  move apart very slowly (“expand”)\cite{Sola:2013gha,Sola:2014tta} -- cf. Appendix \ref{sec:appendixC}.
Although the  total vacuum energy cannot be measured, the distinctive effect associated with the boundaries (the plates)  and their increasing separation with time can be measured. We may establish an analogy with cosmological spacetime. In it, the $4$-dimensional scalar curvature can be expressed  in different equivalent ways as follows:
\begin{equation}\label{eq:4curvature}
\begin{split}
R=&6\left(\frac{a^{\prime\prime}}{a^3}+\frac{K}{a^2}\right)=6\left(\frac{\dot{a}^2}{a^2}+\frac{\ddot{a}}{a}+\frac{K}{a^2}\right)= 6\left(2H^2 +\dot{H}+\frac{K}{a^2}\right)\\
& 
=-8\pi G \ T^\mu_{\ \mu} =8\pi G \sum_{i={\rm vac.}, {\rm mat.}}(\rho_i-3P_i)
\end{split}\,,
\end{equation}
where in the second line we used Einstein's equations to express the result in terms of the energy densities and pressures of vacuum and matter (relativistic and nonrelativistic).
The above expression is obviously zero in strict Minkowskian spacetime (characterized by $a=1$ and $K=0$; and, of course, without any energy component that could modify the spacetime geometry). Some particular cases can be of interest. For example, if $a=1$ but $K\neq 0$, we can still have $4$-dimensional curvature at the expense of the $3$-dimensional one. Having $K\neq0$ would be somehow the analog of the geometric boundaries introduced by the plates in the static Casimir effect. If the plates also move, we have a dynamical Casimir effect with moving boundaries. Finally, if $K=0$ (as expected from inflation), but $H\neq 0$, there are no boundaries, but we still have a dynamical situation in which the $4$-dimensional curvature of spacetime is exclusively associated with its expansion process. The Casimir analog of it would be peculiar since now there are no plates or can just be conceived at infinity. It is nevertheless the closest parallelism to the cosmological picture, since here the $3$-dimensional geometry in between the infinitely separated plates is stretching and therefore creating a $4$-dimensional curvature and a corresponding vacuum energy density filling all the empty space (without plates!).  In the absence of expansion, there could not be such a spacetime curvature in this instance, and hence no vacuum energy either. So, the expansion is a distinctive effect that stands out from the bulk presence of the three-dimensional geometric background and endows it with $4$-dimensional curvature (see more on this analogy in Sec.\,\ref{sec:VEDMinkowski}). It is therefore natural to expect that the presence and evolution of the VED in the expanding universe should be measurable only through the existence of $4$-dimensional curvature carried by the purely geometric contributions proportional to $R$, $R^2$, $R^{\mu\nu}R_{\mu\nu}$ etc., hence to $H^2$ and $\dot{H}$ (and corresponding higher powers). These structures carry the ripples of the spacetime geometry and are therefore responsible for the VED. The purported ZPE contributions in flat spacetime before any ripple of curvature is present are of the order $\sim m^4$ for any quantized matter field, as we have seen in the previous section. But no matter what the value of the ZPE is, the VED must be zero to be consistent with Einstein's equations, see Sec.\ref{sec:VEDMinkowski}. The necessary adjustment to be made between the ZPE and the $\rL$ term to ensure this fact constitutes, however, no fine tuning at all, as it is just a formal (in fact, artificial)  arrangement imposing by fiat that spacetime is flat. It is similar to the use of normal ordering (cf. Sec. \ref{sec:finetuningCCP}), another arbitrary arrangement. They have no physical significance, as flat (Minkowski) spacetime  does not exist as a global spacetime in nature, only as a local approximation in the neighborhood of a given spacetime point. At the tangent Minkowski space of this point (local inertial frame),  we perform that adjustment which is, of course, valid only as an approximation in that flat neighborhood. We can do the same at the next neighborhood, but we cannot compare the two neighborhoods in curved space,  so it is clear that it has no global physical meaning. Only spacetime curvature can have global physical meaning in the universe. Thus, according to Eq.\,\eqref{eq:4curvature}, the existence of a nonvanishing VED is meaningful in the spacetime manifold only if the latter has four-dimensional curvature $R$, and  therefore gravity is present. So gravity is ultimately the source of VED in cosmological spacetime.

To put in a nutshell, the VED is zero in Minkowski space, and the ZPE computed in it cannot have any cosmological significance,  much in the same way as the vacuum energy of the space surrounding the Casimir device is not measurable either (cf. Appendix. \ref{sec:appendixC})). Although the Casimir effect is, of course, just a rough analogy, explicit QFT calculations indeed show that in cosmological spacetime we expect curvature-dependent contributions to the VED of order $m^2 R$ ($\sim m^2 H^2,  m^2 \dot{H}$), in which both the spacetime geometry and the quantum field masses participate, these being much softer than the quartic mass  contributions $\sim m^4$. Obviously, all effects in which the Hubble rate enters explicitly are not harmful at all from the point of view of the  CCP, since $H^2\ll m^2$ at present for any known particle mass (recall that $H_0\sim 10^{-42}$ GeV).  Let us see in practice how these softer vacuum effects emerge explicitly from our QFT framework.

\subsection{Off-shell renormalization in curved spacetime: the RVM}\label{sec:RenormEMT}

The next step in our program, of course,  is to retake the calculation of the  ZPE of the quantized matter field $\phi$  in FLRW spacetime.  We have seen in Sec.\,\ref{eq:RegZPE} that the result is UV-divergent and it must be renormalized. As indicated previously, we adhere to adiabatic renormalization \cite{Birrell:1982ix,Fulling89,Parker:2009uva},  in which physical quantities are organized in the so-called adiabatic orders,  although with a crucial and distinctive characteristic  with respect to the traditional method, to wit: we renormalize the energy-momentum tensor (EMT) `off-shell'.  In fact, following the RVM approach of\cite{Moreno-Pulido:2020anb,Moreno-Pulido:2022phq}, this means that we define its renormalized VEV  (associated with the fluctuations $\delta\phi$  of the fields) by subtracting the vacuum EMT at an arbitrary mass scale $M$, which plays the role of a renormalization point:
\begin{eqnarray}\label{EMTRenormalized}
\langle T_{\mu\nu}^{\delta \phi}\rangle_{\rm Ren}(M)&=&\langle T_{\mu\nu}^{\delta \phi}\rangle(m)-\langle T_{\mu\nu}^{\delta \phi}\rangle^{(0-4)}(M)\,.
\end{eqnarray}
This subtraction recipe defines the `off-shell ARP' for renormalizing the EMT. It is a characteristic inherent to the modern RVM formulation within the QFT context\cite{SolaPeracaula:2022hpd} and was not present in the old RVM formulation\cite{Sola:2013gha}.

In $4$-dimensional spacetime, the above subtraction at the scale $M$ is performed only up to the fourth adiabatic order, this is the meaning of $\langle T_{\mu\nu}^{\delta \phi}\rangle^{(0-4)}(M)$. This prescription suffices to provide a finite renormalized vacuum EMT. The on-shell value $\langle T_{\mu\nu}^{\delta \phi}\rangle(m)$ can be computed at any adiabatic order (bearing, however, in mind the asymptotic character of the adiabatic expansion) or even performing the calculation in an exact form when possible (e.g. in the case of de Sitter spacetime\cite{SolaPeracaula:2026trz} -- see Sec. \ref{sec:RenormEMTdS} --,  but certainly not in a general FLRW spacetime for which no such exact solution exists\cite{Moreno-Pulido:2020anb,Moreno-Pulido:2022phq,Moreno-Pulido:2022upl,Moreno-Pulido:2023ryo}.) We should also note that despite the fact that the  above subtraction prescription  is not manifestly covariant, adiabatic regularization is known to be equivalent to covariant point splitting. This equivalence was proven for zero spatial curvature ($K=0$) in \, \cite{Birrell78} and was generalized for any $K$ in \cite{Anderson:1987yt}.

On applying the above renormalization recipe to the ZPE part of the EMT, as given by  Eq.\,\eqref{EMTFluctuations}, one finds after some calculations the following result\,\cite{Moreno-Pulido:2020anb,Moreno-Pulido:2022phq}:
\begin{equation}\label{Renormalized2}
\begin{split}
&\langle T_{00}^{\delta \phi}\rangle_{\rm Ren}(M)
=\frac{a^2}{128\pi^2 }\left(-M^4+4m^2M^2-3m^4+2m^4 \ln \frac{m^2}{M^2}\right)\\
&-\left(\xi-\frac{1}{6}\right)\frac{3 \mathcal{H}^2 }{16 \pi^2 }\left(m^2-M^2-m^2\ln \frac{m^2}{M^2} \right)+\left(\xi-\frac{1}{6}\right)^2 \frac{9\left(2  \mathcal{H}^{\prime \prime} \mathcal{H}- \mathcal{H}^{\prime 2}- 3  \mathcal{H}^{4}\right)}{16\pi^2 a^2}\ln \frac{m^2}{M^2}+\dots
\end{split}
\end{equation}
Here the dots stand just for higher order adiabatic orders.   According to \eqref{EMTvacuum}, the renormalized full vacuum EMT at the scale $M$ is now obtained from including the contribution from the (renormalized) $\rL$-term in the Einstein-Hilbert action \eqref{eq:EH}:
\begin{equation}\label{RenEMTvacuum}
\langle T_{\mu\nu}^{\rm vac}\rangle_{\rm Ren}(M)=-\rho_\Lambda (M) g_{\mu \nu}+\langle T_{\mu \nu}^{\delta \phi}\rangle_{\rm Ren}(M)\,.
\end{equation}
On the other hand, the renormalized VED just follows from the $00th$-component of the previous expression:
\begin{equation}\label{RenVDE}
\rv(M)= \frac{\langle T_{00}^{\rm vac}\rangle_{\rm ren}(M)}{a^2}=\rho_\Lambda (M)+\frac{\langle T_{00}^{\delta \phi}\rangle_{\rm ren}(M)}{a^2}\,,
\end{equation}
where we should remember that $g_{00}=-a^2$ in the conformal metric.
Notice that Eq.\,\eqref{RenVDE} for the VED originates from treating the vacuum as a perfect fluid, that is, with an EMT of the form
\begin{equation}\label{eq:VaccumIdealFluid}
\langle T_{\mu\nu}^{{\rm vac}}\rangle=\Pv g_{\mu \nu}+\left(\rv+\Pv\right)u_\mu u_\nu\,,
\end{equation}
in which $u^\mu$ is the $4$-velocity ($u^\mu u_\mu=-1$) and no presumption is made on the vacuum EoS.  In conformally flat coordinates and considering  the comoving FLRW  frame, $u^\mu=\left(1/\sqrt{-g_{00}},0,0,0\right)=\left(1/a,0,0,0\right)$ and hence  $u_\mu=(-a,0,0,0)$. The $00th$-component thus yields  $\langle T_{00}^{{\rm vac}}\rangle=-a^2 \Pv+\left(\rv+\Pv\right) a^2=a^2\rho_{\rm vac}$, which holds good irrespective of the EoS relating $\rv$ and $\Pv$.  This leads to Eq.\,\eqref{RenVDE}
by inserting the $00th$-component of \eqref{RenEMTvacuum}\footnote{We may also derive it in more formal terms. Recall that in relativistic mechanics the energy of a particle of $4$-momentum $p_\mu$ as measured by an observer of $4$-velocity $u^\mu$ is given by  $E=-p_\mu u^\mu$. Similarly, in field theory the VED measured by that observer is  $\rv=\langle  T_{\mu\nu}^{\rm vac} u^\mu u^\nu\rangle$. Recalling that in the rest frame of the observer the $4$-velocity is $u^\mu=\left(1/a,0,0,0\right)$,  Eq.\,\eqref{RenVDE} follows at once.}.

In the renormalized theory formulated in our framework, the sum \eqref{RenVDE} provides the physically measurable VED at the scale $M$, which again takes the general form
$  {\rm VED}=\rL+{\rm ZPE}\,.$
More explicitly,
\begin{equation}\label{RenVDEexplicit}
\begin{split}
\rv(M,H)&= \rho_\Lambda (M)+\frac{1}{128\pi^2 }\left(-M^4+4m^2M^2-3m^4+2m^4 \ln \frac{m^2}{M^2}\right)\\
&+\left(\xi-\frac{1}{6}\right)\frac{3{H}^2 }{16 \pi^2}\left(M^2-m^2+m^2\ln \frac{m^2}{M^2} \right)\\
&-\left(\xi-\frac{1}{6}\right)^2 \frac{9\left(6H^2\dot{H}+2H\ddot{H}-\dot{H}^2\right)}{16\pi^2 }\ln \frac{m^2}{M^2}+\cdots
\end{split}
\end{equation}
where in the above equation  we have converted all the terms involving the Hubble rate in conformal time into cosmic time. In particular,  it is easy to show that  $2  \mathcal{H}^{\prime \prime} \mathcal{H}- \mathcal{H}^{\prime 2}- 3  \mathcal{H}^{4}=a^4\left(6H^2\dot{H}+2H\ddot{H}-\dot{H}^2\right)$. In the case  $H=$const. (corresponding to de Sitter spacetime), such a higher order adiabatic contribution  simply vanishes.

\subsection{Renormalized Einstein equations in QFT in curved spacetime}\label{sec:GeneralizedEqs}

As is well-known, GR is perturbatively non-renormalizable in the context of quantum gravity\,\cite{tHooft:1974toh,Goroff:1985sz}. However, by complementing the Einstein–Hilbert action with appropriate higher derivative (HD) geometric terms, specifically quadratic terms in the curvature, gravity can be made perturbatively
renormalizable\cite{Stelle:1976gc}, although we should add that the dust has not yet settled fully in this approach.  Be as it may, we reiterate  that in our RVM framework we will not quantize the gravitational field, only the matter fields. The semiclassical formulation (QFT in curved spacetime) will suffice for our considerations. In it, the HD terms in the action are also indispensable for renormalization.
Thus, in order to implement the renormalization program in our context, we must extend the EH action by considering the HD gravity terms up to the second adiabatic order\,\cite{Birrell:1982ix}. This extension of the vacuum action leads to the renormalized Einstein field equations, which incorporate the quantum effects from the quantized matter fields. If we consider only vacuum terms, the generalized field equations (compare with the original form \eqref{EinsteinEqs}) can be written as follows:
\begin{equation}\label{eq:MEEs}
\frac{1}{8\pi G(M)} G_{\mu \nu}+\rho_\Lambda(M) g_{\mu \nu}+\alpha(M)\ \leftidx{^{(1)}}{\!H}_{\mu \nu}= \langle T_{\mu \nu}^{\delta \phi}\rangle_{\rm ren}(M)\,,
\end{equation}
where $M$ is the renormalization scale introduced before.
Here  $\leftidx{^{(1)}}{\!H}_{\mu \nu}$ is the (covariantly conserved)  HD tensor which appears from the metric functional variation of  the $R^2$-term in the  higher derivative vacuum action for FLRW spacetime, and $\alpha(M)$ is the corresponding renormalized coupling.  We recall the reader that  the HD term obtained from  the variation of $R_{\mu\nu}R^{\mu\nu}$ (the square of the Ricci tensor), called  $\leftidx{^{(2)}}{\!H}_{\mu \nu}$, is not needed in our case since it is not independent of $\leftidx{^{(1)}}{\!H}_{\mu \nu}$  for conformally flat spacetimes (in particular, FLRW)\,\cite{Birrell:1982ix}.

The couplings $G(M),\rL(M)$ and $\alpha(M)$ are the renormalized couplings at the scale $M$, and $\langle T_{\mu \nu}^{\delta \phi}\rangle_{\rm ren}(M)$ is the (adiabatically) renormalized vacuum EMT.  Me way now striaghtforwardly subtract the renormalized Einstein equations \eqref{eq:MEEs}, term by term,  at two scales $M$ and $M_0$, with the result
\begin{equation}\label{eq:SubtractedMEEs}
\frac{1}{8\pi}\delta G^{-1}(m,M,M_0) G_{\mu \nu}+\delta\rL(m,M,M_0) g_{\mu \nu}+ \delta\alpha(M,M_0)\ \leftidx{^{(1)}}{\!H}_{\mu \nu}= \delta\langle T_{00}^{\delta \phi}\rangle^{\rm (0-4)}_{\rm ren}(M,M_0)\,,
\end{equation}
where we have defined
$\delta\langle T_{00}^{\delta \phi}\rangle^{\rm (0-4)}_{\rm ren}(M,M_0)\equiv\langle T_{00}^{\delta \phi}\rangle^{\rm (0-4)}_{\rm ren}(M)-\langle T_{00}^{\delta \phi}\rangle^{\rm (0-4)}_{\rm ren}(M_0)$ and similarly for $\delta\rL$, $\delta G^{-1}$ and $\delta\alpha$.  Upon using \eqref{Renormalized2} for the calculation of $\delta\langle T_{00}^{\delta \phi}\rangle^{\rm (0-4)}_{\rm ren}(M,M_0)$ and the explicit expressions for $G_{00}$ and  $\leftidx{^{(1)}}{\!H}_{00}$ in the conformally flat metric\footnote{These are the following\cite{Moreno-Pulido:2022phq}: $G_{00}=3a^2H^2$ and $\leftidx{^{(1)}}{\!H}_{00}= -18 a^2\left(\dot{H}^2-2H\ddot{H}-6H^2\dot{H}\right)$.}, we obtain the following results:
\begin{equation}\label{SubtractionrL}
\delta\rL(m,M,M_0)\equiv\rL(M)-\rL(M_0)=\frac{1}{128\pi^2}\left(M^4-M_0^{4}-4m^2(M^2-M_0^{2})+2m^4\ln  \frac{M^{2}}{M_0^2}\right)\,,
\end{equation}
\begin{equation} \label{SubtractionG}
\delta G^{-1}(m,M,M_0)\equiv G^{-1}(M)-G^{-1}(M_0)=\left(\xi-\frac{1}{6}\right)\frac{1}{2\pi}\left[M^2 - M_0^{2} -m^2\ln \frac{M^{2}}{M_0^2}\right]
\end{equation}
and
\begin{equation} \label{Subtractionalpha}
\delta\alpha(M,M_0)\equiv\alpha(M)-\alpha(M_0)= -\frac{1}{32\pi^2}\left(\xi-\frac{1}{6}\right)^2 \ln \frac{M^2}{M_0^{2}}\,.
\end{equation}
These relations furnish the scaling laws obeyed by the couplings $\rL(M)$, $G(M)$, $\alpha(M)$ in the renormalized Einstein's equations, i.e. their respective (finite) shifts when we perform a change of scale  (renormalization point)  from $M_0$ to $M$.
We shall encounter these same equations when we address the adiabatic renormalization of the vacum EMT in de Sitter spacetime.

Having obtained the explicit form of the renormalized $\rv(M,H)$ \eqref{RenVDEexplicit} in our approach, the next step is to compute its scaling evolution  in the span mediating in between two cosmic epochs in the near past, say those characterized by the Hubble rate values  $H$ and $H_0$. This is actually the physical result we are aiming at since it corresponds to the main aim of the renormalization theory, namely, to compute the scaling evolution of a quantity between two renormalization points. But for this, we must use our scale setting $M=H$ at the end of the calculation to trace the evolution of the VED with the cosmic expansion.  The result thus follows from Eq.\,\eqref{RenVDEexplicit} by selecting the renormalization points $M$ and $M_0$ precisely at the values of the Hubble rate in those epochs, hence $M=H$ and $M_0=H_0$ respectively. Defining  $\rv(H)\equiv\rv(M=H,H)$ and  $\rv(H_0)\equiv \rv(H_0,H_0)$, we obtain
\begin{equation}\label{DiffVEDphys}
\begin{split}
\rv(H)-\rv(H_0)&=\frac{3\left(\xi-\frac{1}{6}\right)}{16\pi^2}\left[H^2\left(H^2-m^2+m^2\ln\frac{m^2}{H^2}\right)\right.\\
&\left.-H_0^2\left(H_0^2
-m^2+m^2\ln\frac{m^2}{H_0^2}\right)\right]+\cdots\,,
\end{split}
\end{equation}
where we have neglected the contribution from the terms in the last line of \eqref{RenVDEexplicit} as they will not be relevant for the RVM mechanism of inflation (cf. Sec. \ref{sec:H4infl}) and they are also irrelevant for the current universe.  Interestingly enough, the dangerous terms $\sim m^4$ have canceled out in the difference \eqref{DiffVEDphys}. In fact, the entire term in parenthesis at the first line of \eqref{RenVDEexplicit} cancels exactly in the difference against $\rL(M)-\rL(M_0)$, given by Eq.\,\eqref{SubtractionrL}, as can be easily checked. This is a most remarkable and crucial property of our renormalization framework, since because of it the VED evolution becomes free of quartic mass contributions $\sim m^4$, which, if present, would recreate the unnatural fine tuning associated with the CCP.

A final observation in this section may be in order.  We may wonder what would happen if our starting action \eqref{eq:Sphi} were to include an effective potential for the scalar field under SSB (i.e. $\langle\phi\rangle\equiv v\neq 0$). In the context of our off-shell subtraction prescription \eqref{EMTRenormalized}, all constant terms that do not evolve with the expansion cancel out  in the renormalized expression. For the effective potential, the external tails of the gravitational field do not carry external momentum and therefore cannot introduce additional powers of $H$.  So an effective potential can only lead to contributions of order  $\sim m^4$ or $v^4$ (if there is a VEV $v$ for $\phi$). Now, terms with powers of $H$ can still emerge  at the level of the one-loop effective action of the scalar field (which, in contrast to the effective potential, is sensitive to external momenta). However, by pure covariance it can only  bring in corrections of the existing type $\sim m^2H^2$ in the late universe. In fact,  the new terms must adapt to the tensorial structure on the \textit{l.h.s.} of Einstein's equations \eqref{eq:MEEs}, and hence  appear as a linear combination of $g_{\mu\nu}$, $G_{\mu\nu}$ and $ \leftidx{^{(1)}}{\!H}_{\mu \nu}$. In the FLRW metric, they are of order $m^4$, $m^2 H^2$ and $H^4$, respectively.  Therefore, e.g. the $00$th component of the vacuum EMT, $\langle T_{00}^{\delta \phi}\rangle_{\rm ren}(M)$, can receive contributions of the form
\begin{equation}\label{eq:T00additional}
 A \frac{a^2 m^4}{16\pi ^2} \ln \frac{m^2}{M^2} +  \frac{a^2m^2 H^2}{16\pi ^2}\left(B+ C \ln \frac{m^2}{M^2}\right)+{\cal O}(H^4)\,,
\end{equation}
with calculable coefficients $A,B$ and $C$. However, all these terms  are of the same type as those we previously found in the ZPE structure, see Eq.\,\eqref{Renormalized2}.
Therefore, the  $\sim m^4$ (and $v^4$) contributions will cancel against  \eqref{SubtractionrL} when we operate the difference  \eqref{DiffVEDphys}  between two scales, and only $\sim m^2 H^2$ effects will remain in the current universe, exactly as with the ZPE. The upshot is that  there can be no aggravation of the fine tuning issue,  and the evolution of the VED (and vacuum pressure)  should remain qualitatively as smooth as we have found with the ZPE treatment alone.

\subsection{Renormalized VED from subtracting Minkowski vacuum}\label{sec:VEDMinkowski}

It is useful to note that the renormalized VED in our subtraction scheme  can be thought of as being the result of subtracting  the Minkowskian ZPE from the curved spacetime result. Indeed,  let us evaluate
\begin{equation}\label{eq:RenEMTSubtractMink}
\begin{split}
&\rv(M)\equiv\frac{\langle T_{00}^{\delta \phi}\rangle_{\rm ren}(M )}{a^2}-\left[\frac{\langle T_{00}^{\delta \phi}\rangle_{\rm ren}(M )}{a^2}\right]^{\rm Mink}
=\frac{\langle T_{00}^{\delta \phi}\rangle_{\rm ren}(M )}{a^2}-\langle T_{00}^{\delta \phi}\rangle_{\rm ren}^{\rm Mink}(M)\,,
\end{split}
\end{equation}
where in Minkowski spacetime $a=1$, $\cH=H=0$. In the flat background, Eq.\eqref{eq:MEEs} boils down to just  $\rho_\Lambda (M) \eta_{\mu \nu}= \langle T_{\mu\nu}^{\delta \phi}\rangle_{\rm ren}^{\rm Mink}(M)$, or
\begin{equation}\label{eq:T00Minkowski}
\langle T_{00}^{\delta \phi}\rangle_{\rm ren}^{\rm Mink}(M)=-\rL(M)\,.
\end{equation}
As a result, Eq.\,\eqref{eq:RenEMTSubtractMink} yields
\begin{equation}\label{RenVDE2}
\rv(M)=\frac{\langle T_{00}^{\delta \phi}\rangle_{\rm ren}(M)}{a^2}+\rho_\Lambda (M)= \frac{\langle T_{00}^{\rm vac}\rangle_{\rm ren}(M)}{a^2}\,,
\end{equation}
which is precisely our starting formula for the renormalized VED, Eq.\,\eqref{RenVDE}.

This alternative formulation reminds us of our analogy with the Casimir effect made in Sec.\,\ref{sec:CasimirAnalogy}, in the sense that  subtracting the  Minkowskian contribution is similar to obtaining a finite energy density in the Casimir device after removing the background space and only leaving the differential effect caused by the presence of the  plates, which in our case would be represented by the curvature of spacetime. The latter is the key ingredient that provides the departure of the FLRW background from Minkowskian spacetime.
Thus, upon performing such a subtraction, the VED should depend only on the curved geometry and evolve only mildly with the cosmic evolution through a function of the Hubble rate. Such a mildly evolving function is exactly given by the \textit{r.h.s.} of Eq.\,\eqref{DiffVEDphys}.

From the former considerations, in Minkowski space we should expect zero vacuum energy in our context. This result is usually justified using normal ordering of the quantum operators in the canonical formalism.  Here we do not need to appeal to normal ordering, since after all it is an arbitrary prescription to get rid of the ZPE from the quantum fields in flat spacetime.  In our case, the renormalized ZPE in Minkowski space  is the  value of $\langle T_{00}^{\delta \phi}\rangle_{\rm ren}(M)$, given by Eq.\,\eqref{Renormalized2} for $a=1$ and $\cH=0$:
\begin{equation}\label{eq:ZPEMinkowski}
\begin{split}
\langle T_{00}^{\delta \phi}\rangle_{\rm ren}^{\rm Mink}(M)=\frac{1}{128\pi^2 }\left(-M^4+4m^2M^2-3m^4+2m^4 \ln \frac{m^2}{M^2}\right)\,.
\end{split}
\end{equation}
This quantity is purely formal and has no connection with the measured CC term. This situation is similar to our discussion around Eq.\,\eqref{VEDMink}, except that here it is placed in a gravitational framework and therefore such expression is recovered in the flat spacetime limit of our curved spacetime result. In fact, the above quantity cancels exactly against $\rL(M)$. From equations \eqref{eq:T00Minkowski} and \eqref{RenVDE2} we confirm that the VED in Minkowskian spacetime is exactly zero in our renormalization setup:
\begin{equation}\label{eq:VEDMinkowski}
\begin{split}
\rv^{\rm Mink}&=\left[\frac{\langle T_{00}^{\delta \phi}\rangle_{\rm ren}(M )}{a^2}\right]^{\rm Mink}+\rL(M)\\
&=\langle T_{00}^{\delta \phi}\rangle_{\rm ren}^{\rm Mink}(M)+\rL(M)=-\rL(M)+\rL(M)=0\,,
\end{split}
\end{equation}
 for all scales $M$, as it should.

 \subsection{Fine tuning and CCP: a digression on the quantum vacuum}\label{sec:finetuningCCP}

In a non-gravitatinal context (i.e. in QFT in flat spacetime) the ZPE is given by the first two terms on the \textit{r.h.s.} of Eq.\,\eqref{RenVDEexplicit}  i.e.
\begin{equation}\label{RenVDEexplicitMinkow}
\begin{split}
\rv^{\rm Mink}(M)&= \rho_\Lambda (M)+\frac{1}{128\pi^2 }\left(-M^4+4m^2M^2-3m^4+2m^4 \ln \frac{m^2}{M^2}\right)\\
&= \rho_\Lambda (M)+\frac{m^4}{64\pi^2}\left(\ln\frac{m^2}{M^2}-\frac32-\frac{M^4}{2m^4}+\frac{2M^2}{m^2}\right)\,.
\end{split}
\end{equation}
This expression for the renormalized VED of a scalar field in Minkowski spacetime is not required to be zero, in principle, since it is not related to the cosmological constant. The second equality is particularly convenient in order to make a more direct contact with Eq.\,\eqref{VEDMink})  in the modified MSS (for which  $C_{\rm vac}=-3/2$). We can see that the two expressions match for $M^2\ll m^2$, which is reminiscent of the fact that when we move to FLRW spacetime we associate $M$ with $H$, which is certainly much smaller than any known particle mass.

On the other hand, if Minkowski spacetime is to be a particular solution of Einstein's equations, this is only possible if the VED computed in QFT in flat spacetime is exactly zero. So,   Eq.\eqref{RenVDEexplicitMinkow} must vanish:
\begin{equation} \label{eq:rLMMinkowski}
\rL(M)+\frac{1}{128\pi^2 }\left(-M^4+4m^2M^2-3m^4+2m^4 \ln \frac{m^2}{M^2}\right)=0\,.
\end{equation}
This says that $\rv=\rL+$ZPE$=0$ in Minkowski space.
The cosmological constant is then zero and, of course, also the VED, as we have seen in the previous section.  In flat spacetime, the scale $M$ becomes a purely formal artifact devoid of physical meaning,  much in the same way as the artificial mass unit $\mu$ employed in DR, see Eq.\,\eqref{VEDMink}. Given that there is no  gravity dynamics in Minkowski space, nothing  can physically run with the scale $M$  (or $\mu$). Therefore, the above equation \eqref{eq:rLMMinkowski}  is no fine tuning at all, since it carries no relation between measurable quantities; it is just a formal consistency relation aimed at obtaining Minkowski spacetime  as a particular solution of Einstein's equations. There is no more fine tuning in it than there is in the totally \textit{ad hoc} imposition of normal ordering in the canonical formulation of QFT. Fine tuning actually occurs instead in the opposite situation, that is, when one artificially enforces that the flat spacetime formula \eqref{RenVDEexplicitMinkow} should yield the value of the CC. Not too surprising, since such a requirement is physically meaningless.

 At the price of some digression on these matters, which I believe are central for understanding the quantum vacuum,  let us comment briefly here on the situation in Quantum Electodynamics (QED) -- for an extended discussion, see Appendix \ref{sec:appendixC}. In QED, the quanta have zero mass, $m=0$, and the above approach cannot be applied to study the vacuum energy. However, standard QED tells us that the electromagnetic field can be treated as an infinite collection of harmonic oscillators. This eventually leads to an infinite  sum of the vacuum fluctuations of all the modes. The usual practice in QFT in Minkowski spacetime, where only energy differences matter, is to get rid of such an infinite sum (which is after all a constant, even if infinite) through the normal ordering of the ladder operators in the field Hamiltonian $H_F$. That is, one redefines it as follows:
 \begin{equation}\label{eq:redefHamilton}
 \begin{split}
 H_F\rightarrow H_F-\langle 0|H_F|0\rangle &=\frac12\,\hbar\omega \left(a a^\dagger+a^\dagger a\right) -\frac12\hbar\omega\\
 &=\frac12\hbar\omega \left(2a^\dagger a +1\right) -\frac12\hbar\omega\\
 &= \hbar\omega a^\dagger a\,,
 \end{split}
 \end{equation}
where we have used the standard commutation relations \eqref{CommutationRelation}, which in a discrete situation for a single mode just reads $[a,a^\dagger]=1$.  We do this for all modes $\omega_k$, of course,  and sum over all of them. The new Hamiltonian is said to be `normally ordered' since the creation operators $a^\dagger$ appear to the left of the annihilation operators $a$, which ensures that the vacuum energy is zero, $H|0\rangle=0$, due to the fact that the $a$-operators  annihilate the vacuum: $a|0\rangle=0$.  This is fine but, of course, it is based on an arbitrary definition which we are entitled to use  since we do not have an absolute measure of energy in that spacetime.

However, in a GR context, it is not just an option; we are actually enforced to kill the vacuum energy in Minkowski spacetime in order to prevent  a contribution to the CC and hence contradict Einstein's equations. In this sense, it is surprising to see that one might still  insist on calculating the value of $\CC$ in a flat spacetime context after we have already declared that the VED is zero by fiat (i.e. on imposing normal ordering). I suppose that the confusion in the literature must be caused because one still has in mind   Eq.\eqref{RenVDEexplicitMinkow}, corresponding to a massive scalar field, and then one is tempted to compute the finite contribution emerging from  the ZPE part of it for various particle masses. In the QED context, instead, we cannot use that formula, of course, since the photon is massless. Still,  the Hamiltonian acting on the vacuum state (before operating normal ordering) is infinite and for QED there does  not exist a counterpart to the  finite renormalized  term given by Eq.\eqref{RenVDEexplicitMinkow}, except a zero renormalized ZPE. In QED  the situation might look simpler in that we can  switch on normal ordering and then  all vacuum effects become somehow hidden under the rug. But this is actually not true, the situation is actually much subtler. Even after imposing normal ordering,  QED would be inconsistent without the vacuum field ${\bf E}_0(t, {\bf x})$. This is well-known, but sometimes is forgotten! If we would ignore it,  the canonical commutation relations of Quantum Mechanics would break down. For example, the canonical commutator  $[{\bf x}(t), {\bf p}(t)]$ would collapse to zero after a very short time rather than taking the steady value $i\hbar$ that we expect at all times; the reason being that for ${\bf E}_0(t, {\bf x})=0$ the differential equation for ${\bf x}(t)$ would render an exponentially damped solution. Such a behavior is caused by the so-called `radiative (or radiation) reaction field',  a well-known effect  described in the theory of radiation in classical Electrodynamics,  which would exponentially damp the system in the absence of the vacuum field ${\bf E}_0(t, {\bf x})$ (see  Appendix \ref{sec:appendixC} for an extended discussion).

In curved spacetime, in contrast, there is a vacuum fluctuating field  and we do \textit{not} have to enforce vanishing ZPE through  normal ordering nor any other artifact. We have actually computed the VED in detail  for a non-minimal scalar field model and the result  is given by Eq.\,\eqref{RenVDEexplicit}. This defines a physical $\CC$-term, which turns out to be dynamical. The different pieces involved on the \textit{r.h.s.} of Eq.\,\eqref{RenVDEexplicit}  cannot be isolated. In particular, we are not entitled to assume that Eq.\,\eqref{eq:rLMMinkowski} holds now. Here, the scale $M$ is no longer a formal quantity, since we identify it with $H$, which is now relevant since we are in FLRW spacetime. Therefore, all terms in that equation are involved and are nonzero, but cannot be isolated piece by piece  At this point, let us imagine (in the manner of a  Gedankenexperiment)  a process in which we gradually `turn off' the curvature of spacetime by formally taking the limit $a\to 1$  ($H\to 0$), which implies vanishing curvature of spacetime, e.g. $R\to 0$ (for $K=0$ in Eq.\,\eqref{eq:4curvature}) as well as the curvature tensors. The \textit{l.h.s.} of \eqref{RenVDEexplicit} must go to zero, since in this limit we have to obtain Minkowski space as a solution of the field equations. At the same time, the \textit{r.h.s.}  must go to zero too. The terms that are explicitly dependent on the Hubble rate obviously go to zero; and hence the sum of the first two terms (which do not depend on $H$) must also go to zero. In this way we reach the flat spacetime condition Eq.\,\eqref{eq:rLMMinkowski} as a limit of the curved spacetime result; that is to say, we consistently retrieve the Minkowskian space result from the curved spacetime case \eqref{RenVDEexplicit}.  However, at any intermediate stage of this limit we cannot determine $\rL(M)$  separately from the ZPE,  as only the sum is physically meaningfull: this sum defines the dynamical vacuum energy density in curved spacetime.

This raises the question as to whether or not the VED can be predicted quantitatively at the point $M=H$, which would be tantamount to solving the old CCP. Strictly speaking, the answer is no, since we must repeat once more that the main aim of the renormalization group is to use data on a quantity (in this case, the VED $\rv$) at a given scale, say $M=H_0$, so as to find its value at another scale, $M=H$; in other words, the aim is to predict the RG evolution of that quantity,  not the value itself of the quantity  at a particular scale. All in all,  a clue to a possible solution of the CCP in our renormalization framework  is suggested.

Indeed, the dynamics of the VED  through its scale dependence on $M$, and subsequent identification of $M$ with $H$, expresses in a rigorous mathematical way the original point of view of the RVM framework from the renormalization group approach \cite{Sola:2013gha,SolaPeracaula:2022hpd}. But, in fact, it expresses more than that. It suggests that there is no fine tuning in the renormalized value of the VED (and hence in the physical value of the cosmological term) since, as previously remarked, the first two terms on the \textit{r.h.s.} of Eq.\,\eqref{RenVDEexplicit} vanish in the flat spacetime limit in order that the \textit{l.h.s.} vanishes too. However, when --  in a reversed version of the previous Gedankenexperiment -- we start from Minkowski space and gradually `turn on' the four-dimensional curvature (letting the universe evolve), these terms will be switched on and  by continuity their  contribution must remain as small as the remaining  terms on the \textit{r.h.s.}, which start getting small values  of  ${\cal O}(H^2)$ and ${\cal O}(H^4)$.  Therefore, while we cannot compute the value of the VED at $M=H$,  we can still see from the smoothness of the renormalization procedure that any two points are gently related  and that there is no fine tuning in the structure of the VED.

The fact that this property can be achieved in the off-shell renormalization scheme is  due to the subtraction recipe defined by Eq.\,\eqref{EMTRenormalized}.  This cannot be accomplished in schemes where countertems are used to pick out particular combinations of UV-divergent terms plus additional (arbitrary) constants (e.g. MSS renormalization). In our procedure, we get rid of all additive terms, since the subtraction of UV-divergent quantities includes their full finite parts.  Notice also that the dynamical character of $M=H$ is essential to fulfill this purpose, since it is the clue to performing subtractions at different points of the cosmological evolution, these subtractions being completely smooth (absence of large quartic mass contributions $\sim m^4$).  Put another way, in the RVM framework presented here there are no harmful effects of the type originally proposed by Zeldovich when he first formulated the CCP in the context of particle physics and quantum field theory (see the Introduction)! An appropriate renormalization of the theory can smooth them out. Among a large variety of alternative options to escape from these contributions, there are the unimodular theories of gravity\cite{Carballo-Rubio:2022ofy,Erdem:2026wfk}, in which the VED vanishes while the cosmological constant survives as an integration constant. However, this constant is arbitrary, so there is
no explanation for its particular phenomenological value.

\subsection{$\beta$-function of the physical running vacuum}\label{sec:RVMbetafunction}

The aforementioned cancellation of the $m^4$ terms in the VED evolution can be reformulated in renormalization group (RG) terms, namely in terms of the $\beta$-function driving its running. This function receives only `soft' contributions of the kind $\sim m^2 H^2$  rather than `hard' contributions $\sim m^4$. This is the key feature protecting our renormalization framework from the fine tuning calamity associated with the CCP. It is a more technical and precise way to rephrase the fine tuning discussion made in the last section.  This crucial property was first noticed in\cite{Moreno-Pulido:2020anb,Moreno-Pulido:2022phq}.  Let us put it in quantitative terms.


In fact, a chief result that follows  from equations \eqref{Renormalized2} and  \eqref{RenVDE} is the expression for the  $\beta$-function driving the RG-running of the VED:
\begin{equation*}
\begin{split}
\beta_{\rv}(M,H)&=M\frac{\partial\rv(M,H)}{\partial M}=M\frac{d\rL(M)}{dM}-\frac{1}{32\pi^2}\left(M^2-m^2\right)^2\\
&+\left(\xi-\frac{1}{6}\right)\frac{3 {H}^2 }{8 \pi^2}\left(M^2-m^2\right)-\left(\xi-\frac{1}{6}\right)^2 \frac{9\left(6H^2\dot{H}+2H\ddot{H}-\dot{H}^2\right)}{8\pi^2}
\end{split}
\end{equation*}
\begin{equation}\label{eq:RGEVED1}
\phantom{aaaa}=\left(\xi-\frac{1}{6}\right)\frac{3 {H}^2 }{8 \pi^2}\left(M^2-m^2\right)-\left(\xi-\frac{1}{6}\right)^2 \frac{9\left(6H^2\dot{H}+2H\ddot{H}-\dot{H}^2\right)}{8\pi^2}\,,
\end{equation}
where in the last step we have used that the $\beta$-function for the renormalized parameter $\rL$ in the EH-action is
\begin{equation}\label{eq:betarhoLambda}
  \beta_{\rL} (M)=M \frac{\partial \rL(M)}{\partial M}=\frac{1}{32\pi^2}(M^2-m^2)^2\,.
\end{equation}
The latter follows immediately  from Eq.\eqref{RenVDEexplicitMinkow}, which is  RG-invariant since it corresponds to  the VED in Minkowski spacetime. In fact, in this spacetime the VED is a bare quantity in which $\rL$  has been split into a renormalized term plus a counterterm that cancels the divergence of the unrenormalized ZPE.   The result \eqref{eq:betarhoLambda} can be compared with the one obtained from \eqref{VEDMink}  in the MSS within DR. Using the fact that the \textit{l.h.s.} of \eqref{VEDMink}  is a  bare quantity (and hence RG-invariant) we find:
\begin{equation}\label{eq:betarhoLambdaMS}
  \beta_{\rL}^{\rm MS}\equiv\mu \frac{\partial \rL(\mu)}{\partial \mu}=\frac{m^4}{32\pi^2}\,,
\end{equation}
which is indeed Eq.\eqref{beta4}, with $\hbar=1$ hereafter. Once more we find that this result in the MSS coincides with \eqref{eq:betarhoLambda} for $M^2\ll m^2$. The difference between the two situations is clear: in  Minkowski spacetime there is nothing else in the vacuum action apart from the bare term $\rL$. By contrast, in curved spacetime we have also the curvature  scalar  plus the geometric HD terms, with corresponding bare couplings.  In this case,  only the full effective action (which involves the classical part plus the nontrivial quantum vacuum effects) is RG-invariant. Finally, since $M$ eventually will be set to $H$ in \eqref{eq:RGEVED1}, which is currently negligible in front of any particle mass scale $m$, we have
\begin{equation}\label{eq:betarhoLambda2}
  \beta_{\rv}=\left(\xi-\frac{1}{6}\right)\frac{3 {H}^2 }{8 \pi^2}\left(M^2-m^2\right)\simeq  -\left(\xi-\frac{1}{6}\right)\frac{3 m^2{H}^2 }{8 \pi^2} \,,
\end{equation}
upon neglecting the ${\cal O}(H^4)$  terms in the present universe. Thus,  $\beta_{\rv}\sim m^2 H^2$.  This softer behavior of the $\beta$-function in the curved case  ($m^2 H^2\ll m^4$) explains the cancellation of the quartic mass terms in the previous sections. We note that the expression \eqref{eq:RGEVED1}, which is our starting point,  is exact and does not depend on the fact that \eqref{Renormalized2} was computed only up to  4$th$ adiabatic order. The reason is that  the higher order terms (order 6$th$ and above)  are finite and hence independent of $M$.

Equation  \eqref{eq:RGEVED1} gives the running of the VED with the scale  $M$ at fixed $H$, i.e. the partial derivative. Since in the end we must put $M=H$, we need to compute the total derivative with respect to $H$ of the VED expression \eqref{DiffVEDphys}, where the scale setting $M=H$ has already been implemented. Therefore, we obtain
\begin{equation}\label{eq:DerivTotal3}
\begin{split}
\beta_{\rm vac}(H)\equiv& H\frac{d\rv(H)}{dH}=\left(\xi-\frac{1}{6}\right)\frac{3 m^2H^2 }{8 \pi^2 }\left(-2+\ln\frac{m^2}{H^2}\right)\\
&\simeq \left(\xi-\frac{1}{6}\right)\frac{3 m^2H^2 }{8 \pi^2 }\ln\frac{m^2}{H_0^2}
\end{split}
\end{equation}
where again we have neglected the ${\cal O}(H^4)$ contributions in the late universe and in the log we have used the current value of the Hubble rate, $H_0$, and the fact that for any known particle $\ln\frac{m^2}{H_0^2}\gg 1$.  We can infer the previous result on more formal terms. Let us assume that $M=M(H)$ is a sufficiently smooth function of $H$, with an inverse $H=H(M)$. The $\beta$-function of the full running with $H$ is obtained from  the total derivative of the VED with respect to $M$, at $M=H$.  The total derivative reads
\begin{equation}\label{eq:DerivTotal}
\begin{split}
\beta_{\rm vac}(M,H)&\equiv M\frac{d\rv(M,H(M))}{dM}=M\left(\frac{\partial\rv}{\partial M}+\frac{\partial\rv}{\partial H}\frac{dH}{dM}\right)\\
&=\beta_{\rv}+M\frac{\partial\rv}{\partial H}\frac{dH}{dM}\,.
\end{split}
\end{equation}
Using equations \eqref{RenVDEexplicit}  and \eqref{eq:RGEVED1}, and considering only the ${\cal O}(H^2)$ contributions  for the present universe, we find:
\begin{equation}\label{eq:DerivTotal2}
\begin{split}
\beta_{\rm vac}(M,H)=&\left(\xi-\frac{1}{6}\right)\frac{3 H^2 }{8 \pi^2 }\left(M^2-m^2\right)\\
&+M\left(\xi-\frac16\right)\frac{3H}{8\pi^2}\left(M^2-m^2+m^2\ln\frac{m^2}{M^2}\right)\frac{ dH}{dM}\,.
\end{split}
\end{equation}
This equation  holds for any function $M=M(H)$. If we  specialize to $M=H$, we obtain  once more Eq.\,\eqref{eq:DerivTotal3} within the same approximation (q.e.d.).
This `total' $\beta$-function for the VED running has a different sign (and coefficient) as compared to the partial one  in Eq.\eqref{eq:betarhoLambda2}, since the former explores the full evolution (total derivative) with respect to $M$ and not just the variation of $\rv$ with $M$ at fixed $H$ (as in the latter). The above equation \eqref{eq:DerivTotal3}  tells us that if $\xi>1/6$ (resp. $\xi<1/6$) the VED decreases (resp. increases) with expansion, and therefore the VED effectively behaves as quintessence (resp. phantom DE).  Interestingly enough, we find that the quantum vacuum can mimic dynamical DE\cite{Moreno-Pulido:2022phq,Moreno-Pulido:2022upl}. See Sec.\,\ref{eq:RVMPheno} for more details.

The above discussion should make clear once and for all that the physical running of the VED obeys a $\beta$-function with the general behavior $\beta_{\rv}\sim m^2 H^2$, which is much softer than that of the $\rL$ term, $\beta_{\rho_\CC}\sim m^4$. As noted, the latter has been repeatedly discussed in the literature\cite{Brown:1992db,Akhmedov:2002ts,Ossola:2003ku,Martin:2012bt,Koksma:2011cq,Visser:2016mtr,Gorbar:2002pw}, but in fact does not reflect the physical behavior of the VED running (nor that of $G$). While in some cases the reason is that the approach to the calculation of the VED is Minkowskian \textit{ab initio}, in others the approach is made in curved spacetime but without being sensitive to the background geometry for the relevant terms under discussion, this sensitivity being necessary, of course,  so as to obtain the VED as a function of the expansion rate. In all these situations, one is left again and again with different forms of the same result \eqref{VEDMink}, which is completely blind to the effects of the cosmological background.

The proper way to access the relevant physical effects is to undertake the vacuum EMT in curved spacetime  by  introducing the off-shell character of the renormalization procedure, since after all this is in the spirit of the RG; and, at the same time,  one has to correctly identify from the very beginning the VED  as being the sum of the CC term in the action and the ZPE in curved spacetime (plus additional SSB contributions, if present), not just through the computation of the coupling renormalization of the term in the action. This is what we have done here in the RVM formulation and we have obtained the physical running of the relevant quantities. An alternative approach within the RVM that fully corroborates our result is provided in  Sec.\,\ref{sec:effActionVED}, which makes use of the effective action method introduced in the next section.

\section{Effective action approach in the RVM framework}\label{sec:effAction}

The renormalized EMT considered in Sec.\,\ref{sec:RenormEMT} is related to the (renormalized)  effective action of vacuum\cite{DeWitt:1967ub},  $W$,  which describes the vacuum fluctuations of the quantized matter fields of QFT in FLRW spacetime:
\begin{equation}\label{eq:DefW}
\langle T^{\delta\phi}_{\mu\nu}\rangle=-\frac{2}{\sqrt{-g}} \,\frac{\delta W}{\delta g^{\mu\nu}}\,.
\end{equation}
This functional relation allows a nontrivial cross-check for the computation of the renormalized EMT.  We have previously computed the latter directly  by expanding the solution of the Klein-Gordon equation $(\Box-m^2-\xi R)\phi=0$  in Fourier modes and letting the creation and annihilation  operators to act on the adiabatic vacuum with the usual commutation relations. We may alternatively compute it using the above effective action  $W$ through the DeWitt-Schwinger expansion\cite{{DeWitt1975,ParkerCargese1978},Birrell:1982ix,Fulling89,Parker:2009uva}. In the RVM approach, this implies correcting the standard  DeWitt-Schwinger coefficients of such an expansion by taking into account the off-shell effects at the scale $M$, and finally use Eq.\,\eqref{eq:DefW} to retrieve with this method the renormalized EMT at the renormalization point $M$ \cite{Moreno-Pulido:2022phq}.

The vacuum effective action $W$ is computed from the trace of the logarithm of the inverse Green's function in curved spacetime\cite{SolaPeracaula:2022hpd}.  More specifically,
\begin{equation}\label{eq:EAW}
\begin{split}
W= &\frac{i\hbar}{2}Tr \ln (-G_F^{-1})=\frac{i\hbar}{2}Tr \ln (-G_F)^{-1}= -\frac{i\hbar}{2}Tr \ln (-G_F)\\
=&  -\frac{i\hbar}{2} \int d^4 x \sqrt{-g}\lim\limits_{x\to x'} \ln\left[ -G_F(x,x')\right]\equiv \int d^4 x \sqrt{-g}\, L_W\,.
\end{split}
\end{equation}
The integrand in the last term defines the  Lagrangian density $\sqrt{-g}\, L_W$ associated with the quantum vacuum effective action.  In the above expression,  we have  kept $\hbar$ explicitly so as to emphasize the pure quantum nature of the above action. In subsequent expressions, we set $\hbar=1$.

For the explicit calculation of $L_W$ one follows the method based on the  DeWitt-Schwinger expansion of the effective action. However, the standard method must be corrected for so as to take into account that the expansion is not on-shell in our case. Therefore, we need to compute the off-shell curved spacetime propagator $G_F$ involved in the previous formulas. Such a propagator satisfies the equation
\begin{equation}\label{KGPropagatorOffShelltext}
\left(\Box_x-M^2-\Delta^2-\xi R(x)\right)G_F(x,x^\prime)=-\left(-g(x)\right)^{-1/2}\delta^{(n)}(x-x^\prime)\,,
\end{equation}
where $\delta^{(n)}$ is the Dirac $\delta$-distribution in $n$ spacetime dimensions. It may be  convenient to keep $n$ general if one is (optionally) interested in using DR to regularize UV divergences. The renormalization itself does not depend on that, cf.\cite{Moreno-Pulido:2020anb,Moreno-Pulido:2022phq}  An exact solution of \eqref{KGPropagatorOffShelltext} is not generally possible, but $G_F$ can be determined by means of an adiabatic expansion of the solution, in complete analogy with the method we have used before to compute the vacuum EMT from the mode expansion of the fields (cf. Sec. \ref{sec:AdiabaticVacuum}).  Again, the key to the off-shell method is the presence of the floating scale $M$ and the quantity $\Delta^2\equiv m^2-M^2$, which also appeared in the mode expansion method.
We see once more that  $\Delta^2$ must be treated as being of  adiabatic order $2$, since it goes together with the $R$ term in the above propagator equation. We cannot combine $M^2$ and $\Delta^2$ in it, since they belong to different adiabatic orders, and this fact must be respected when solving the equation for $G_F$ by adiabatic expansion. The procedure is a bit cumbersome. The calculation in the on-shell case can be found in \cite{Birrell:1982ix,Fulling89,Parker:2009uva}.

In the off-shell situation one proceeds similarly, but with due account of the new ingredients related to the scale $M$, which will produce additional terms non-existing in the traditional on-shell expansion. The final result for the effective Lagrangian of vacuum defined in Eq.\eqref{eq:EAW} in the off-shell case is the following\cite{Moreno-Pulido:2022phq} (see Appendix \ref{sec:appendixB} for a summary of the calculation):
\begin{equation}\label{eq:effLagrangian}
\begin{split}
L_W &=\frac{\mu^{4-n}}{2(4\pi)^{n/2}} \sum_{j=0}^\infty \hat{a}_j (x) \int_0^\infty (is)^{j-1-n/2}e^{-iM^2 s}ids\\
&=\frac{1}{2(4\pi)^{2+\frac{\varepsilon}{2}}}\left(\frac{M}{\mu}\right)^{\varepsilon}\sum_{j=0}^\infty \hat{a}_j (x) M^{4-2j}\Gamma \left(j-2-\frac{\varepsilon}{2}\right)\,.
\end{split}
\end{equation}
This result constitutes the  heat-kernel expansion of the propagator with the DeWitt-Schwinger technique\,\cite{DeWitt1975}.
Here we use DR to regularize the divergences and define $\varepsilon\equiv n-4$. The limit $\varepsilon \rightarrow 0$ is understood. As usual,  $\Gamma$ is Euler's gamma function and $\mu$ is  the aforementioned 't Hooft's mass unit to keep the effective Lagrangian with natural dimension  $+4$ of energy  in $n$ spacetime dimensions. The final results must be independent of $\mu$, which is an unphysical parameter\footnote{Our final renormalized result depends on $M$ only, but not on $\mu$.  This is in contrast to the approach of\,\cite{KohriMatsui2017}, which lacks of our subtraction off-shell prescription at $M$, and the final results  still carry explicit $\mu$-dependence and the renormalized results lead to  $\sim m^4$ contributions responsible for extreme fine-tuning in the CCP.}.   The sum over   $j=0,1,2,...$  involves even adiabatic orders only.  However, as warned, the presence of $\Delta^2$ modifies the  DeWitt-Schwinger coefficients $\hat{a}_i (x)$. Up to fourth adiabatic order, the new coefficients read as follows:

\begin{equation}\label{eq:ModifDWScoeff}
\begin{split}
&\hat{a}_0 (x)=1=a_0 (x),\\
&\hat{a}_1 (x)=a_1(x)-\Delta^2=-\left(\xi-\frac{1}{6}\right)R-\Delta^2 ,\\
&\hat{a}_2 (x)=a_2(x)+\frac{\Delta^4}{2}+\Delta^2 R \left(\xi-\frac{1}{6}\right)=\frac{1}{2}\left(\xi-\frac{1}{6}\right)^2R^2+\frac{\Delta^4}{2}\\
&\phantom{XXXX}+\Delta^2 R \left(\xi-\frac{1}{6}\right)-\frac{1}{3}Q^\lambda_{\ \lambda}\,,
\end{split}
\end{equation}
with
\begin{equation}\label{eq:traceQ1}
\frac{1}{3}{Q^\lambda}_\lambda \equiv-\frac{1}{120}C^2+\frac{1}{360}E+\frac{1}{6}\left(\xi-\frac{1}{5}\right)\Box R\,,
\end{equation}
in which $E$ is the Euler density (whose action defines the Gauss-Bonnet term) and $C^2$ is the square of the Weyl tensor.
The coefficients  $a_i(x)$ are the ordinary  DeWitt-Schwinger coefficients for  $\Delta=0$ (on-shell expansion).   The effective Lagrangian \eqref{eq:effLagrangian} is manifestly UV-divergent  since Euler's $\Gamma$-function is divergent for $j=0,1,2$  in  $n=4$ spacetime dimensions. Therefore, renormalization is required.  We avoid using the MSS in this context; instead, we utilize the off-shell ARP  adapted to the effective action method.  In fact,  we define the renormalized vacuum effective Lagrangian at the
renormalization point M as follows:
\begin{equation}\label{eq:LWrenormalized}
L_W^{\rm ren}(M )= L_W (m)-L_W^{(0-4)}(M)\equiv  L_W (m)-L_{\rm div} (M)\,,
\end{equation}
where  $L_{\rm div}(M)\equiv  L_W^{(0-4)}(M)$ is the divergent part.  Notice that  $L_W (m)$  and $L_{\rm div} (M)$ are both divergent, but the former may involve the full DeWitt-Schwinger expansion at any desired order, whereas the latter is computed only up to order $4$.  This subtraction prescription, which is  performed at the level of the effective vacuum Lagrangian, is the exact analog of the off-shell ARP definition \eqref{EMTRenormalized}  for the renormalized vacuum EMT and it suffices to make $L_W^{\rm ren}(M)$ finite.  Using Eqs.\,\eqref{eq:effLagrangian}-\eqref{eq:ModifDWScoeff}, and after some relatively lengthy algebra, one can check that the pole terms (which appear in the limit $\varepsilon\to0$) exactly cancel out in Eq.\,\eqref{eq:LWrenormalized}, leaving the following finite result:
\begin{equation}\label{eq:LWrenM}
\begin{split}
L_W^{\rm ren}(M)=\delta \rho_\Lambda(M)-\frac{1}{2}\delta\MPl^2(M) R-\delta \alpha_Q(M) \frac{{Q^\lambda}_\lambda}{3}-\delta \alpha_2(M) R^2+\cdots\,,
\end{split}
\end{equation}
The dots stand for higher order adiabatic contributions that decouple at large $m$, and
\begin{equation}\label{eq:deltacouplings}
\begin{split}
&\delta\rL(M)=\frac{1}{8\left(4\pi\right)^2}\left(M^4-4m^2M^2+3m^4-2m^4 \ln \frac{m^2}{M^2}\right),\\
&\delta\MPl^2(M) =\frac{\left(\xi-\frac{1}{6}\right)}{(4\pi)^2}\left(M^2-m^2+m^2\ln \frac{m^2}{M^2}\right),\\
&\delta \alpha_Q(M)=-\frac{1}{2(4\pi)^2}\ln\frac{m^2}{M^2},\\
&\delta{\alpha_2}(M)=\frac{\left(\xi-\frac{1}{6}\right)^2}{4(4\pi)^2}\ln\frac{m^2}{M^2}.
\end{split}
\end{equation}
 These quantities  are finite renormalization effects which are generated in the  subtraction \eqref{eq:LWrenormalized}. We have defined $\delta\rL(M)=\rL(M)-\rL(m)$ and $\delta\MPl^2(M)= \MPl^2(M)-\MPl^2(m)$, where  $\MPl^2(M)=\frac{G^{-1}(M)}{8\pi}=\frac{\mpl^2 (M)}{8\pi}$ is the reduced Planck mass squared.   In general, for two arbitrary values $M_1$ and $M_2$ of the scale, we have
\begin{equation}\label{eq:rLdif}
\rho_\Lambda(M_2)-\rho_\Lambda(M_1 )=\frac{1}{8(4\pi)^2}\left(M_2^4-M_1^4-4m^2(M_2^2-M_1^2)-2m^4\ln \frac{M_1^2}{M_2^2}\right)\,,
\end{equation}
\begin{equation}\label{eq:MPLdif}
\MPl^2(M_2)-\MPl^2(M_1)=\frac{\left(\xi-\frac{1}{6}\right)}{(4\pi)^2}\,\left(M_2^2-M_1^2+m^2\ln \frac{M_1^2}{M_2^2}\right)\,.
\end{equation}
Similarly with the dimensionless coefficients $\alpha_2$ and $\alpha_Q$ of the HD terms.  We confirm from the above results that the dependence on  $\mu$ has canceled altogether along with the poles. It should be emphasized that the subtracted term $L_{\rm div}(M)$ at the scale $M$  in \eqref{eq:LWrenormalized} involves not just the UV-divergences but also the full expression obtained up to adiabatic order 4 ($j=0,1,2$)  in the DeWitt-Schwinger expansion\,\eqref{eq:effLagrangian}, and therefore it includes their finite parts. This is consistent with the off-shell ARP procedure \eqref{EMTRenormalized}  which we used before from the direct mode expansion of the EMT. Thus, as promised, no arbitrary additive constants are left within the ARP procedure, in contrast to e.g. MSS renormalization.

\subsection{RG-invariance and running couplings}\label{sec:RGinvariance}

While the vacuum effective Lagrangian $L_W^{\rm ren}(M)$ computed in the previous section  is not RG-invariant (since it is $M$-dependent), the full vacuum effective Lagrangian involving also the extended classical part (EH part plus HD geometric terms) must be RG-invariant, of course.  In fact, the classical part of the Lagrangian reads
\begin{equation}\label{eq:LEHHD}
\begin{split}
L_G^{\rm cl.}=L_{EH}+L_{HD}=&-\rho_\Lambda +\frac{1}{2}\MPl^2R+\alpha_Q \frac{{Q^\lambda}_\lambda}{3}+\alpha_2 R^2\\
=&-\rho_\Lambda+\frac{1}{2}\MPl^2  R+\alpha_1 C^2+\alpha_2 R^2+\alpha_3 E+\alpha_4\Box R\,,
\end{split}
\end{equation}
where the coefficients $\alpha_i$  for $i=1,3,4$ in the second expression can be easily related to those of \eqref{eq:traceQ1} using standard formulae\cite{Moreno-Pulido:2022phq}.
Thus, the full vacuum effective Lagrangian finally reads
\begin{equation}\label{eq:Full-Leff1}
\begin{split}
L_{\rm eff}&=L_G^{\rm cl.}(M)+L_W^{\rm ren}(M)=-\rL(M)+\frac{1}{2}\MPl^2(M)  R+\alpha_1(M) C^2+\alpha_2(M) R^2+\alpha_3(M) E\\
&+\alpha_4(M)\Box R+\delta \rL(M)-\frac{1}{2}\delta\MPl^2(M) R-\delta \alpha_Q(M) \frac{{Q^\lambda}_\lambda}{3}-\delta \alpha_2(M) R^2+\cdots\\
&=\left[-\rL(M)+\delta\rL(M)\right]+\frac12\left[\MPl^2(M)-\delta\MPl^2(M)\right] R+\left[\alpha_1 (M) +\frac{1}{120} \delta\alpha_Q(M)\right] C^2\\
&+\left[\alpha_3 (M)-\frac{1}{360}\delta\alpha_Q(M)\right] E+\left[\alpha_4 (M)-\frac{1}{6}\left(\xi-\frac15\right) \delta\alpha_Q(M)\right]\Box R\\
&+\left[\alpha_2 (M)-\delta\alpha_2(M)\right] R^2+\cdots\\
\end{split}
\end{equation}
where we have used Eq.\,\eqref{eq:traceQ1} in the second equality, and the dots stand for the same higher order adiabatic contributions which have not been included in \eqref{eq:LWrenM}, i.e. the subleading finite pieces emerging from the DeWitt-Schwinger expansion \,\eqref{eq:effLagrangian}.

The above Lagrangian $L_{\rm eff}$ is the full effective Lagrangian which is RG-invariant\,\cite{Moreno-Pulido:2022phq}, since the dependence on $M$ cancels among the various terms. In fact, we have already paired the terms appropriately so that each expression in square brackets defines a coupling whose derivative with respect to $M$ must be zero.  Therefore, defining the corresponding $\beta$-functions as
\begin{equation}\label{eq:BetaFunction}
  \beta_i=M \frac{\partial \lambda_i(M)}{\partial M}
\end{equation}
for each of the couplings  $(\lambda_i=\rL,\MPl^2,\alpha_1,...,\alpha_4)$, and using the relations \eqref{eq:deltacouplings},  we find the renormalization group equation (RGE) for each one of them:
\begin{equation}\label{eq:BetaFunctionrL}
\beta_{\rL} (M)=\frac{1}{2(4\pi)^2}(M^2-m^2)^2
\end{equation}
\begin{equation}\label{eq:BetaFunctionMPl}
\beta_{\MPl^2} (M)=\frac{\left(\xi-\frac{1}{6}\right)}{8\pi^2} (M^2-m^2)
\end{equation}
and
\begin{equation}\label{eq:BetaFunctions12}
\begin{split}
\beta_{\alpha_1}=-\frac{1}{120(4\pi)^2}\ \ \ \ \ \
 \beta_{\alpha_2}=-\frac{\left(\xi-\frac{1}{6}\right)^2}{2(4\pi)^2}
 \end{split}
\end{equation}
 \begin{equation}\label{eq:BetaFunctions34}
\begin{split}
  \beta_{\alpha_3}=\frac{1}{360 (4\pi)^2}\ \ \ \ \ \ \
   \beta_{\alpha_4}=\frac{\xi-\frac15}{6(4\pi)^2}\,.
\end{split}
\end{equation}
Of particular interest are the first two RGE's since they involve the parameter $\rL$ and the reduced Planck mass squared  $\MPl^2(M)=\frac{G^{-1}(M)}{8\pi}$.  The first of them leads to a RGE that coincides with Eq.\,\eqref{eq:betarhoLambda}, as it should, confirming that the effective action approach is consistent with the direct EMT computation. Integrating this equation  between the scales $M_0$ and $M$ we retrieve  Eq.\eqref{SubtractionrL}. On the other hand, Eq.\,\eqref{eq:BetaFunctionMPl} leads to the following RGE for the inverse Newton's coupling:
\begin{equation}\label{eq:RGinvG}
M\frac{d}{d M}\left(\frac{1}{16\pi G}\right)=\frac{(\xi-\frac16)}{(4\pi)^2}\,(M^2-m^2)\,.
\end{equation}
It can also be integrated  to compute the scaling evolution of the gravitational coupling between  $M_0$ and $M$ and one finds Eq.\eqref{SubtractionG} --  another consistent result with the direct EMT method. It can also be recast
\begin{equation}\label{eq:RGENewton}
G(M)=\frac{G(M_0)}{1+\frac{\left(\xi-\frac{1}{6}\right)}{2\pi}G(M_0)\left(M^2-M_0^2-m^2\ln \frac{M^2}{M_0^2}\right)}\,.
\end{equation}
 As for the $\beta$-function coefficients for the HD terms \eqref{eq:BetaFunctions12}-\eqref{eq:BetaFunctions34}, they are  well-known to be correct\cite{Nelson:1984sy}, see also \cite{Sola:2013gha}.  We must emphasize once more that $\beta_{\rL}$ given by \eqref{eq:BetaFunctionrL} is not the running of the physical cosmological term, but just that of a coupling in the EH action. As noted previously, this has been repeatedly discussed in the literature\cite{Brown:1992db,Akhmedov:2002ts,Ossola:2003ku,Martin:2012bt,Koksma:2011cq,Visser:2016mtr,Gorbar:2002pw} and in some cases may have proven confusing.  From our point of view, one should never speak of the running of the cosmological constant when referring to $\rL(M)$, or at least not without adding that the physical running of the vacuum energy density $\rv(M)$ is governed by a completely different equation. The physical running of the cosmological term  stems from the running of the renormalized VED, which we have clearly identified in Sec.\,\ref{sec:RVMbetafunction}. Finally, in contrast to other results\cite{Gorbar:2002pw}, we do also find the running law of $G$ with the scale, as given by Eq.\,\eqref{eq:RGENewton}\cite{Moreno-Pulido:2022phq}. From it, the physical running can  be obtained by evaluating the change of $G$ between the two cosmic epochs $M_0=H_0$ and $M=H$, which yields the result
 \begin{equation}\label{eq:runGH}
G(H)=\frac{G_N}{1-\frac{\left(\xi-\frac{1}{6}\right)}{2\pi}\frac{m^2}{m^2_{\rm Pl}}\ln \frac{H^2}{H_0^2}}=\frac{G_N}{1-\epsilon\ln \frac{H^2}{H_0^2}}\,,
\end{equation}
in which  $G_N$ defines the local gravity value usually associated with the inverse Planck mass squared:  $ G(H_0)=G_N=1/m^2_{\rm Pl}$ (in natural units), and we have neglected $H^2-H_0^2$ versus the logarithmic term $m^2\ln \frac{H^2}{H_0^2}$, as the ratio between the first and second  is of order $H_0^2/m^2\ll 1$, for $H$ close to $H_0$. The predicted running of $G$ with $H$ is, therefore,  very mild, since it is logarithmic and with a very tiny coefficient in front of it:
\begin{equation}\label{eq:epsilonparameter0}
\epsilon\equiv\frac{1}{2\pi}\,\left(\xi-\frac{1}{6}\right)\,\frac{m^2}{\mpl^2}\,.
\end{equation}
The result \eqref {eq:runGH} will be reconfirmed using a completely different reasoning path involving the calculation of the vacuum pressure, see Sec.\,\ref{sec:RunningG}. As with the VED, it will be shown that the running of $G$ ultimately stems from the renormalization of the EMT  (the most basic physical quantity in our approach) and not just from mere coupling renormalization. This can already be foreseen from the considerations in Sec.\ref{sec:GeneralizedEqs}  and the derivation of Eq.\eqref{SubtractionG} from the generalized Einstein's equations, which involve the full renormalized form of the EMT.

\subsection{Running of the VED from the vacuum effective action}\label{sec:effActionVED}

Once we are secured as to the delicate issues of RG-invariance of the full effective action and the RGE's of the various couplings, let us finally derive the vacuum EMT directly from the renormalized vacuum effective action \eqref{eq:EAW} using the vacuum effective Lagrangian \eqref{eq:LWrenM}:
\begin{equation}\label{eq:effActionLren}
\begin{split}
W_{\rm ren}(M)=\int d^4 x\sqrt{-g} \left( \delta \rL(M)-\frac{1}{2}\delta\MPl^2(M) R-\delta \alpha_Q(M) \frac{{Q^\lambda}_\lambda}{3}-\delta \alpha_2(M) R^2\right)\,.
\end{split}
\end{equation}
The renormalized vacuum EMT now follows from functionally differentiating the above expression according to  Eq.\eqref{eq:DefW}.
Notice that we may drop  the contribution  from the Euler density  $E$ in ${Q^\lambda}_\lambda$, Eq.\,\eqref{eq:traceQ1}, since the Gauss-Bonnet term associated with it is a topological invariant  in $n=4$ spacetime dimensions. Similarly we may drop also the total derivative term $\Box R$ at the level of the action, and also that of $C^2$ which vanishes for the FLRW metric (a conformally flat spacetime). So there is virtually no contribution to the EMT from the terms in \eqref{eq:traceQ1}.  The remaining terms in \eqref{eq:effActionLren}  are easily handled, and from a straightforward calculation, we find
\begin{equation}\label{eq:TrenMm0FLRWMm}
\begin{split}
\langle T_{\mu\nu}^{\delta \phi}\rangle_{\rm ren}(M)=
& \delta \rL(M) g_{\mu\nu}+\delta\MPl^2(M) G_{\mu\nu}+\delta\alpha(M)\leftidx{^{(1)}}{\!H}_{\mu\nu}\,,
 \end{split}
\end{equation}
where, in particular, we have used the standard metric variations
\begin{equation}\label{eq:Gmn}
\frac{1}{\sqrt{-g}}  \frac{\delta}{\delta g^{\mu\nu}}  \int d^4x  \sqrt{-g} R= R_{\mu\nu}-\frac12 R g_{\mu\nu}=G_{\mu\nu}
\end{equation}
and
\begin{equation}\label{eq:H1munu}
\frac{1}{\sqrt{-g}}  \frac{\delta}{\delta g^{\mu\nu}}  \int d^4x  \sqrt{-g} R^2= \leftidx{^{(1)}}{\!H}_{\mu\nu}
\end{equation}
as well as the relation $2\delta\alpha_2=\delta\alpha$ (where $\alpha$ was defined in \eqref{eq:MEEs}). We recall that the coefficients of the various tensor terms on the \textit{r.h.s.} of Eq.\,\eqref{eq:TrenMm0FLRWMm} are given explicitly  by equations\,\eqref{eq:deltacouplings}.
Finally, the ZPE is the $00th$-component of the vacuum EMT. Therefore, expressing the result in terms of the cosmic time and the corresponding Hubble function $H=\dot{a}/a$, we find the renormalized ZPE:
\begin{equation}\label{eq:T00Integrated}
\begin{split}
&\left\langle T_{00}^{\delta \phi}\right\rangle_{\rm ren}(M)
=\frac{a^2}{128\pi^2}\left(-M^4+4m^2M^2-3m^4+2m^4\ln \frac{m^2}{M^2}\right)\\
&- \left(\xi-\frac{1}{6}\right)\frac{3 a^2 H^2}{16\pi^2}\left(m^2-M^2-m^2\ln \frac{m^2}{M^2}\right)\\
&+\left(\xi-\frac{1}{6}\right)^2\frac{9a^2}{16\pi^2}\left(6H^2\dot{H}+2H\ddot{H}-\dot{H}^2\right)\ln \frac{m^2}{M^2}+\mathcal{O}\left(\frac{H^6}{m^2}\right)\,,
\end{split}
\end{equation}
where we have used again the relations in the footnote on p.22. The notation $\mathcal{O}(H^6 / m^2)$ schematically denotes the terms of adiabatic order 6. The result \eqref{eq:T00Integrated} obtained from the effective action approach  coincides exactly with the result \,\eqref{Renormalized2} obtained from the mode expansion method upon converting the Hubble rate in conformal time to cosmic time. The two methods are therefore fully consistent.

The following comment is in order to close this section. It aims to clarify the relationship between our off-shell renormalization approach in QFT in curved spacetime and other formulations in which QG is taken into account as the fundamental framework. The two formulations are of course quite different, but the resemblance is particularly clear in the context of the effective action language of this section. Since we are working in the semiclassical formulation, our use of the renormalization group is \textit{not} based on seeking, say,  a scale-dependent, Wilsonian type, average gravitational action $\Gamma_k[g_{\mu\nu}]$ defining an effective field theory of QG at a certain infrared energy cutoff, $k$, below which the running of the VED, $G$ and other quantities stops\cite{Reuter:1996cp,Bonanno:2001xi}. Notwithstanding, there is some analogy between the two approaches,  as could be expected.  In the Wilsonian RG formulation,  only frequencies with momenta larger than $k$ are integrated out in the path integral while those with momenta smaller than $k$ are not included. In fact, $\Gamma_k$ is a functional integral similar to the ordinary effective action in which the
contributions of all field modes with momenta smaller than $k$ are suppressed. Thus, $\Gamma_k$ interpolates between the classical action, say the EH action (for $k\to \infty$) and the ordinary effective action (for $k\to 0$). Not surprisingly, however, as in any other formulation, the physical identification of the cutoff $k$ is ambiguous  and hence the procedure is ultimately phenomenological; for example, one can propose  $k\sim 1/t$ ($t$ being the cosmic time)\cite{Reuter:1996cp}. One could also take  $k\sim H$ (which is essentially equivalent for most of our past and again is a viable option in the future if one moves to $k\sim \sqrt{\Lambda}\sim H$)\cite{Bonanno:2007wg} . The scale setting option $k=H$ is the closest to ours and it was  first proposed in \cite{ShapSol}.  In this case, the truncated effective action $\Gamma_k$ involves only physics above the cutoff scale $k\sim H$, not below it   In an analogous way,  in our off-shell RVM approach,  we obtain a renormalized theory  at a floating scale $M$, which we eventually set to $H$ to connect with phenomenology. This defines the infrared cutoff of our vacuum effective action in the semiclassical approximation\footnote{As proposed in \cite{Fritzsch:2012qc} -- see also Sec.\,\eqref{eq:GeneralRVM} -- couplings can have a double running pattern (ordinary and cosmological), $\alpha_i=\alpha_i(\mu,H)$,  and therefore can be sensitive to a large number of unsuspected light d.o.f's that might lurk in the far infrared\,\cite{Shapiro:1999zt}. In this sense we cannot exclude the existence of important renormalization effects which are not captured by the above coarse-graining procedure.}.  The expansion history of the universe above that value gets connected through the renormalization group, although with a proviso: the running is  triggered by the quantum effects from the matter fields only, which are the only ones involved in the semiclassical approach. Whether this is just an approximation or not depends, of course, on whether QG exists as a more accurate theory.

\subsection{Renormalized vacuum pressure}\label{sec:RenVacuumPressure}

Another important quantity in our QFT analysis of the RVM is the vacuum pressure. We should not assume \textit{a priori} (as frequently done in the literature) that it satisfies the canonical equation of state (EoS) $\Pv=-\rv$ since quantum effects can introduce corrections to it.  The vacuum pressure in our QFT context can be obtained using the isotropy of the vacuum and the assumption of it being a perfect fluid.  An observer of $4$-velocity $ u^\mu$ measures a  vacuum pressure $\Pv=\frac13\left(g^{\mu\nu}+u^\mu u^\nu\right) \langle T_{\mu\nu}^{\rm vac}\rangle$. In the rest frame of the observer, the $4$-velocity is $u^\mu=\left(1/\sqrt{-g_{00}},0,0,0\right)=(1/a,0,0,0)$, expressed in the conformal metric,  and we have  $g^{00}+u^0 u^0=0$. Thus, using  Eq.\,\eqref{RenEMTvacuum}, the renormalized vacuum pressure is
\begin{equation}\label{eq:VacuumPressureDef}
\Pv(M)=\frac13 g^{kk}\langle T_{kk}^{\rm vac}\rangle_{\rm ren}(M)= \frac{\langle T_{ii}^{\rm vac}\rangle_{\rm ren}(M)}{a^2}= -\rho_\Lambda (M)+ \frac{\langle T_{ii}^{\delta \phi} \rangle_{\rm ren}(M)}{a^2}\,,
\end{equation}
(for any fixed $i=1,2,3$, and summation over $k=1,2,3$ is understood).
The isotropy of vacuum has been used in our assumption that each spatial component counts equal. The same property implies $\langle T^{\delta \phi} \rangle_{\rm ren}(M)=g^{\mu\nu}\langle T^{\delta \phi}_{\mu\nu} \rangle_{\rm ren}(M)=\left(-\langle T_{00}^{\delta \phi} \rangle_{\rm ren}(M)+3\langle T_{ii}^{\delta \phi} \rangle_{\rm ren}(M)\right)\frac{1}{a^2}$ for the trace of the vacuum EMT. Thus,
\begin{equation}\label{eq:VacuumT11}
\frac{\langle T_{ii}^{\delta \phi} \rangle_{\rm ren}(M)}{a^2}=\frac{1}{3}\left(\langle T^{\delta \phi} \rangle_{\rm ren}(M)+\frac{\langle T_{00}^{\delta \phi} \rangle_{\rm ren}(M)}{a^2}\right)\,.
\end{equation}
Finally,  using  Eq.\,\eqref{RenVDE} for the renormalized VED to eliminate $\rL(M)$ in favor of $\rv(M)$, the vacuum pressure \eqref{eq:VacuumPressureDef} can be written \cite{Moreno-Pulido:2022phq}:
\begin{equation}\label{eq:VacuunPressure}
    P_\mathrm{vac}(M)=-\rho_\mathrm{vac}(M)+\frac{1}{3}\left(\langle T^{\delta\phi}_\mathrm{ren}\rangle(M)+4\frac{\langle T_{00}^{\delta\phi}\rangle_\mathrm{ren}(M)}{a^2}\right) \,.
\end{equation}
It is obvious from this formula that the effective EoS of vacuum may deviate from the naive one $\Pv=-\rv$, provided the quantity in parentheses in the above expression is nonvanishing. To compute it, we use the renormalized ZPE given by  Eq.\,\eqref{eq:T00Integrated} and take the trace of the renormalized vacuum EMT, Eq.\,\eqref{eq:TrenMm0FLRWMm}. The trace is found to be
\begin{equation}\label{eq:TraceIntegrated}
\begin{split}
&\left\langle T^{\delta \phi}\right\rangle_{\rm ren}(M)
=\frac{1}{32\pi^2}\left(3m^4-4m^2M^2+M^4-2m^2\ln \frac{m^2}{M^2}\right)\\
&+\left(\xi-\frac{1}{6}\right)\frac{3}{8\pi^2} \left(2H^2+\dot{H}\right) \left(m^2-M^2-m^2\ln \frac{m^2}{M^2}\right)\\
&-\left(\xi-\frac{1}{6}\right)^2\frac{9}{8\pi^2}\left(12H^2 \dot{H}+4\dot{H}^2+7H\ddot{H}+\dddot{H}\right)\ln \frac{m^2}{M^2}+\mathcal{O}\left(\frac{H^6}{m^2}\right)\,.
\end{split}
\end{equation}
Substituting \eqref{eq:T00Integrated} and \eqref{eq:TraceIntegrated} into \eqref{eq:VacuunPressure} the vacuum pressure up to fourth adiabatic order can be delivered in a remarkably simple form:
\begin{equation}\label{eq:fullpressure}
\begin{split}
\Pv(M,H)=&-\rv(M,H)+\left(\xi-\frac{1}{6}\right)\frac{1}{8\pi^2}\dot{H}\left(m^2-M^2-m^2\ln\frac{m^2}{M^2}\right)\\
&-\frac{3}{8\pi^2}\left(\xi-\frac{1}{6}\right)^2\left(6\dot{H}^2+3H\ddot{H}+\dddot{H}\right)\ln \frac{m^2}{M^2}\,.
\end{split}
\end{equation}
We should note that in deriving this equation  there is once more an exact cancellation of the quartic mass terms $\sim m^4$ introduced by the quantum corrections on the vacuum pressure. In fact,  one can check explicitly that the sum of the two terms in parentheses on the \textit{r.h.s.} of Eq.\eqref{eq:VacuunPressure} is free from these unwanted quartics.  This is remarkable. It follows that the scaling evolution of the renormalized vacuum pressure is, as the renormalized VED itself,  free from quartic mass dependencies.  In addition, we can see in a manifest way that the net quantum correction to the classical vacuum EoS  vanishes identically  for $H=$const. as it is entirely dependent on time derivatives of $H$.  This feature will play an important role in the next section, when we discuss the inflationary epoch.

Before closing this section, the following observation may be appropriate. With the above formulas connecting the renormalized vacuum energy density and pressure,  we can now understand the issue mentioned in Sec. \,\ref{sec:AdiabaticVacuum} as to why the full renormalized vacuum EMT in the RVM (defined in Sec. \ref{sec:RenormEMT}) generally does \textit{not} satisfy the relation  $\langle T_{\mu\nu}^{\rm vac}\rangle=-g_{\mu\nu} \rv$,  which is nevertheless generally taken for granted in many places of the literature. In fact, this relation clearly hinges on the assumption of the validity of $\Pv=-\rv$ as a `canonical' definition of vacuum. However, it is not generally warranted in the presence of quantum effects that can modify it; and, in fact, we have just seen that it is the case in the RVM, since quantum corrections to this `classical' EoS introduce a small deviation, which can be significant in some cases. For the RVM, this relation is satisfied only during the de Sitter regimes ($H=$const.) occurring in the remote past and in the remote future, as is apparent from Eq.\,\eqref{eq:fullpressure}. See also our discussion of the de Sitter scenario in Sec.\,\ref{sec:RenormEMTdS}.

\section{RVM in the very early universe: inflation without inflaton}\label{sec:RVMinflation}

Inflation is necessary to fix a number of serious inconsistencies of the standard $\CC$CDM model. Without inflation, we could not understand the homogeneity and isotropy of the observed CMB, for instance, or the high level of spatial flatness of the universe at present without fine-tuning.  Furthermore. in the absence of inflation, we could not figure out the origin of structure formation nor the large amount of entropy today: $S\sim 10^{88}$ (in natural units).  Since the standard model $\CC$CDM cannot solve any of these problems, they are usually `fixed' \textit{ad hoc} by postulating the existence of a cosmic scalar field\cite{Guth:1980zm,Linde:1981mu} (later called the ``inflaton''\cite{Nanopoulos:1983up}) which purportedly takes care of these arrangements early on in the cosmic history\cite{KolbTurner,LiddleLyth,Kallosh:2025ijd}. Even if this patch-up  can be efficient and is considered acceptable by many cosmologists, we must admit that it is not very natural, since it is added up without explaining the origin of the inflaton field.  A much more appealing proposal should be to have a unified  theory of the cosmic evolution based on fundamental physics, that is,  an overarching approach accounting for the expansion history from the very early times till today.

\subsection{$H^4$-inflation from RVM}\label{sec:H4infl}

As we shall see, the structure of the RVM involves not only the necessary ingredients to shape  the physics of the current cosmological era through a slightly time-evolving VED, but also the physics of the very early epoch characterized by  a super-fast time evolution. Indeed, it predicts a new mechanism of exponential inflation, called `RVM-inflation', which is characterized by a short period where $H\simeq H_I=$const.  The value $H_I$ must  be very large, presumably around a characteristic Grand Unified Theory (GUT) scale just below the Planck mass.  The matter fields in the GUT constitute the driving force for this framework, nothing else. Therefore,  no \textit{ad hoc} scalar field potential for an inflaton field is required at all, in stark contrast to the traditional approach\,\cite{KolbTurner,LiddleLyth,Kallosh:2025ijd}. In our framework, it is instead the pure work of gravity fueled by the quantum matter effects of the semiclassical theory.

As said, RVM-inflation is characterized by the existence of a short period in the very early cosmic history in which  $H$ remains constant.  However, to trigger inflation with this mechanism, we need powers of $H$ greater than  $H^2$. If we take into account that only even powers of the Hubble rate are allowed by the general covariance of the effective action, the canonical option is $H^4$. We can see that $H^4$-powers are indeed present in our renormalized VED\eqref{DiffVEDphys}. In passing, this explains why the terms in the third line of Eq.\,\eqref{RenVDEexplicit} were not considered in \eqref{DiffVEDphys}, since they do not contribute to RVM-inflation, for which $H=$const. Neglecting also the terms related to the value $H_0$ pertaining to the current universe,  we find that the VED triggering inflation in the primeval universe is as follows:
\begin{equation}\label{eq:VEDinfl}
\rho_\mathrm{vac}(H)\simeq\frac{3\left( \xi-\frac{1}{6}\right)}{16\pi^2}\left[ H^4+H^2m^2\left(\ln\frac{m^2}{H^2}-1\right)\right]\,.
\end{equation}
An important feature can be appreciated from this formula: RVM-inflation is not possible for minimal coupling ($\xi=0$) since the coefficient of the dominant term $H^4$ during inflation cannot be negative. Nor can we assume the conformal limit ($\xi=1/6$). Besides, if many scalar fields are involved with the same mass and non-minimal coupling, a  multiplicity factor can be easily included. If, however, the masses and/or minimal couplings are different, one has to sum over the various contributions in an obvious way.  For simplicity, we shall stick to one scalar field at the moment.

Equation \eqref{eq:VEDinfl} can be  rewritten by making a clear separation between the leading and subleading power of the Hubble rate:
\begin{equation}\label{eq:VEDinf2l}
\rho_\mathrm{vac}(H)=\frac{3\left( \xi-\frac{1}{6}\right)}{16\pi^2}\, H^4+\frac{3{\nu}(H)}{8\pi G_N}\,H^2\,,
\end{equation}
where
\begin{equation}\label{eq:nuH}
{\nu}(H)\simeq\epsilon \left(-1+\ln\frac{m^2}{H^2}\right)\,,
\end{equation}
with $\epsilon$ as defined previously in Eq.\,\eqref{eq:epsilonparameter0}.
Parameter $\epsilon$ is very small since $m^2\ll \mpl^2$, even for particle masses around the GUT scale $M_X\sim 10^{16}$ GeV. It is involved both in the running of $G$ and of the VED.  We shall, however, not consider the running effect of $G$ for the analysis of inflation,  since it is completely negligible. On the one hand the running of $G$ is logarithmic, as evident from Eq.\,\eqref{eq:runGH},  and on the other hand from \eqref{eq:VEDinf2l} we can see that it would lead to  ${\cal O}(\epsilon^2)$ effects.
During the inflationary period, $H$ remains essentially constant around a large value $H_I\gg m$. We shall confirm this fact after solving the cosmological equations in the next section.  Thus, in the inflationary stage we can approximate $\nu(H)$ by its value at $H=H_I$, given by
\begin{equation}\label{eq:nuI}
\nu_I\equiv\nu(H_I)=\frac{1}{2\pi}\,\left(\xi-\frac16\right)\,\frac{m^2}{\mpl^2}\left(-1+\ln \frac{m^2}{H_I^{2}}\right)\,.
\end{equation}
To determine this parameter, we need to estimate $H_I$. This will be done in the next section. When the inflationary period is over, $H$ evolves towards values much smaller than $H_I$, and $\nu(H)$ increases according to Eq.\eqref{eq:nuH}. So when we reach the present universe, $H=H_0$ and the value of $\nu(H_0)$ is much greater than $\epsilon$.  It is useful to realize that at low energies we can replace $H$ with $H_0$ in the log term of \eqref{eq:nuH}, and then the relevant coupling for the late universe is given by
\begin{equation}\label{eq:nueffAprox2bis}
\nueff\equiv\nueff(H_0)=\frac{1}{2\pi} \left( \xi-\frac{1}{6}\right) \frac{m^2}{m_\mathrm{Pl}^2}\,\ln\frac{m^2}{H_0^2}\,.
\end{equation}
This effective parameter is the one that can be fitted by phenomenological analyzes of the RVM, such as\cite{SolaPeracaula:2021gxi,SolaPeracaula:2023swx,deCruzPerez:2025dni}. Because $H_0^2\lll m^2$ for any particle mass, the  low energy coupling \eqref{eq:nueffAprox2bis} is much larger than the high energy one \eqref{eq:nuI}.  We discuss the RVM prediction for the current universe in Sec.\,\ref{eq:RVMPheno} .

\subsection{Solution of the $H^4$ inflationary scenario}\label{sec:solutions}

To solve the cosmological equations with the vacuum energy density given by  Eq.\,\eqref{eq:VEDinf2l},  let us assume that ${\nu}(H)$ in that equation can be approximated by the previously defined constant value $\nu_I$ in \eqref{eq:nuI}. This introduces a great simplification and in fact provides a rather accurate analytical solution, as will be confirmed by comparing with the exact numerical approach.  For constant $\nu(H)$, the VED given by \eqref{eq:VEDinf2l} can be considered a particular case of a more general family of models
\begin{equation}\label{eq:rvm1}
\rv(H)=\frac{3}{\kappa^2} \left( c_0+\nu H^2+ \frac{H^4}{H_I^2} \right)\,,
\end{equation}
in which  $\kappa^2=8\pi G_N=8\pi/\mpl^2$. The coefficients of the terms $H^2$ and $H^4$ can be identified by comparing the above equation with \eqref{eq:VEDinf2l}. In particular, $\nu$ is taken as in \eqref{eq:nuI} within the approximation mentioned above. The identification of $H_I$ will be discussed later.  As for $c_0$, it has no analog in \eqref{eq:VEDinf2l}, but this is because we neglected the terms related to the current universe in the more general  Eq.\eqref{DiffVEDphys}. This is justified for the early universe, of course. The neglected terms determine $c_0$. In the absence of the powers of  $H$ in \eqref{eq:rvm1}, the cosmological constant that we have measured today would just be $\CC=3c_0$. We neglect this constant to discuss the inflationary period, but we still keep the $H^2$ term as it represents the leading subdominant contribution, which will be important in the post-inflationary regime.
Thus, the cosmological equations to solve are the following Friedmann-like equations:
\begin{equation}
\begin{split}\label{eq:friedmann}
3H^2&=\kappa^2 (\rho_\mathrm{rad}+\rho_\mathrm{vac}(H)) \; , \\
3H^2+2\dot{H}&=-\kappa^2 (P_\mathrm{vac}(H)+\frac{1}{3}\rho_\mathrm{rad})\,.
\end{split}
\end{equation}
In these equations, the EoS of matter is $w_{\rm rad}=1/3$ since matter is relativistic in the very early epochs. In fact, matter appears from the decay of the primeval (and highly energetic) vacuum, as we shall show in a moment. Moreover,  $\rv(H)$ is the function of the Hubble rate given by \eqref{eq:rvm1} with $c_0=0$.
For the vacuum EoS, we can use the canonical form  $\Pv(H)=-\rv(H)$ during the very short inflationary period at $H=$const. This is fully justified  in view of  Eq.\eqref{eq:fullpressure}.
This fact allows us to combine the above equations into a single differential equation for the Hubble rate:
\begin{equation}\label{eq:Hdiffeq}
\dot H+2H^2=2(\nu H^2+\frac{H^4}{H_I^2})\,.
\end{equation}
It is apparent from it that there exists an inflationary solution for $H=$const.,  specifically for $H=H_I\sqrt{1-\nu}\simeq H_I$, which is the starting point for inflation.  However, there is an evolution of the Hubble rate from this point onward, which can be computed by solving Eq.\,\eqref{eq:Hdiffeq}. For $\nu=$const an exact analytical solution can be obtained  in terms of the scale factor (recall that $d/dt=a H d/da$), with the following outcome:
\begin{equation}\label{eq:solHubble}
H(a)=\sqrt{1-\nu} \frac{H_I}{\sqrt{1+Da^{4(1-\nu)}}}\,.
\end{equation}
Substituting this result into equations \eqref{eq:friedmann} we can also analytically solve for the energy densities of vacuum and radiation.  The results are readily found:
\begin{align}\label{solDensities}
\rho_\mathrm{vac}(a)&=\frac{3H_I^2(1-\nu)(1+\nu D a^{4(1-\nu)})}{\kappa^2\ (1+D a^{4(1-\nu)})^2}\,,\\
\rho_\mathrm{rad}(a)&=\frac{3H_I^2(1-\nu)^2D a^{4(1-\nu)}}{\kappa^2 (1+D a^{4(1-\nu)})^2}\,.
\end{align}
The integration constant $D$ is determined by the equality condition between vaccum energy density and radiation:  $\rho_\mathrm{vac}(a_\mathrm{eq})=\rho_\mathrm{rad}(a_\mathrm{eq})$, with $a_{\mathrm{eq}}$  the equality point. The latter defines the incipient radiation era.  Thus,   $D=\frac{1}{1-2\nu}a_\mathrm{eq}^{-4(1-\nu)}\equiv a_*^{-4(1-\nu)}$. The above cosmological solution holds good for a VED of the generic form \eqref{eq:rvm1} and for  $c_0$  zero or negligible. Previous analyses addressed that form on pure phenomenological grounds without connection with QFT calculations, see \cite{Lima:2013dmf,Perico:2013mna,Basilakos:2013xpa,Lima:2014hia,Lima:2015mca ,Sola:2015rra,Sola:2015csa,Sola:2014tta,Yu2020}.
Instead, here we have derived the VED form \eqref{eq:rvm1} in the context of QFT in curved spacetime,

It is convenient to rephrase the above solution in terms of the rescaled variables $\hat{a}=a/a_*$ and $\Tilde{H}_I=\sqrt{1-\nu}H_I$, as follows:
\begin{align}
H(\hat{a})&=\frac{\Tilde{H}_I}{\sqrt{1+\hat{a}^{4(1-\nu)}}}\,,\label{eq:infaprox}\\
\rho_\mathrm{vac}(\hat{a})&={\rho}_I\frac{1+\nu\hat{a}^{4(1-\nu)}}{[1+\hat{a}^{4(1-\nu)}]^2}\label{eq:infaprox2}\,,\\
\rho_\mathrm{rad}(\hat{a})&={\rho}_I(1-\nu)\frac{\hat{a}^{4(1-\nu)}}{[1+\hat{a}^{4(1-\nu)}]^2}\label{eq:infaprox3}\,,
\end{align}
where  ${\rho}_I=\frac{3}{\kappa^2}\Tilde{H}_I^2$  is the total energy density at the early start of the inflationary period.  As can be seen,  the initial point $a=0$ is nonsingular in this framework since the Hubble function and the energy densities are functions taking finite values  in it: $H(0)=\Tilde{H}_I$, $\rho_\mathrm{vac}({0})={\rho}_I=\frac{3}{\kappa^2}\Tilde{H}_I^2$ and $\rho_\mathrm{rad}(0)=0$. Thus, we find that RVM-inflation is characterized by a non-singular de Sitter phase.

\begin{figure}[t!]
  \begin{center}
      \resizebox{1.10\textwidth}{!}{\includegraphics{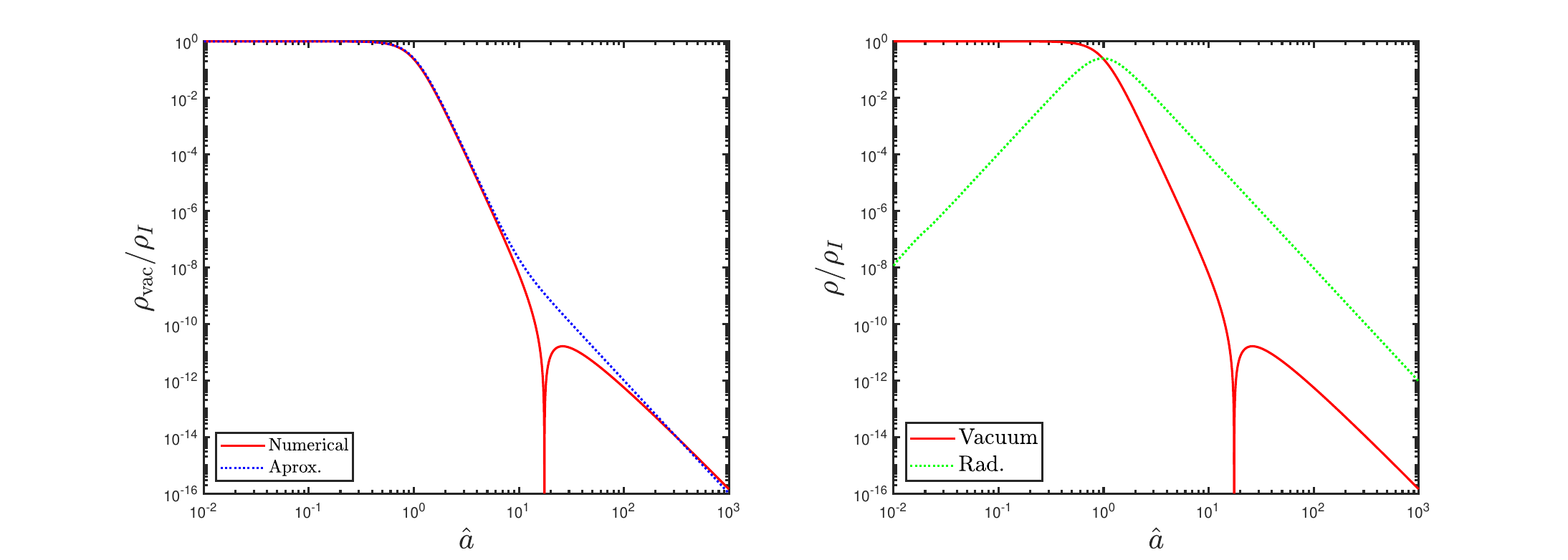}}
\caption{Transition from inflation to the radiation epoch. On the left it is shown the numerical solution of the VED  (solid line) versus the approximate analytical solution.  On the right the radiation energy is also included. In both cases we assume  $\nu=10^{-4}$, and $m$ is of order of a typical GUT scale.  See \cite{SolaPeracaula:2025yco}
for more details.}
\label{Fig:NumSolution}
  \end{center}
\end{figure}


It is also worthwhile to note from  $\rho_\mathrm{rad}(\hat{a})$ above that once we leave inflation behind and penetrate deep into the radiation-dominated epoch ( $\hat{a}\gg 1$, equivalently  $a\gg a_*$) we obtain the  asymptotic behavior $\rho_\mathrm{rad}(\hat{a})\simeq {\rho}_I(1-\nu) \hat{a}^{-4(1-\nu)}\simeq \rho_I a_*^4 \,a^{-4}$ for small $|\nu|\ll1$.  This is essentially the standard evolution  law of the radiation energy density:  $\rho_\mathrm{rad}=\rho_\mathrm{rad}^0{a}^{-4}$. From this fact, we can obtain a useful estimate for the crossover point $a_*$ (the point where $\hat{a}=1$), which depends on a combination of low-energy and high-energy parameters:

\begin{equation}\label{eq:astar}
a_{*}\sim
\left(\Omega_{\rm rad}^0\,\frac{\rco}{\rho_I}\right)^{\frac{1}{4}}\,.
\end{equation}
This relation will be used later on to estimate the numerical value of $a_{*}$ once we estimate the order of magnitude of $\rho_I$.
From equations \eqref{eq:infaprox2} and \eqref{eq:infaprox3} we find that during the radiation-dominated epoch, the vacuum energy is suppressed by the small factor $|\nu|\ll1$ (because $\rho_\mathrm{vac}/\rho_\mathrm{rad}\sim \nu$) and therefore the primordial BBN period can proceed standard without being altered, see the BBN bound on $\nu$ from \cite{{Asimakis:2021yct}}.

As promised, we now turn to the estimate of the inflationary scale ${H}_I$. It can be found by comparing equations \eqref{eq:VEDinf2l} and \eqref{eq:rvm1}:
\begin{equation}\label{eq:HIvalue}
H_I=\sqrt{\frac{2\pi}{\xi-\frac{1}{6}} }\ m_\mathrm{Pl}=\frac{m}{\sqrt{\epsilon}}\,,
\end{equation}
where the second equality follows from our definition of $\epsilon$ in Eq.\,\eqref{eq:epsilonparameter0}. For $0<\epsilon\ll 1$, which can be considered the normal regime for implementing RVM-inflation,  we have $H_I>m$. The case $\epsilon=0$ is clearly not allowed since $H_I$ is not well defined; it corresponds to the conformal limit ($\xi=1/6$). This was also expected from the fact that, in this limit, the coefficient of $H^4$ in Eq.\,\eqref{eq:VEDinf2l}would be zero and there would be no running of the VED to account for inflation. It is also worth noticing that we cannot accommodate $\xi=0$ (minimal coupling) either, for which $\epsilon<0$. In hindsight, this was an important reason to assume $\xi\neq0$ from the beginning.  But not any nonvanishing $\xi$ is admissible, we need $\xi>1/6$, i.e. $\epsilon>0$. Only if this condition is met, the coefficient of the leading power $H^4$ can be positive to trigger inflation. Now, for multiple scalar fields with different non-minimal couplings $\xi_i$ (hence different $\epsilon_i$), these conditions for inflation could be somewhat relaxed.

\begin{figure}[t!]
  \begin{center}
      \resizebox{0.5\textwidth}{!}{\includegraphics{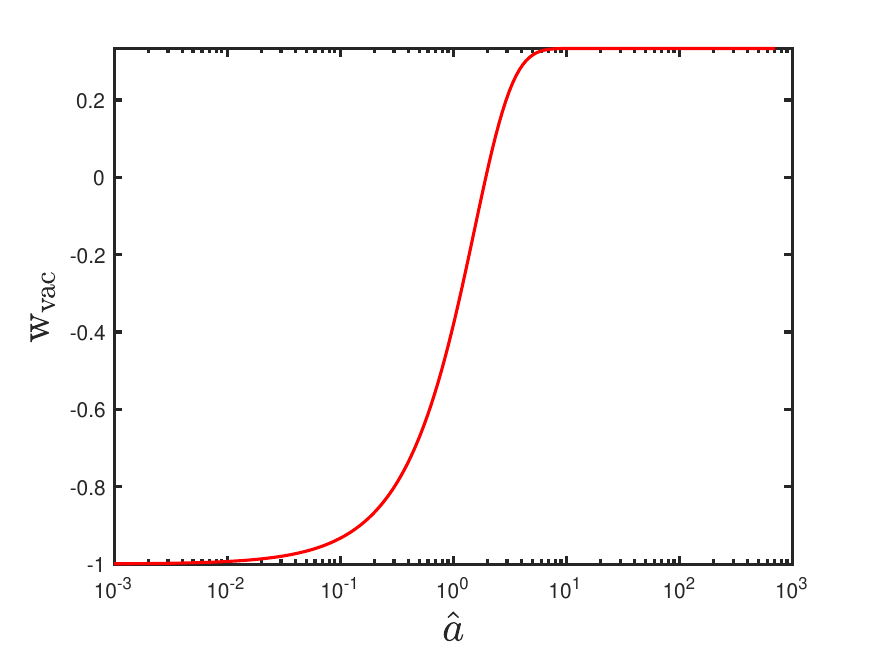}}
\caption{Equation of state parameter  of the running vacuum in the very early universe, as a function of $\hat{a}=a/a_*$. Its value changes from $\wv=-1$) during inflation to $\wv=1/3$ in the radiation-dominated epoch.  }
\label{Fig:EoSvacuumInfl}
  \end{center}
\end{figure}

Let us now compare the numerical analysis of the energy densities in Fig. \ref{Fig:NumSolution} with the results of the analytical solution. From that figure we can see that in the beginning ($\hat{a}=0$) there is no radiation at all while the vacuum dominates and its energy density is the total energy density of the universe, $\rho_I$. This fact is also reflected in the approximate analytical solution given by equations \eqref{eq:infaprox2} and \eqref{eq:infaprox3}. The vacuum decays very fast into radiation and  the inflationary period  goes through a `graceful exit' transition into the standard FLRW radiation-dominated epoch.  When the universe exits the inflationary phase (for $\hat{a}\gg 1$) both energy densities scale approximately as $\rho\sim a^{-4}$, but the VED is
suppressed by a tiny coefficient $\nu$,  which in the exact numerical solution is actually a slowly varying function $\nu(H)$, Eq.\,\eqref{eq:nuH}. Because of this dependence on $H$, this transition can only be handled through numerical analysis. The EoS of the vacuum further evolves and it changes from $-1$ during the inflationary period into $w_\mathrm{vac}\xrightarrow{}1/3$, i.e. it adopts the same EoS as that of radiation -- cf.  Sec.\ref{sec:EoSRVM} for more details. The evolution of the vacuum EoS in the very early universe can be appreciated in Fig. \ref{Fig:EoSvacuumInfl}.

The RVM-inflationary epoch connects in a continuous way with the canonical description of the late universe in the RVM,  which remains very close to the standard FLRW picture and has a mildly evolving VED as a fundamental distinctive feature  (see Sec.\,\ref{eq:RVMPheno}). The RVM account of the early cosmic history, therefore, is able to perform a graceful exit transition from the inflationary era into the radiation epoch. The tiny amount of VED left out during the radiation-dominated epoch is highly suppressed in front of the energy density of radiation. This feature is welcome, of course, in that RVM-inflation does not spoil the  successful explanation  of the BBN  by the standard model of cosmology\,\cite{Asimakis:2021yct}. Overall, it is reassuring to confirm that the analytical and numerical analyzes of the cosmological transition from the inflationary phase into the standard radiation regime convey consistent conclusions.

\subsection{$H^4$-inflation from unstable de Sitter vacuum}\label{sec:RenormEMTdS}
This scenario has recently been discussed in great detail in \cite{SolaPeracaula:2026trz}. De Sitter spacetime has been amply studied in the literature; see e.g. \cite{Chernikov:1968zm,Candelas:1975du,Dowker:1975tf,Brown:1976wc,Bunch:1978yq,Ford:1984hs,Mottola:1984ar,Dolgov:2005se,Kamenshchik:2021tjh,Landete:2013axa,Firouzjahi:2022xxb,Firouzjahi:2023wbe} and the textbooks  \cite{Birrell:1982ix,Fulling89,Parker:2009uva}, and references therein.  De Sitter spacetime is well-known to be the maximally symmetric solution to the vacuum Einstein field equations $R_{\mu\nu}=\Lambda g_{\mu\nu}$, $R=4\Lambda$,  with positive curvature and hence positive cosmological constant, $\CC>0$. Its symmetry group is $SO(1,n)$ for n spacetime dimensions, therefore enjoying the same degree of symmetry as  Minkowski spacetime ($10$ Killing vectors for $n=4$). Because of this, analytical QFT methods can be applied with relative simplicity.

The Hubble rate of de Sitter spacetime is constant, since it is fixed by the primordial vacuum energy density $\rho_I$, which is assumed to remain constant during the early stages in order to trigger inflation:
\begin{equation}\label{eq:Hexp}
   H^2= \left(\frac{\dot{a}}{a}\right)^2=\frac{8\pi G\rho_I}{3}\equiv H_I^2={\rm const.}
\end{equation}
The energy density  in the inflationary phase is  typically associated with the VED $\rv$ of some GUT scale $M_X$ near the Planck mass, hence  $\rho_I=\rv\sim M_X^4$.
As a result, the scale factor grows exponentially, and it leads indeed to a super-fast inflationary epoch :
\begin{equation}\label{eq:aexp}
a(t)\propto e^{H_It}= e^{\sqrt{\frac{\CC_I}{3}} \,t}\,,
\end{equation}
where  $\CC_I=8\pi G\rho_I$  is the (huge) cosmological constant in that epoch.
Such an exponential inflation flattens virtually any trace left of three-dimensional curvature, if it was nonvanishing to start with. Indeed, for any value of $K$ the last term in the first line  of Eq.\,\eqref{eq:4curvature} is virtually zero, and the curvature of de Sitter spacetime simplifies to
\begin{equation}\label{eq:RdeSitter}
    R=6(2H^2+\dot{H})=12H^2\,.
\end{equation}
Being $H$ constant, de Sitter spacetime is an exactly soluble QFT. Despite its simplicity, solving the model is not as trivial as one might naively think; in fact, it is rather cumbersome.
We cannot provide a minimally self-contained discussion of de Sitter spacetime here, but the main features are worth mentioning owing to the close similarities with the RVM framework. For full details, see \cite{SolaPeracaula:2026trz}. In both scenarios, RVM and de Sitter, inflation is triggered by the power $H^4$ of the Hubble rate and includes subleading $H^2$ powers as well. Of course, $H$ cannot remain constant all the time, as otherwise inflation would never end.  It suffices to assume that there is  a very short period where $H$ remains approximately constant and very large in the early universe: $H\simeq H_I$. The mode functions discussed in Sec.\,\ref{sec:AdiabaticVacuum} then obey the differential equation
\begin{equation}\label{eq:varphiequation}
       h_k''+\left[k^2+\frac{m^2+\left(\xi-1/6\right) R}{\tau^2H^2}\right] h_k=0\,,
\end{equation}
where again $\tau$ is the conformal time. The above equation can be transformed into a  Bessel equation by an appropriate change of variables.
For convenience, we can express the solution in terms of Hankel functions of the first and second kind $\mathbb{H}_\varsigma^{(1)} = J_\varsigma+iY_\varsigma$ and $\mathbb{H}_\varsigma^{(2)} = J_\varsigma-iY_\varsigma$, where $\varsigma$ is the order of the Bessel equation, which is given by
\begin{equation}\label{eq:nudefinition}
  \varsigma^2=\frac94-\mu^2-12\xi\,.
\end{equation}
We have defined $\mu\equiv m/H$ (which should not be confused with the 't Hooft mass unit $\mu$ in DR since we do not use it in this context).
By imposing the Bunch-Davies vacuum limit for $\tau\xrightarrow{}-\infty$ (corresponding to $a\to 0$ in the very early epoch\,\cite{Bunch:1978yq}), the Hankel function of the second kind does not participate and we obtain
\begin{equation}\label{eq:hkmodesdeSitter}
    h_k=\frac{\sqrt{\pi}}{2} |\tau|^{1/2}  e^{\frac{i\pi}{2}\left( \varsigma+\frac{1}{2}\right)} \ \mathbb{H}_\varsigma^{(1)} (k|\tau|) \; .
\end{equation}
Inserting this expression in Eq.\,\eqref{eq:vacuumEMT} and performing some calculations, we find
\begin{equation}\label{eq:T00mexactk}
    \begin{split}
        &\langle T_{00}^{\delta\phi}\rangle^{\rm dS}(m)=\frac{e^{-\pi{\rm Im}\varsigma} }{16\pi a^2}\int dk k^2 \left\{ \frac{1}{4{|\tau}|}{|\mathbb{H}_\varsigma^{(1)}}|^2+{|\tau|} k^2 {{|\mathbb{H}_\varsigma^{(1)}}'}|^2\right.\\
        &\left.\phantom{XXXXXX}+\frac{k}{2} \left( \mathbb{H}_\varsigma^{(1)} {\mathbb{H}_\varsigma^{(1)*}}'+\mathbb{H}_\varsigma^{(1)*}{\mathbb{H}_\varsigma^{(1)}}' \right)
        +\omega_k^2 {|\tau|} {|\mathbb{H}_\varsigma^{(1)}}|^2\right.\\
       &\left. +\left(\xi-\frac{1}{6}\right)\left[-\frac{6}{{|\tau|}} {|\mathbb{H}_\varsigma^{(1)}}|^2-\frac{6}{{|\tau|}} \left({|\mathbb{H}_\varsigma^{(1)}}|^2+k{|\tau|}\left( \mathbb{H}_\varsigma^{(1)} {\mathbb{H}_\varsigma^{(1)*}}'+\mathbb{H}_\varsigma^{(1)*}{\mathbb{H}_\varsigma^{(1)}}' \right)\right) \right] \right\}\,.
    \end{split}
\end{equation}
Here, ${\rm Im}\varsigma$ denotes the imaginary part of $\varsigma$.
One can show that this expression is divergent and therefore renormalization is required, as in the RVM case. So once more we subtract the first four adiabatic orders to it, according to the prescription Eq.\eqref{EMTRenormalized} within the off-shell ARP. The above integrations with Hankel functions are nontrivial, and special formulas are needed; see \cite{SolaPeracaula:2026trz} (Appendix A.3) for details.  After a significant amount of calculations one can finally produce the off-shell adiabatically renormalized ZPE at the scale $M$ (which is free from divergences since they cancel in the subtraction):
\begin{equation}\label{eq:RenormZPEdeSitter}
\begin{split}
    \langle T_{00}^{\delta\phi}\rangle_\mathrm{Ren}&(M)=\frac{a^2}{128\pi^2}\left\{ \left(M^2-3m^2+36H^2\left(\xi-\frac16\right)\right)\left(m^2-M^2+12H^2\left(\xi-\frac{1}{6}\right)\right)\right.\\
    &+\left.2m^2\left(m^2+12H^2\left(\xi-\frac{1}{6}\right)\right)\left(\psi\left[\frac{3}{2}-\varsigma(\mu)\right]+\psi\left[\frac{3}{2}+\varsigma(\mu)\right]-\ln \frac{M^2}{H^2}\right)\right\}\\
    &+\frac{a^2H^4}{960\pi^2}-\frac{a^2H^2m^2}{96\pi^2}-\frac{3a^2H^2(m^2-M^2)\left(\xi-\frac{1}{6}\right)}{8\pi^2}-\frac{9a^2H^4\left(\xi-\frac{1}{6}\right)^2}{2\pi^2}\,.
\end{split}
\end{equation}
In it,  $\psi(z)=d\ln\Gamma(z)/dz=\Gamma'(z)/\Gamma(z)$ is the standard digamma function, i.e. the logarithmic derivative of Euler's gamma function. Now,  given the $00$th component of the vacuum  EMT in de Sitter spacetime, the maximal degree of symmetry of this spacetime allows one to generate the full vacuum EMT as follows\cite{SolaPeracaula:2026trz}\footnote{Notice that if we substitute Eq.\eqref{eq:full EMT1} in the full vacuum EMT \eqref{RenEMTvacuum} and then use our definition of VED,  Eq.\,\eqref{RenVDE}, we immediately find  $\langle T_{\mu\nu}^{\rm vac}\rangle=-g_{\mu\nu} \rv$,  since here we are in de Sitter space. However, as warned in Sec.\,\ref{sec:RenVacuumPressure}, such a relation is not fulfuilled in general spacetimes.}
\begin{equation}\label{eq:full EMT1}
\langle T_{\mu\nu}^{\delta \phi}\rangle_\mathrm{Ren}(M)=-g_{\mu\nu}\,\frac{\langle T_{00}^{\delta \phi}\rangle_\mathrm{Ren}(M)}{a^2}\,,
\end{equation}
which can be explicitly written in a convenient way:
\begin{equation}\label{eq:fullRenARP}
\begin{split}
    \langle T_{\mu\nu}^{\delta\phi}\rangle_\mathrm{Ren}&(M)=-\frac{g_{\mu\nu}}{64\pi^2}\left\{m^2\left[ m^2+\left(\xi-\frac{1}{6}\right) 12H^2\right]\right.\\ &\left.\times\left[\psi\left(\frac{3}{2}+\varsigma\right)+\psi\left(\frac{3}{2}-\varsigma\right)- \ln\left(\frac{M^2}{H^2}\right)\right]\right.\\
    &-\frac{2}{3}m^2H^2-\frac{3}{2}m^4-\frac{1}{2}M^4+2m^2M^2\\
    &+\left.\left(\xi-\frac{1}{6}\right)\left[ 12M^2H^2-24m^2H^2\right]-72H^4\left(\xi-\frac{1}{6}\right)^2 +\frac{H^4}{15} \right\}\,.
\end{split}
\end{equation}
From this expression one can easily check  that in the on-shell situation ($M=m$) it boils down to
\begin{equation}\label{eq:BDresult}
\begin{split}
    \langle T_{\mu\nu}\rangle_\mathrm{BD}&=-\frac{g_{\mu\nu}}{64\pi^2}\left\{ m^2\left[m^2+\left(\xi-\frac{1}{6}\right)12H^2\right]\right.\\
    &\left.\left[\psi\left(\frac{3}{2}+\varsigma\right)+\psi\left(\frac{3}{2}-\varsigma\right)-\ln \left(\frac{m^2}{H^2}\right)\right] \right.\\
    &-\left.m^2\left(\xi-\frac{1}{6}\right)12H^2-\frac{2m^2H^2}{3}-72H^4\left(\xi-\frac{1}{6}\right)^2+\frac{H^4}{15}\right\}\,,
\end{split}
\end{equation}
which is precisely the well-known Bunch-Davies result that was  obtained in \cite{Bunch:1978yq} using the manifestly covariant point-splitting regularization procedure. See also the approach of \cite{Dowker:1975tf}.  This shows the consistency of our off-shell result.

Using \eqref{eq:RenormZPEdeSitter} in the definition of VED, Eq.\,\eqref{RenVDE}, we find the vacuum energy density in the early universe (during inflation) in the de Sitter scenario\cite{SolaPeracaula:2026trz}:
\begin{equation}\label{eq:VED_Ren_final2}
\begin{split}
    &\rho_\mathrm{vac}^{\rm infl.}(H)=\frac{3m^2 H^2}{16\pi^2}\left(\xi-\frac{1}{6}\right)\left(\psi\left[\frac{3}{2}-\varsigma(\mu)\right]+\psi\left[\frac{3}{2}+\varsigma(\mu)\right]\right)\\
    &+\frac{1}{128\pi^2}\left\{ H^4\left[\frac{2}{15}+24\left(\xi-\frac{1}{6}\right)-144\left(\xi-\frac{1}{6}\right)^2\right]-m^2H^2\left[\frac{4}{3}+48\left(\xi-\frac{1}{6}\right)  \right]\right\}\\
    &+\frac{m^4}{64\pi^2}\,\left[\psi\left(\frac{3}{2}-\varsigma(\mu)\right)+\psi\left(\frac{3}{2}+\varsigma(\mu)\right)-\ln \mu^2 \right]\,.
\end{split}
\end{equation}
To derive the above expression we have proceeded as in the RVM case by performing the difference of the VED between two scales,  say $M_0=H_0$ (which represents the current universe) and another scale $M=H$ (representing in this case the early universe) using the scaling relations discussed in Sec.\,\ref{sec:GeneralizedEqs}. In fact, denoting once more $\rv(M,H)$ at $M=H$ simply as $\rv(H)$, and similarly for $\rv(H_0)$, the above result follows upon neglecting the contribution from the current universe ($H=H_0$) since the VED in much smaller than in the inflationary time.

In an analogous manner as in the RVM, see Eq.\,\eqref{eq:VEDinf2l},  we can rewrite the above result  for the renormalized VED in the de Sitter scenario as follows:
\begin{equation}\label{eq:VEDdeSitterl}
\rho^{\rm infl.}_\mathrm{vac}(H)=\frac{\alpha} {2\pi^2}\, H^4+\frac{3{\tnu}(H)}{8\pi}\, \mpl^2H^2\,,
\end{equation}
where we have neglected the last term of \eqref{eq:VED_Ren_final2} since it is subleading during the inflationary stage ($H^4\gg m^4$). We have also defined the coefficient
\begin{equation}\label{eq:alphadS}
    \alpha=\frac{1}{480}+\frac38 \left(\xi-\frac16\right)- \frac94 \left(\xi-\frac16\right)^2\,.
\end{equation}
Notice that the expression \eqref{eq:VEDdeSitterl} in the de Sitter case also adapts to the template \eqref{eq:rvm1}, provided we use a constant value of $\nu$. This values is taken to be   $\tilde{\nu}_I\equiv\tilde{\nu}(H_I)$, with
\begin{equation}\label{eq:nutildeH}
\begin{split}
    \Tilde{\nu}(H) &= \frac{m^2}{2\pi m_\mathrm{Pl}^2}\left(\xi-\frac{1}{6}\right)\left(\psi\left[\frac{3}{2}-\varsigma(\mu)\right]+\psi\left[\frac{3}{2}+\varsigma(\mu)\right]\right)-\frac{m^2}{\pi m_\mathrm{Pl}^2}\left[\frac{1}{36}+\left(\xi-\frac{1}{6}\right)  \right] \, .
\end{split}
\end{equation}
\begin{figure}[t]
    \centering
    \includegraphics[width=0.8\linewidth]{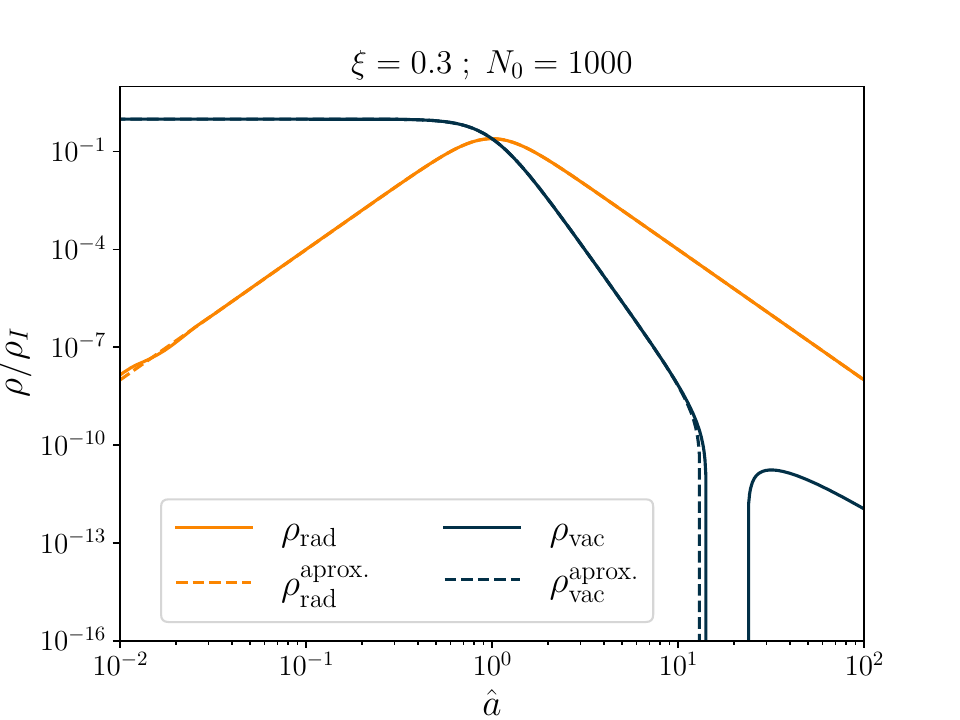}
    \caption{Unstable de Sitter vacuum decay: shown are the energy densities for vacuum and radiation for $\xi=0.3$, as a function of  the scale factor normalized to the transition point, $\hat{a}=a/a_*$. The value $N_0$ is the field multiplicity. The numerical solution  (solid lines) is compared with the (RVM-like) analytical approximation (dashed lines) given by Eqs.\,\eqref{eq:infaprox2}-\eqref{eq:infaprox3}. See \cite{SolaPeracaula:2026trz} for more details. }
    \label{fig:ved_greater}
\end{figure}
It follows that the (approximate) analytical solution \eqref{eq:infaprox}-\eqref{eq:infaprox3} can also be used here with the new identification of $\nu$ in the de Sitter case.
The coefficient of the $H^4$ power, Eq.\, \eqref{eq:alphadS}, now has a richer structure than in the RVM. Again, it must be positive for inflation to occur. This condition implies that for the de Sitter inflation, the range of $\xi$ turns out to be much narrower than in the RVM case (see Sec.\,\ref{sec:H4infl}). In fact, the necessary condition  $\alpha>0$ to have inflation implies the following (approximate) numerical range for $\xi-1/6$:
\begin{equation}\label{eq:Physregionximinus16}
    -0.00538< \left(\xi-\frac{1}{6}\right) < 0.172
\end{equation}
or, equivalently, for $\xi$:
\begin{equation}\label{eq:inflation_limit}
    0.161<\xi<0.339\,.
\end{equation}
As can be seen, for a single component, the non-minimal coupling  is fairly well determined. Notice that since $\xi>0$ there cannot be instabilities in the equation of motion of $\phi$\cite{Arbuzova:2025xkm}.
Furthermore, in the context of a typical GUT with many scalar fields $N_0$ (which may also involve supersymmetric particles, of course), each with different nonminimal couplings ${\xi}_i$, the condition for inflation becomes
\begin{equation}\label{eq:condinflationNo}
 A\equiv\frac{N_0}{480}+\frac{3}{8} \sum_i^{N_0} \left(\xi_i-\frac16\right)-\frac{9}{4}\sum_i^{N_0} \left(\xi_i-\frac16\right)^2>0 \,.
\end{equation}
The value of \eqref{eq:condinflationNo} determines the parameter $H_I$ defined in \eqref{eq:rvm1}. In the de Sitter case, we have
\begin{equation}\label{eq:HIA}
    H_I=\sqrt{\frac{3\pi}{4 G A}}=\sqrt{\frac{3\pi}{4A}}\ \mpl \, ,
\end{equation}
where in the second equality we have expressed the result in terms of the Planck mass, $\mpl$.

For the unstable de Sitter scenario, the numerical and analytical solutions are very similar to those shown in Fig.\,\ref{Fig:NumSolution} for the RVM.  In fact, an explicit example of de Sitter vacuum decaying into radiation can be seen in Fig.\,\ref{fig:ved_greater} for a choice of parameters in the physical region; for more details, see \cite{SolaPeracaula:2026trz}.  In both inflationary scenarios (RVM and de Sitter) we have an initial situation dominated by a huge vacuum energy density of order $\sim H^4$ that stays approximately constant for a short time and then decays into radiation, as shown in the figures. The subsequent radiation-dominated epoch obtained in both cases represents nothing more than the primeval FLRW regime of the $\CC$CDM model up to small corrections of order $\nu$. Therefore, there is a smooth connection with the standard thermal history, except that (as a bonus) the existence of an inflationary period of the sort we have described can explain the large entropy of the universe at present, see Sec.\ref{sec:Thermoidynamics}.  Let us, however,  mention a significant difference between the two $H^4$ inflationary scenarios. The equation of state (EoS) of vacuum is not the same.  We deal with this subject in the next section and consider the implications for the current universe.

\section{Running vacuum and equation of state in the current universe}\label{eq:RVMPheno}

Consider first the RVM scenario. In the late universe, the ${\cal O}(H^4)$ terms of the VED formula given by Eq.\,\eqref{DiffVEDphys} are negligible and one can easily show that the leading evolution of the vacuum energy  can be expressed as follows \cite{Moreno-Pulido:2020anb,Moreno-Pulido:2022phq}:
\begin{equation}\label{eq:RVMform}
\rv(H)\simeq \rvo+\frac{3\nu(H)}{8\pi}\,(H^2-H_0^2)\,\mpl^2\,,
\end{equation}
where $\rvo\equiv\rv(H_0)$ is identified with today's value of the VED through the observed CC, i.e. $\rvo=\CC_{\rm obs}/(8\pi G)$.
In the formula above, we have introduced the effective coupling
\begin{equation}\label{eq:nueff2}
\nu(H)\equiv\frac{1}{2\pi}\,\left(\xi-\frac16\right)\,\frac{m^2}{\mpl^2}\left(-1+\ln \frac{m^2}{H^{2}}-\frac{H_0^2}{H^2-H_0^2}\ln \frac{H^2}{H_0^2}\right)\,.
\end{equation}
This coupling is the low-energy counterpart of \eqref{eq:nuH}. Owing to the log behavior, function $\nu(H)$ changes very slowly with the Hubble rate and the effect from the last term becomes quickly suppressed for higher values of $H$ above $H_0$. In addition, $\ln \frac{m^{2}}{H^2}\simeq \ln \frac{m^{2}}{H_0^2}\gg1$ for the late universe, and hence $\nu(H)$ can be approximated by letting $H\to H_0$, i.e. by the effective parameter
\begin{equation}\label{eq:nueffAprox2}
\nueff\equiv\nueff(H_0)=\frac{1}{2\pi} \left( \xi-\frac{1}{6}\right) \frac{m^2}{m_\mathrm{Pl}^2}\left(-2+\ln\frac{m^2}{H_0^2}\right)\simeq\epsilon\ln\frac{m^2}{H_0^2}\,,
\end{equation}
where
\begin{equation}\label{eq:epsilonparameter}
\epsilon=\frac{1}{2\pi}\,\left(\xi-\frac{1}{6}\right)\,\frac{m^2}{\mpl^2}\,,
\end{equation}
is the same small parameter that we have found for the running of $G$, see Eq.\,\eqref{eq:runGH}.
Parameters $\nueff$ and $\epsilon$ are both small ($|\nueff|, |\epsilon|\ll 1$), but $\epsilon\ll\nueff\ll1$ since $\ln\frac{m^2}{H_0^2}={\cal O}(100)$ for virtually any known particle mass (recall that $H_0\sim 10^{-42}$ GeV).
For practical purposes, we can write \eqref{eq:RVMform} as follows:
\begin{equation}\label{eq:RVMcanonical}
\rv(H)=\rvo+\frac{3\nueff}{8\pi G_N}\,(H^2-H_0^2)\,.
\end{equation}
This equation is indeed the canonical form of the running VED at low energies in the RVM context, see \cite{Sola:2013gha,SolaPeracaula:2022hpd}. The characteristic in it is the presence of `soft vacuum effects' of the type $\sim m^2 H^2$, which can give a measurable contribution in the current universe (see the next two sections).  There are no `hard vacuum effects' of the Zeldovich type (cf. Sec.\,\ref{sec:introduction} ), i.e. no contributions of the form $\sim m^4$.  Therefore, the RVM seems to provide the kind of smoothly renormalized theory of the quantum vacuum which is free from troublesome effects prompting to extreme fine tuning of the cosmological parameters. This is probably one of the main achievements of the RVM approach.

Consider now the de Sitter scenario. We have previously discussed the relation between the renormalized pressure and energy density of vacuum, see Eq.\,\eqref{eq:VacuunPressure}.  It turns out that the term in parentheses on the right hand side of that equation varnishes identically for the renormalized quantities in the de Sitter case\,\cite{SolaPeracaula:2026trz}. As a consequence, the renormalized pressure and density satisfy the canonical vacuum EoS:
\begin{equation}\label{eq:renormEoS}
    P_\mathrm{vac}(M)=-\rho_\mathrm{vac}(M)\ \ \ \ \ \ (\wv=-1\,, \ \textrm{for de Sitter})\,.
\end{equation}
This is a remarkable result which, though expected, is still worth mentioning since it is a QFT result. In fact. it is a nontrivial property of the de Sitter vacuum since it is not satisfied by the unrenormalized quantities, only by the renormalized ones.  In view of this result, we might be led to think that this scenario just mimics a cosmological term, but this is not exactly so. Remarkably enough, the late time behavior of the decaying de Sitter vacuum  is very similar to that of the RVM, which means that the  VED evolution is exactly as in Eq.\,\eqref{eq:RVMcanonical}. We skip details here (see\cite{SolaPeracaula:2026trz}) and just summarize the main steps leading to this result. At low energies,  we can neglect the $\sim H^4$ contribution in Eq.\,\eqref{eq:VEDdeSitterl} and we can just concentrate on the second term of that expression. But we should not forget to restore the last term of \eqref{eq:VED_Ren_final2} since it is no longer subleading at low energies. Despite the presence of the $m^4$ factor in it, expansion of the digamma function for $\mu\gg1$ (i.e for $H\ll m$)  shows that the mentioned term  yields  soft contributions of the type $m^2 H^2$  in the current universe\cite{SolaPeracaula:2026trz}. Additionally, when we perform the subtraction of the VED between points $H$ and $H_0$ of the cosmic expansion,  now we must also restore the low-energy terms that depend on $H_0$  (recall that the latter were neglected at the inflationary epoch, but they must be now kept at low energies). The final result is, as announced, that the VED evolution law for the de Sitter scenario at low energies is formally coincident with the RVM one given by Eq.\,\eqref{eq:RVMcanonical} and with the same coefficient \eqref{eq:nueffAprox2}.
The only proviso, of course,  is the more restricted range for $\xi$ that holds for the  de Sitter scenario compared to the RVM.

The net outcome is that  both QFT pictures of the cosmological evolution, viz. running vacuum and unstable de Sitter vacuum, turn out to converge to the same prediction for the late time vacuum dynamics, i.e. Eq.\,\eqref{eq:RVMcanonical}, notwithstanding the fact that in the de Sitter scenario the vacuum EoS parameter is strictly constant, $\wv=-1$ -- Eq.\,\eqref{eq:renormEoS} --  whereas in the RVM case $\wv=\wv(z)$ is evolving with the expansion (see the next section).  It goes without saying that the fundamental parameter $\nueff$ in the common VED formula  \eqref{eq:RVMcanonical} of both scenarios is subject to the same phenomenological bounds $\nueff\lesssim 10^{-4}-10^{-3}$ obtained in the literature\cite{SolaPeracaula:2021gxi,SolaPeracaula:2023swx,deCruzPerez:2025dni}.
Overall, the formal resemblance between the RVM and de Sitter is notable, since both QFT pictures of the cosmological evolution converge to one and the same picture for the late time universe  in which there is  dynamical DE in the form of running vacuum energy. This is significant in view of the current evidence of dynamical DE\cite{DESI:2024mwx,DESI:2024aqx,DESI:2025zgx,DESI:2025fii}.


\subsection{Equation of state of the quantum vacuum in the RVM}\label{sec:EoSRVM}

In Sec.\,\ref{sec:RenVacuumPressure} we have elaborated on the  pressure $\Pv$ of the quantum vacuum and related it to the VED, $\rv$. Both quantities receive independent quantum corrections and, in general, we find that the canonical result $\Pv=-\rv$ receives  quantum corrections which make it depart from the canonical value. In the RVM context, the leading correction term deviating from the canonical result, which is relevant for the present universe, is given as follows  (see Eq.\eqref{eq:fullpressure}):
\begin{equation}\label{eq:pressureLateUniv}
\Pv(M)=-\rv(M)+\frac{\left(\xi-\frac{1}{6}\right)}{8\pi^2}\dot{H}\left(m^2-M^2-m^2\ln\frac{m^2}{M^2}\right)\,.
\end{equation}
Setting $M=H$, neglecting terms higher than ${\cal O }(H^2)$, using also the running VED formula at low energies, Eq.\,\eqref{eq:RVMcanonical}, and expressing these results in terms of the redshift, the equation of state (EoS) of vacuum can be rewritten in a more practical way for phenomenological analysis\,\cite{Moreno-Pulido:2022upl}:
\begin{figure}[t]
\begin{center}
\includegraphics[scale=0.55]{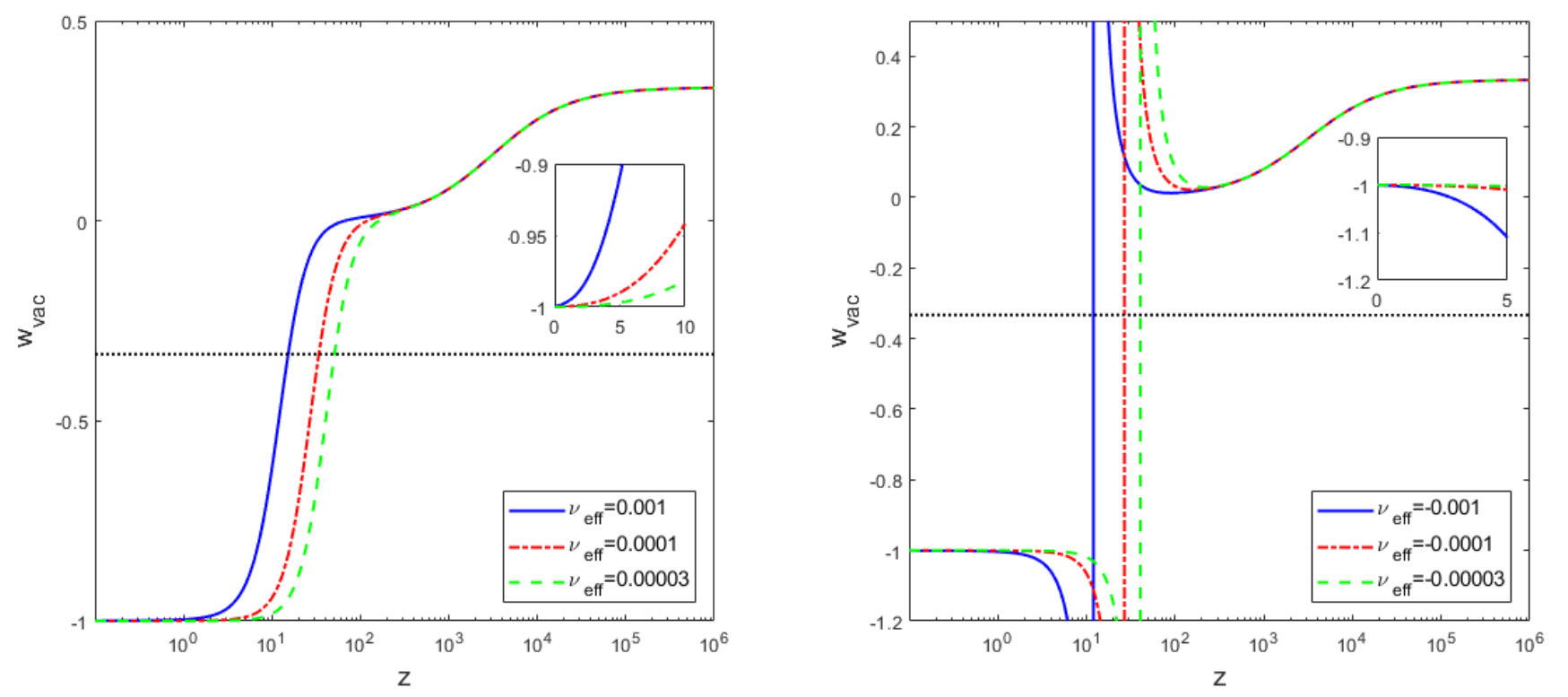}
\end{center}
\caption{The running vacuum EoS \eqref{ApproximateEos2} as a function of the  redshift\cite{Moreno-Pulido:2022phq,Moreno-Pulido:2022upl,Moreno-Pulido:2023ryo}. It mimics quintessence for $\nueff>0$ (left) and phantom DE for $\nueff<0$ (right).  In the last case the VED vanishes at some point in the past, and this appears as a vertical asymptote in the EoS plot, although no physical singularity occurs in any observable.}
\label{Fig2}
\end{figure}
%
\begin{equation}\label{ApproximateEos2}
\wv=\frac{\Pv}{\rv} \simeq -1+\nu_{\rm eff} \,\frac{\Omega_{\rm m}^0 (1+z)^3+\frac{4\Omega_{\rm r}^0}{3}(1+z)^4}{\Omega_{\rm vac}^0+\nu_{\rm eff} \left(-1+E^2 (z)\right)}\,,
\end{equation}
up to corrections ${\cal O}\left(\nueff^2\right)$, with $ E^2 (z)\equiv \Omega_{\rm vac}^0+\Omega_{\rm m}^0 \left(1+z\right)^3+\Omega_{\rm r}^0 \left(1+z\right)^4$\,. Notice that  the important coefficient $\nu_{\rm eff}$  was defined in the previous section and, in general, it embodies the combined effects from fermions and bosons, see\cite{Moreno-Pulido:2023ryo} for the detailed formula.  From the above EoS formula it is evident that  for $\nueff>0$ ($\nueff<0$) the quantum vacuum mimics quintessence (phantom DE) near our time (cf. Fig.\ref{Fig2}). This is particularly transparent upon approximating Eq.\,\eqref{ApproximateEos2} linearly in $\nueff$ for low redshift ($z\ll1$):
\begin{equation}\label{EqStateScalar}
w_{\rm vac}(z)\simeq -1+\nu_{\rm eff}\frac{\Omega_{\rm m}^0}{\Omega_{\rm vac}^0}(1+z)^3\,.
\end{equation}
However, the more accurate formula \eqref{ApproximateEos2} shows a peculiar feature of the RVM, which is highlighted  in Fig.\,\ref{Fig2}. In it, we display the EoS of the running vacuum in the extended redshift domain comprising the matter-dominated epoch up to the radiation-dominated epoch. One can appreciate in a manifest way that at higher and higher redshifts the vacuum equation of state tends to adopt the EoS of the dominant matter component (namely, first $\wv\simeq 0$ and subsequently $\wv\simeq 1/3$).  This curious `chameleonic nature' of the quantum vacuum, referred to in \cite{Moreno-Pulido:2022upl}, is caused by the quantum effects on the classical description. If detected, it can provide evidence of the QFT-RVM approach.  Despite the fact that $\wv\simeq -1$ at present, as it is obvious from \eqref{EqStateScalar}, the quantum corrections induce a deviation of the standard expectations, which leads to an effective quintessence or phantom-like behavior (depending on the sign of $\nueff$). The canonical form $w=-1$ is recovered again, with great precision, in the very early universe where inflation occurs. We have shown this feature previously in Fig.\ref{Fig:EoSvacuumInfl} when we discussed the inflationary period.

Finally, we should clarify that the fact that the EoS of the de Sitter scenario (studied in Sec.\,\ref{sec:RenormEMTdS}) remains canonical does not mean that it is perceived as such by the usual methodology of the cosmological observations since the VED of the de Sitter scenario remains dynamical. The practical methodology for scrutinizing the DE properties makes use of a variety of phenomenological parameterizations of the DE  \cite{DESI:2024mwx,DESI:2024aqx,DESI:2025zgx,DESI:2025fii} which allows to capture potential evidence of its dynamical character \cite{Park:2024vrw,Park:2024pew,deCruzPerez:2026mkg}; and it  may even provide a model-agnostic reconstruction of the background quantities associated with dark energy and
late-time cosmic expansion -- cf. \cite{Gonzalez-Fuentes:2025lei,Gonzalez-Fuentes:2026rgu} and references therein. These purely phenomenological methods are nonetheless far from providing a satisfactory explanation of its ultimate nature. Therefore, nothing prevents that a positive sign of DE dynamics captured by these parameterizations can actually be the direct reflect of dynamical vacuum energy of the sort studied here.  In fact the evolution of the VED is not necessarily linked to a departure of the fundamental EoS from $-1$. We have mentioned that the VED of the de Sitter scenario is indeed evolving at low energies, and it does it in an identical form to that of the RVM, a remarkable result. Given this fact, it should be noticed that it has long been shown that the dynamics of the VED implies automatically an effective EoS behavior departing from $\wv=-1$ when described in terms of standard parameterizations of the DE, see \cite{Sola:2005et,Sola:2005nh,Das:2005yj, Basilakos:2013vya}. Therefore, the dynamical DE that we may have detected in modern times may still correspond to either the RVM or even to the de Sitter unstable vacuum scenario that we have described.

\subsection{Confronting the running vacuum  with cosmological data}\label{sec:RVM today}

\begin{figure}[t]
\begin{center}
\includegraphics[scale=0.8]{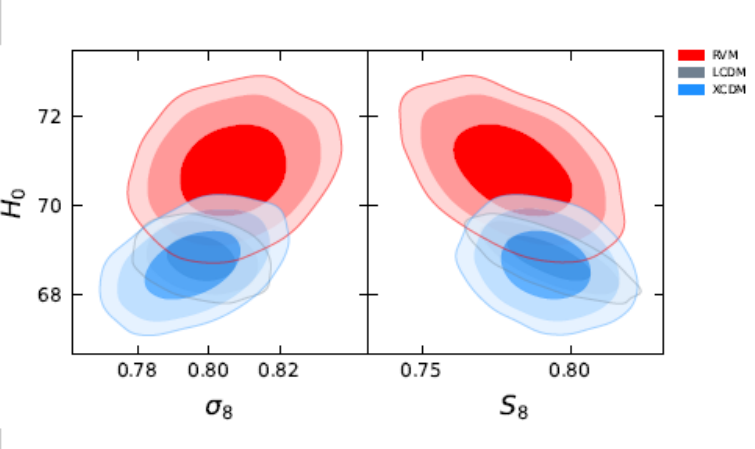}
\end{center}
\caption{Easing the $H_0$ and $\sigma_8$-tensions. Contours  at $1\sigma$,  $2\sigma$ and $3\sigma$ c.l.  in the ($\sigma_8$-$H_0$) and ($S_8$-$H_0$) planes, where $S_8 \equiv \sigma_8\sqrt{\Omega_{\rm m}^0/0.3}$. The considered models are RVM,  $\CC$CDM and the generic DE parameterization XCDM (or $w$CDM) for the data sets  SnIa+BAO+$H(z)$+LSS+CMB used in \cite{SolaPeracaula:2023swx}.  Only the RVM alleviates the $H_0$ tension and at the same time it reduces  the $S_8$ one.}
\label{contours}
\end{figure}

In this short section, we illustrate the practical performance of the RVM with a direct phenomenological application to fitting the observational data. We do this by reviewing the analysis of the RVM confronted with the $\CC$CDM and also with the generic DE parameterization XCDM\cite{Turner:1997npq} (also called $w$CDM) for the data sets  SnIa+BAO+$H(z)$+LSS+CMB used in \cite{SolaPeracaula:2021gxi,SolaPeracaula:2023swx}.
We have shown that the simplest form of the VED in the RVM for the current universe is given by Eq.\,\eqref{eq:RVMcanonical}. This canonical version should suffice for a phenomenological analysis, but it can be generalized.  Being both  $H^2$ and $\dot{H}$  homogeneous quantities of adiabatic order two, one expects that the VED at low energy should be in general a function of both terms carrying independent coefficients\cite{Sola:2015rra}. In fact, this was later confirmed in the QFT approach, where one can show that in a  more general context the VED can take the extended form\, \cite{Moreno-Pulido:2020anb,Moreno-Pulido:2022phq}:
\begin{equation}\label{eq:RVMvacuumdadensity}
\rv(H) = \frac{3}{8\pi G_N}\left(c_{0} + \nueff{H^2+\tilde{\nu}_{\rm eff}\dot{H}}\right)+{\cal O}(H^4)\,.
\end{equation}
Again, the ${\cal O}(H^4)$ terms can be neglected for the present universe. The two coefficients $\nueff$ and $\tilde{\nu}_{\rm eff}$ must be fitted to the observational data.  However, following\cite{SolaPeracaula:2023swx,SolaPeracaula:2021gxi} we explore the simplest option $\tilde{\nu}_{\rm eff}=\nueff/2$, as this leads to a dynamical component of the VED which is proportional to the Ricci scalar  ${R} = 12H^2 + 6\dot{H}$:
\begin{equation}\label{eq:RRVM}
\rv(H) =\frac{3}{8\pi{G_N}}\left(c_0 + \frac{\nueff}{12} {R}\right)\equiv \rv({ R})\,.
\end{equation}
This form has the advantage that the dynamical part is negligible in the radiation-dominated epoch, where $R\propto -T^\mu_\mu= \rho-3P=(\rho_r+\rho_m)-3(P_r+P_m)=\rho_m\ll \rho_r$ (since $\rho_r-3P_r=0$ and $P_m=0$); therefore the VED influence in the radiation-dominated period is essentially nullified and the basic facts of the thermal history during the radiation-dominated epoch are preserved, in particular  the primordial BBN features and the bounds derived from them \cite{Asimakis:2021yct}.  The contours in Fig.\ref{contours}, obtained from the fitting analysis to the observational data \cite{SolaPeracaula:2023swx,SolaPeracaula:2021gxi}, are based on the VED form \eqref{eq:RRVM}.  One can see from that figure that the $H_0$ and $\sigma_8$ tensions are highly alleviated within the RVM; see the references mentioned above for a detailed discussion. One finds $\nueff$ in the ballpark of $10^{-4}-10^{-2}$ depending on particular assumptions.

We should also mention that the specific scenario considered above assumes that matter is conserved in the standard way  ($\rho_m\sim a^{-3}$). In order to preserve the Bianchi identity in a matter conserving scenario and a variable VED, a simultaneous running of Newton's coupling $G$ must occur.  The variation of $G$ is found to be logarithmic with the Hubble rate, hence very mild.  In point of fact, we know that this is a generic QFT prediction of the RVM (cf. Sec.\ref{sec:RGinvariance}), and in the next section an extended discussion is provided. The details and explicit solution of the model are given in  \cite{SolaPeracaula:2021gxi,SolaPeracaula:2023swx}, where other variants of the RVM are considered. For example, we may assume that $G$ remains fixed at the expense of an interaction between vacuum and dark matter. One may even consider that $\nueff$ vanishes above a given redshif (implying a true CC above that redshift), or even that $\nueff$ flips sign in a particular redshift interval (the so-called `flipped RVM')\cite{deCruzPerez:2025dni}.  We refer the reader to the mentioned works for these alternative implementations of the RVM. They are in part phenomenological, but can also be theoretically motivated in stringy RVM (StRVM) scenarios\cite{Mavromatos:2020kzj,PhantomVacuum2021,NickJoan_PR}.

\subsection{Running of Newton's $G$ with the cosmic expansion}\label{sec:RunningG}

From the running law of  the VED at late time given by Eq.\,\eqref{eq:RVMform}, and with the help of the vacuum pressure equation that we have inferred in Sec.\,\ref{sec:RenVacuumPressure}, we can reconfirm the important running law \eqref{eq:runGH} for the gravitational coupling that we derived in Sec.\,\ref{sec:RGinvariance}\cite{Moreno-Pulido:2022phq}. Although the derivation of Eq.\,\eqref{eq:runGH} in that section might be considered as just a formal running coupling renormalization  similar to the case of the bare parameter $\rL$ in the EH action, it is not  so.  The difference is that in the computation of the VED one needs to put together two pieces at a time to construct the physical result: $\rv=\rL+$ZPE. None of these pieces makes physical sense in an isolated way, only the sum represents the physical observable (after renormalization).  In the case of $G$ we do not meet such a situation, and the running law obtained for $G$ is physically meaningful after renormalization. In fact, the two running laws are intimately connected  by a generalized form of the Bianchi identity that also takes into account the quantum effects on the vacuum pressure.  This alternative derivation, which we undertake here, can therefore be viewed as manifestly more physical, as it stems from the local conservation of the full vacuum EMT, which is the most relevant physical quantity  in our discussion.  To proceed, we first multiply both sides of the generalized Einstein's equations \eqref{eq:MEEs} by $8\pi G(M)$ and subsequently compute the covariant derivative of the resulting expression. In doing this, we use the fact that the HD tensor is covariantly conserved ($(\nabla^\mu\, \leftidx{^{(1)}}{\!H}_{\mu\nu}=0$); and, of course, we also borrow the Bianchi identity of the Einstein tensor ($\nabla^\mu G_{\mu\nu}=0$).  Finally, using the definition of the full vacuum EMT in renormalized form, Eq.\,\eqref{RenEMTvacuum}, and taking  the $\nu=0$ component of the obtained result, we reach after some calculations the following expression:
\begin{equation}
\begin{split}\label{MixedConservation}
\dot{G}(M)\left(\rho_\phi+\rv\right)&+G(M)\dot{\rho}_{\rm vac}+3HG(M)\left(\rv(M)+P_{\rm vac}(M)\right)\\
&=\left( \alpha(M)\dot{G}(M)+G(M)\dot{\alpha}(M)\right) \frac{\leftidx{^{(1)}}{\!H}_{00}}{a^2}\,,
\end{split}
\end{equation}
where dots stand, as usual, for cosmic time differentiation. In it, we can see that the scaling evolution of $G(M)$ is explicitly involved.  For completeness, in the above equation we have included  the contribution from the  background matter field $\phi_b$ defined in Sec.\,\ref{sec:AdiabaticVacuum} (which we have just denoted $\phi$ here for simplicity). The present discussion is not affected by its presence since  we assume that it is covariantly conserved, $\nabla^\mu T_{\mu\nu}^\phi=0 $, a relation that holds identically provided that there is no exchange of matter with the vacuum. This conservation law  can be straightforwardly verified using the explicit form of the EMT in Eq.\,\eqref{EMTScalarField} and the equation of motion for $\phi$ (viz. the free KG equation).  Since we perform our analysis in the post-inflationary epoch and, in particular, in the current universe, we may disregard the HD contribution  on the \textit{r.h.s.}  of Eq.\,\eqref{MixedConservation}. Therefore, we are left with the simplified form
\begin{equation}\label{ModifiedConsLaw}
\dot{G}(H)\left(\rho_\phi+\rv(H)\right)+G(H)\dot{\rho}_{\rm vac}(H)+3HG(H)\left(\rv(H)+P_{\rm vac}(H)\right)=0\,,
\end{equation}
where we have set $M=H$, according to our prescription to make contact with physics.  This equation can be conveniently put as follows,
\begin{equation}\label{eq:NonConserVEDandG}
\dot{\rho}_{\rm vac}+3H\left(\rv+P_{\rm vac}\right)=-\frac{\dot{G}}{G}\,(\rho_\phi+\rv)=- \frac{\dot{G}}{G}\,\frac{3H^2}{8\pi G}\,,
\end{equation}
where in the last step, we used Friedmann's equation. From the above generalized conservation law  we can see that the vacuum pressure can play an important role, since the last term on the \textit{l.h.s} is generally non-zero (except in de Sitter spacetime).  Furthermore, Eq.\eqref{eq:NonConserVEDandG}  shows that the vacuum is not conserved if $\dot{G}\neq 0$. Thus, in general,  we have  $\nabla^\mu \langle T_{\mu\nu}^\phi\rangle_{\rm ren}\neq0$ (in contrast to $\nabla^\mu T_{\mu\nu}^\phi=0 $ for the background field).
It should not be surprising,  since a dynamical vacuum cannot satisfy the Bianchi identity unless there is a compensation for its variation through a corresponding evolution of $G$ (or through interaction with matter, or both things at a time). This observation is the key to finding the running of $G$ with the scale, which is seen to be directly related to the vacuum conservation law.
In fact, the \textit{l.h.s.} of Eq.\,\eqref{eq:NonConserVEDandG} can be explicitly computed using  the equations for the VED and the vacuum pressure that we have obtained in previous sections. To this end, we insert the late time expression for the VED, $\rv(H)$, from \eqref{eq:RVMform} in the above equation,  and use the explicit expression for $\rv(H)+P_{\rm vac}(H)$ relevant for the current universe, namely Eq.\eqref{eq:pressureLateUniv}, which is proportional to $\dot{H}$.  As always, we neglect the higher order terms ${\cal O}\left(H^4\right)$  generated in intermediate calculations, which are irrelevant after the inflationary epoch.  All in all, Eq.\,\eqref{eq:NonConserVEDandG} can be put into a differential equation for $G$ as follows:

\begin{equation}\label{ModifiedConsLaw2}
\begin{split}
& 3\frac{d}{dt}\left[\frac{\nueff (H)}{8\pi G_N}(H^2-H_0^2)\right]+3H\frac{\left(\xi-\frac{1}{6}\right)}{8\pi^2}\dot{H}m^2\left(1-\ln \frac{m^2}{H^2}\right)+\frac{3H^2 \dot{G}(t)}{8\pi G^2(t)}=0\,,
\end{split}
\end{equation}
where we recall that  $G_N$  is constant, whereas $G(t)=G(H(t))$  is the function that we wish to determine by solving the above equation.  Notice that there is a pending time derivative to be computed in the first term of that equation, which involves calculating $\dot{\nu}_{\rm eff}(H)=\frac{d}{dt}{\nu}_{\rm eff}(H)$ using the exact expression \eqref{eq:nueff2}  rather than just a constant approximation.  At this point it is important to keep all terms since,  in general,  expressions that are neglected are not necessarily negligible after being differentiated. After some calculations we reach the following beautifully simple and compact expression:
\begin{equation}\label{MixedConservationApprox3}
\frac{d{G}}{G^2}=\frac{\left(\xi-\frac{1}{6}\right)}{\pi}m^2\frac{dH}{H}\,,
\end{equation}
in which we have replaced the time derivatives $\dot{G}=dG/dt$ and $\dot{H}=dH/dt$  by just the differentials $dG$ and $dH$  since $dt$ cancels on both sides.
Finally, integrating by simple quadrature Eq.\,\eqref{MixedConservationApprox3}  from the present time $(H_0,G(H_0)$, where $G(H_0)=G_N\equiv 1/\mpl^2$, up to an arbitrary moment around our past $(H,G(H))$,  we get the final result we are after:
\begin{equation}\label{eq:runGH2}
G(H)=\frac{G_N}{1-\frac{\left(\xi-\frac{1}{6}\right)}{2\pi}\frac{m^2}{m^2_{\rm Pl}}\ln \frac{H^2}{H_0^2}}\,.
\end{equation}
In this way we have proven once more Eq.\,\eqref{eq:runGH} following a completely different reasoning road, which was now based directly on the local conservation laws (q.e.d.)  The obtained formula is our QFT prediction for the physical running of the gravitational coupling with the cosmic expansion. We emphasize that the running law for $G$ that we have inferred is directly connected to that of the VED through the generalized Bianchi identity \eqref{ModifiedConsLaw}. Notice that for $\xi<1/6$, the gravitational coupling  $G(H)$ increases logarithmically with the expansion of the universe and hence was smaller in the past; that is,   $G(H)$ is asymptotically free. For $\xi>1/6$ it does the opposite, decreases with the expansion and was bigger in the past.

An even more powerful result follows from using the expressions for $\rv(M,H)$ and $\Pv(M,H)$ that we have obtained in the previous sections, before implementing the scale setting $M=H$, in order to express the conservation law satisfied by the VED as a function of the  renormalization scale $M$. We obtain\cite{Moreno-Pulido:2022phq,Moreno-Pulido:2022upl}:
\begin{equation}\label{eq:NonConserVED1}
\begin{split}
\dot{\rho}_{\rm vac}+3H\left(\rv+P_{\rm vac}\right)&=\frac{3\dot{M}}{8\pi^2 M}\left[\left(\xi-\frac{1}{6}\right)H^2(M^2-m^2)\right.\\
&\left.+3\left(\xi-\frac{1}{6}\right)^2\left(\dot{H}^2-2H\ddot{H}-6H^2\dot{H}\right)\right]\\
&=\frac{\dot{M}}{M}\,\beta_{\rv}\,.
\end{split}
\end{equation}
Remarkably, as the last equality shows,  the \textit{r.h.s} of this equation is proportional to $\beta_{\rv}$ (the $\beta$-function for the physical running of the VED), which we obtained in Eq.\,\eqref{eq:RGEVED1}. The results obtained previously can be recovered as a particular case of that more powerful formula. For example, on inserting  Eq.\,\eqref{eq:runGH2} into Eq.\eqref{eq:NonConserVEDandG}, one finds
\begin{equation}\label{eq:NonConserVED2}
\dot{\rho}_{\rm vac}+3H\left(\rv+P_{\rm vac}\right)=-\frac{3}{8\pi^2}\left(\xi-\frac{1}{6}\right) H \dot{H} m^2\,,
\end{equation}
which is the same result emerging from  Eq.\eqref{eq:NonConserVED1} when we consider the current universe and implement the setting $M=H$, see Eq.\,\eqref{eq:betarhoLambda2}. The above equation tells us unambiguously that in an expanding universe ($H\neq 0$) the VED is generally dynamical. In the de pure Sitter case, $H=$const. and $\Pv=-\rv$,  and thus $\dot{\rho}_{\rm vac}=0$, which implies that $\rv=$const. in the short period when $H$ is strictly constant.  But for FLRW spacetime, both  $H\neq 0$ and $\dot{H}\neq 0$, and at the same time $\Pv\neq -\rv$ (cf. Eq.\eqref{eq:pressureLateUniv}); hence $\dot{\rho}_{\rm vac}\neq0$.   Recall our analogy with the Casimir effect in Sec.\,\ref{sec:CasimirAnalogy}) when the plates are at infinity and the space is stretching (expanding). There is, in fact,  a nonvanishing (and dynamical) $4$-dimensional curvature ($\dot{R}\neq 0$) associated with the vacuum being dynamical ($\dot{\rho}_{\rm vac}\neq0$.).

The beauty of Eq.\,\eqref{eq:NonConserVED1} is that it embodies the essence of the RVM, for it says that the non-conservation of the VED, and hence its dynamical character,  is due to both the running of $\rv$ with $M$  (caused by $\beta_{\rv}\neq 0$)  and also due to the cosmic time dependence of the renormalization scale $M$ (viz. the fact that $\dot{M}\neq 0$, which in our case is implemented as $M=H(t)$). Both features are necessary, not just one.  This remarkable property of the VED in curved spacetime is in contradistinction to ordinary gauuge theories of strong and EW interactions, and enables us to probe the effect of the (cosmic) time-dependence of the running couplings and masses in the particle and nuclear physics world (see the final part of this review,  Sec.\,\ref{eq:GeneralRVM}).

\subsection{Solving the entropy problem in $H^4$-Inflation }\label{sec:Thermoidynamics}

The implications of the RVM account of cosmic evolution diversify in manyfold aspects. In particular,  the thermodynamic history of RVM-inflation is different from the usual inflationary paradigm based on the hypothetical existence of an inflaton\cite{KolbTurner,LiddleLyth,Kallosh:2025ijd}. The highly non-adiabatic event called `reheating', which is characterized by superheavy massive particles decaying into conventional particles in the early universe after inflation and causing the birth of the radiation epoch, does not occur in the RVM case. There is instead a super-fast heating-up process in which vacuum decays into radiation in a continuous way. In this sense, the RVM is once more different from the Starobinsky type of inflation\,\cite{Starobinsky:1980te}, in which, as already remarked, $H$ never goes through an inflationary regime characterized by $H=$const. and where, in addition,  there is an intermediate state where the scalaron field decays into relativistic particles.  The huge energy of the RVM inflation process stems from the $H^4$-nonlinearities imprinted on the external gravitational field of the very early universe by the quantum fluctuations of the matter fields, not by the presence of, say,  a  $f(R)$ function imposed \textit{ad hoc} in the classical action. These quantum effects are the genuine driving force of RVM-inflation. It is also the reason why the RVM has no horizon problem, for $H\simeq H_I$ remains approximately constant during inflation. It follows that a light pulse starting in
the remote past at $t=t_1\gtrsim t_i$ (beginning of inflation)  will have traveled until the end of the inflation epoch,
$t_f$, a physical distance\begin{equation}\label{eq:horizonInfinite}
d_{H}(a_f)= a(t_f)\int_{a_1}^{a_f}\frac{d
a'}{a'^2\,H}=\frac{a_f}{H_I}\left(\frac{1}{a_1}-\frac{1}{a_f}\right)\simeq
\left(\frac{a_f}{a_1}\right)\,H_I^{-1}\,.
\end{equation}
Being $a_f$ exponentially larger than
$a_1$, we have $a_f/a_1\ggg 1$ and hence the
above integral (called the particle horizon) can be as big as desired. All points in the past are therefore connected by light signals and  the entropy production is causal. This is in stark contrast to the standard $\CC$CDM  model for which $d_H(a)/a\rightarrow 0$ for $a\to 0$ (both in the radiation and matter-dominated epochs), which means that observers have been isolated in the past, and hence they have been unable to exchange information.

Let us explain how the entropy problem and \textit{a fortiori} the horizon problem are resolved in the RVM. They are essentially the same problem\cite{KolbTurner}. Let us start considering a most crucial thermodynamical quantity, which is the temperature of the heat bath of radiation following from the decay of the primeval vacuum in the RVM. It can be obtained from equating the radiation density $\rho_{\rm rad}(a)$, given by Eq.\,\eqref{eq:infaprox3}, to the black-body formula $(\pi^2/30) g_*T^4$. Here  $g_*$ is the number of active degrees of freedom (d.o.f.)  at the given temperature, typically of order $100-1000$ in a GUT (for instance, $g_*=160.75$ in non-supersymmetric $SU(5)$).  We find
\begin{equation}\label{eq:TempRad}
T_\mathrm{rad}(\hat{a})=T_I\,(1-\nu)^{1/4} \frac{\hat{a}^{(1-\nu)}}{\left[1+\hat{a}^{4(1-\nu)}\right]^{1/2}}\,,
\end{equation}
where we have defined the temperature $T_I$ through  $\rho_I=\frac{\pi^2}{30}\,g_\ast\,T_I^4$.
The maximum of the above radiation temperature is readily found to be at the crossover point $\hat{a}=1$ ($a=a_*$) and takes the value (cf. the notation in Sec.\ref{sec:solutions})
\begin{equation}\label{eq:Tmax}
T_\mathrm{max}=\frac{T_I}{\sqrt{2}}\,(1-\nu)^{1/4} \sim \left(\frac{45 m_\mathrm{Pl}^2 m^2}{16\pi^3g_* \epsilon} \right)^{1/4}=\left(\frac{45}{8\pi^2 g_{*}\left(\xi-1/6\right)}\right)^{1/4}\mpl\,.
\end{equation}
We have neglected corrections of order $|\nu|\ll 1$ in the final result. From the  estimates on $g_*$ and $\xi$ mentioned above, one can easily check that this temperature remains at most about one order of magnitude below $\mpl$ (the Planck mass), which is consistent with our field theory approach. On the other hand,   for $\hat{a}\gg1$ (i.e., $a\gg a_{*}$, which means deep into the radiation-dominated epoch) the scaling law of the temperature with the scale factor is the following:
\begin{equation}
T_\mathrm{rad}\, a^{1-\nu}={\rm const.}
\end{equation}
Again this is meaningful since $|\nu|\ll 1$ and hence we recover in very gpod approximation the canonical scaling law  of the radiation temperature in the adiabatic regime: $T_\mathrm{rad}\propto 1/a$.

Finally, consider now  the issue of entropy production in the RVM context. In our case, entropy refers only to the comoving radiation entropy associated with the above heat-bath temperature, and therefore the traditional notion of entropy inherent to the co-moving volume $V\sim a^3$. We do not address here the entropy contribution from the horizon, which is associated with the Generalized Second Law. For a proper discussion, this would require taking into consideration the so-called cosmological apparent horizon. The interested reader, however,  may wish to look up Ref.\cite{Yu2020} for its calculation in the RVM case. In the mentioned reference it is explicitly demonstrated  that the RVM fully satisfies the Generalized Second Law.

Coming back to the (comoving) radiation entropy corresponding to the radiation temperature \eqref{eq:TempRad}, it can be easily computed from
$S_{r}= \left(4\rho_r/3T_{\rm rad}\right) a^3$\,\cite{KolbTurner}. Therefore,
\begin{eqnarray}\label{eq:SrRVM}
S_{r}(\hat{a})=\frac{2\pi^2}{45}\,g_\ast T_{\rm rad}^3a^3=\frac{2\pi^2}{45}\,g_{*}\,T_I^3\,\astar^3\,(1-\nu)^{3/4}\,\frac{\ha^{6-3\nu}}{\Big[1+\ha^{4(1-\nu)}\Big]^{3/2}}\,.
\end{eqnarray}
Since $|\nu|\ll1$, it is easy to see that this formula saturates into an approximate plateau for a sufficiently large value of $\hat{a}$.  During the heating-up period the comoving entropy increases as the sixth power of the scale factor, $S\sim \hat{a}^{(6-3\nu)}\sim \hat{a}^6$, until the aforementioned saturation regime occurs. The plateau signals the entrance in the  radiation-dominated phase and the (causally produced)  entropy injected into this phase is
\begin{equation}\label{eq:SrSaturation}
  S_r(\ha\gg1)\simeq \frac{2\pi^2}{45}\,g_{*}\,T_I^3\astar^3(1-\nu)^{3/4}\ha^{3\nu}
  \equiv S_{r0} \ha^{3\nu}\,.
  \end{equation}
  This entropy stands for  the (approximate)  asymptotic value of the comoving entropy.
 For  $\nu=0$ the expansion is assumed to be isentropic and the quantity $g_{*}T_{\rm rad}^3 a^3$ remains invariable (i.e. entropy is conserved)  during the adiabatic phase  and therefore must   equal the current value
$g_{s,0}\,T_{\gamma 0}^{3}\,a_0^{3}$. Note that $T_{\gamma 0}\simeq 2.725\,$K$\,\simeq  2.3\times 10^{-13}$GeV is the CMB temperature now, and  $g_{s,0}=2+6\times
(7/8)\left(T_{\nu,0}/T_{\gamma 0}\right)^3\simeq 3.91$ is a well-known entropy factor corresponding to the number of light d.o.f. today, calculated from the ratio of the present neutrino and photon temperatures\cite{KolbTurner}. We are now in position to check if  the huge entropy enclosed in the horizon today, $H_0^{-1}$, namely
\begin{equation}\label{eq:S0}
S_{0}=
\frac{2\pi^2}{45}\,g_{s,0}\,T_{\gamma 0}^3\,\left(H_0^{-1}\right)^3\simeq
2.3 h^{-3} 10^{87}\sim 10^{88} \ \ \ \ \ \ (h\simeq 0.7),
\end{equation}
can be explained from the predicted value \eqref{eq:SrSaturation}  in the RVM\footnote{This formula is to be understood as follows. The physical volume of the universe is related to the comoving volume by $V = a^3 L^3$, where $L^3$
is the coordinate volume at present ($a_0=1$).  In the previous fomulas we implicitly assumed that $L^3 = 1$. Now the current horizon is $L=H_0^{-1}$, and hence the physical volume at present is $V_0=\left(H_0^{-1}\right)^3$, which leads to Eq.\,\eqref{eq:S0}.}.
We can perform the check within order of magnitude. In fact, it is easy to convince oneself that to reproduce  the above entropy value in the context of RVM-inflation, the following condition must be met:   $g_*T_I^3 a_*^3\sim g_{s,0} T_{\gamma 0}^3 a_0^3$ (with $a_0=1$ at present).
It is a nontrivial condition to be satisfied, since  $T_{\gamma 0}$ is an experimentally measured quantity at present, while $T_I$ and $a_*$ are theoretical parameters predicted by the the RVM scenario in the very early universe. To check if the above condition is fulfilled,  we need to numerically estimate $a_*$ from Eq.\,\eqref{eq:astar}. The latter depends on $\rho_I=3H_I^2\mpl^2/8\pi$ and ultimately on $H_I$, which is given by Eq.\,\eqref{eq:HIvalue}.  If we take into account the estimate  $\nueff\sim  10^{-4}-10^{-2}$  given in Sec.\,\ref{sec:RVM today}, which is obtained from fitting the low energy data (i.e. from the current cosmological observations\cite{SolaPeracaula:2021gxi,SolaPeracaula:2023swx}), we find from \eqref{eq:nueffAprox2} that $\epsilon \sim 10^{-6}-10^{-4}$.  Thus, for particle masses near the GUT scale, $m\sim M_X\sim 10^{16}$GeV,  Eq.\,\eqref{eq:HIvalue} yields  $H_I\sim 10^{17-18}$ GeV and as a result  $\rho_I=\frac{3H_I^2 \mpl^2}{8\pi} \sim 10^{73}$ GeV$^4$.

Combining the previous  estimate with the measured quantities $\Omega_\mathrm{rad}^0\sim 10^{-4}$ and $\rho_c^0\sim 10^{-47}$ GeV$^4$ in   Eq.\,\eqref{eq:astar},  we find
\begin{equation}\label{eq:astar2}
a_{*}\sim
\left(10^{-4}\,\frac{10^{-47}}{10^{73}}\right)^{1/4}\sim 10^{-31}\,.
\end{equation}
From the aforementioned matching condition for the entropy, and using the previous numerical values we find the following estimate for the temperature at which the bulk of the radiation entropy was produced:
\begin{equation}\label{ee:TIestimate}
T_I\sim \left(\frac{g_{s,0}}{g_*}\right)^{1/3} \frac{T_{\gamma 0}}{a_*}
\lesssim 0.1\, \mpl\,,
\end{equation}
where $\mpl\simeq 1.22\times 10^{19}$ GeV.   This temperature agrees in order of magnitude with the  theoretically predicted value in  Eq.\eqref{eq:Tmax}, which is also of order $0.1 \mpl$ at most.
 This demonstrates the numerical consistency of our result within the rough order of magnitude, and hence the ability of the RVM framework to accommodate the huge entropy value \eqref{eq:S0} causally, whence solving the entropy and horizon problems of the standard model.

 Needless to say, a more detailed calculation should take into account the multiplicity of fields in the theory and other considerations. It is nevertheless  rewarding to see that the order of magnitude is in place.  Similar successful calculations can be carried out for the unstable de Sitter vacuum scenario\cite{SolaPeracaula:2026trz} using the formulae of Sec.\ref{sec:RenormEMTdS}, thus providing full consistency of these unification models between the physics of inflation and the predicted entropy in the current universe.

We should emphasize that the above discussion of RVM inflation has been derived within the general  QFT formulation of the running vacuum framework\cite{SolaPeracaula:2022hpd}, which goes well beyond previous phenomenological considerations on these matters. Our approach provides a theoretical basis for a possible solution to these cosmological problems within fundamental physics. As noted in \cite{Sola:2015csa}, the clues to the cosmological problems of the present may well have profound roots in the past!


\section{Running vacuum and cosmic drift of the fundamental `constants'}\label{eq:GeneralRVM}

Almost  a decade and a half ago, in the work\cite{Fritzsch:2012qc} a possible explanation for the running or `drift' (as it is called in the field)  of the fundamental constants of Nature was first pointed out with the help of the theoretical framework of the running vacuum model (RVM). The author had the privilege to make these intriguing and exciting considerations with the invaluable collaboration of the late Harald Fritzsch. I summarize very briefly these findings here.

The history of this recurrent scientific subject traces back mainly to Dirac's pioneering work in the thirties, when he formulated the ``large number hypothesis''\,\cite{Dirac:1937ti}, and also to Jordan's work in the same year on the possible variation of both the QED fine structure (Sommerfeld's) constant $\alpha=e^2/4\pi$ and $G$\cite{Jordan1937}. About a quarter of a century later the Brans-Dicke (BD) approach to gravity\,\cite{Brans:1961sx} was formulated, in which General Relativity (GR) was extended to accommodate a dynamical scalar field $\varphi(t)$ playing the role of time-evolving $1/G(t)$. It spurred further speculations by Gamow \,\cite{Gamow:1967zza} and others on the
possible variation of $\alpha$, which continued till the present time\footnote{Unfortunately, we cannot make justice here to the extensive literature on the subject; see e.g. \cite{Fritzsch:2012qc,Fritzsch:2015lua,Fritzsch:2016ewd} for essential bibliography particularly related to the approach proposed here, and \cite{Uzan:2010pm,Chiba:2011bz,Martins:2017yxk,Uzan:2024ded,Safronova:2017xyt,Barontini:2021mvu,Bass:2023hoi,Ginges:2003qt} (and references therein) for  general reviews of the subject.}.  However, for all the exciting phenomenology that these developments may entail a fundamental framework is needed to encompass a possible time variation of the fundamental `constants'.  For example, the BD approach can, of course, provide a deeper meaning to the possible time variation of $G$, but it is ultimately a classical approach and, moreover, there is no clue in it as to the meaning of $\CC$ since there is no $\CC$ in it\cite{Brans:1961sx}. As a matter of fact, BD extensions with constant $\CC\neq 0$ (and even variable $\CC$) are possible, see e.g. \cite{SolaPeracaula:2019zsl,SolaPeracaula:2020vpg,deCruzPerez:2023wzd,SolaPeracaula:2018dsw,deCruzPerez:2018cjx} and references therein. Besides, the relation between $\CC$ and quantum vacuum energy is a major issue for the CCP, as we have seen in the previous chapters. This can have groundbreaking implications for the other fundamental `constants' of Nature.

In fact, it may well be that there are no fundamental constants at all in an strict sense. A rich program of investigations on the possible  time (and space) variation of the physical `constants' (VPC) is under way since long,
both in the lab (through high precision quantum optic techniques applied to atomic and molecular clocks, see e.g.\cite{Safronova:2017xyt,Barontini:2021mvu,Bass:2023hoi,Ginges:2003qt}), as well as in the sky, i.e. in the astrophysical domain (e.g. using absorption systems in the spectra of
distant quasars.\cite{Safronova:2017xyt}). In the last decades various astrophysical observations of this sort have suggested positive evidence on the cosmic time evolution of the fine structure constant\,\cite{Murphy:2000pz}, i.e. Sommerfeld's
$\alpha$. At the same time, indications have appeared
 of a possible spatial variation of $\alpha$\cite{Webb:2010hc}.  Claims in the literature have been made about such driftings in time and space at a level of a few standard deviations, although unfortunately the confirmation by other groups is lacking at the moment\cite{Safronova:2017xyt}. There are also claims about the drifting of the
dimensionless ratio $\mu\equiv m_p/m_e$ (between proton and electron
masses). This ratio has been accurately monitored (using quantum optic techniques in the lab and quasar absorption lines). Significant relative time variations ($\dot{\mu}/\mu$) have been reported in some cases\,\cite{Reinhold:2006zn}, but remain to be confirmed by independent observations\cite{Ubachs:2015fro}.
Despite this inconclusive situation,  exciting news are foreseen in this enthralling field of research at some point in the future. If so, it would lead to a  significant modification of our current scientific views and paradigms on the `constants' of Nature. Needless to say, experiments in this field are difficult and must be contrasted by independent observational teams.

The large number of fundamental constants can be felt as unsatisfactory from the theoretical point of view, especially when we consider our most cherished theories such as the standard model (SM) of particle physics, the scientific domain of excellence in our understanding of the microscopic laws of Naure. Nevertheless, we do not have at present an explanation for the large variety of couplings, masses, and mixing angles in the SM.  The total number of parameters in it is quite  large (in total, 27, if the neutrino masses are Majorana masses, and hence two additional CP-violating phases are added). If we were to apply `von Neumann's elephant criterion', the ability of the SM to describe the fundamental interactions would become somewhat diminished\footnote{As recounted by F. Dyson in conversations with E. Fermi, J. von Neumann made the famous claim that  `With four parameters I can fit an elephant, and with five I can make him wiggle his trunk'\cite{Dyson2004}. }.  However, if cosmic evolution enters,  there is plenty of motivation for the VPC approach to  the fundamental `constants', since they might all be running parameters with the cosmic expansion.


It is expected that the underlying physical laws put restrictions to the drift of the physical `constants'. For example, GR is widely accepted as a fundamental theory for describing the geometric properties of spacetime. Thus, we should expect that if such VPC occurs, variations should be  correlated in such a way as to preserve the principle of general covariance. This was actually the guiding principle of Jordan's approach to the subject almost nine decades ago\cite{Jordan1937}.
Consider the field equations of GR in the presence of the
cosmological term:
\begin{equation}
R_{\mu \nu }-\frac{1}{2}g_{\mu \nu }R+\,g_{\mu\nu}\CC=8\pi G\,T_{\mu\nu}\,, \label{EE}
\end{equation}
with $T_{\mu\nu}$ the energy-momentum tensor of matter in the universe, treated as a perfect fluid.  If we assume that the Cosmological Principle is satisfied\cite{peebles:1993}, the possibility that the parameters $G=G(t)$ and $\CC=\CC(t)$ could be functions of the
cosmic time -- as it is the case with the scale factor $a=a(t)$ of the FLRW metric -- is perfectly compatible with that principle.

Whether constant or time-evolving, the (physical) $\CC$ term  on the \textit{l.h.s.} of Einstein's equations can be absorbed on the  \textit{r.h.s.} after
introducing the quantity $\rv=\CC/(8\pi G_N)$, which represents the VED associated to the cosmological term, as noted in the introduction. Einstein's equations
can then be rewritten as follows:
\begin{equation}
R_{\mu \nu }-\frac{1}{2}g_{\mu \nu }R=8\pi G\,\tilde{T}_{\mu\nu}\,, \label{EE2}
\end{equation}
i.e. formally replacing
the ordinary EMT of matter by the total EMT, which includes  both matter and  vacuum energy:
\begin{equation} \label{tildeEMT}
{T}_{\mu\nu}\to \tilde{T}_{\mu\nu}\equiv T_{\mu\nu}-g_{\mu\nu}\,\rv
=  (P_m-\rv)\,g_{\mu\nu}+\big(\rmr+P_m\,\big)\,u_{\mu}\,u_{\nu}\,,
\end{equation}
$\rmr$ and $\pmr$ being the proper density and pressure of the homogeneous and
isotropic cosmic matter fluid, and $u_{\mu}$  its $4$-velocity  ($u^\mu u_\mu=-1$).  Because the Cosmological Principle is still to be preserved, the above total EMT must maintain the form of a perfect fluid.

Notice that in \eqref{tildeEMT} we have assumed that the equation of state (EoS) of the vacuum is the traditional one, $\Pv=-\rv$. Although this equation is subject to quantum effects in the QFT approach \, \cite{Moreno-Pulido:2022upl}, as we have seen in Sections \ref{sec:RenVacuumPressure} amd \ref{sec:EoSRVM},  for the sake of simplicity we do not modify the canonical vacuum EoS at this point, since it does not change the main conclusions within the present VPC context.

Let us describe a few scenarios within the VPC phenomenology that preserve the general covariance of GR. We assume flat three-dimensional space ($K=0$).
Friedmann's equation and the acceleration equation are not altered by the possible time evolution of the vacuum energy, $\rv$,  nor of $G$ (Newton's `constant'):
\begin{equation}\label{Friedmann}
\left(\frac{\dot{a}}{a}\right)^2\equiv H^2=\frac{8\pi G}{3}(\rmr+\rv)
\end{equation}
and
\begin{equation}\label{acceleration}
\frac{\ddot{a}}{a} = H^2+\dot{H}=-\frac{4\pi\,G}{3}\,(\rmr+3\pmr-2\rv)\,.
\end{equation}
In the current universe, the pressure of matter is negligible ($P_m\simeq 0$) and  its energy density dilutes in the standard form $\rmr\sim a^{-3}$. In the long run, $\rv$
dominates over $\rmr$ and this causes the universe to speed up ($\ddot{a}>0$)  for $\rv>0$. This holds irrespective of whether
 $\rv$ is constant or mildly variable. In addition, the Bianchi identity
$\nabla^{\mu}G_{\mu\nu}=0$, involving the Einstein tensor on the
\textit{l.h.s.} of the field equations, implies the following generalized conservation law for
the full EMT times Newton's $G$:
\begin{equation}\label{GBI}
\nabla^{\mu}\left(G\,\tilde{T}_{\mu\nu}\right)=\nabla^{\mu}\,\left[G\,(T_{\mu\nu}-g_{\mu\nu}\,\rv)\right]=0\,.
\end{equation}
This equation can be evaluated using the FLRW metric, leading to the following ``mixed'' local conservation law:
\begin{equation}\label{BianchiGeneral}
\frac{d}{dt}\,\left[G(\rmr+\rv)\right]+3\,G\,H\,(\rmr+P_m)=0\,,
\end{equation}
in which both $G$ and/or $\rv$ can be both functions of the cosmic time. Only for relativistic matter $P_m\neq 0$ (namely, $P_r=1/3$) in \eqref{BianchiGeneral}. Although the
above equation is not independent of (\ref{Friedmann}) and
(\ref{acceleration}), it is useful to understand the possible transfer of
energy between vacuum and matter. For example, if the VED is not constant ({$\drv\neq 0$}), the matter is not conserved, since the
vacuum could decay into matter, or the matter could disappear into the vacuum. In the process, one could also have  a possible contribution from a variable $G$ (if
$\dot{G}\neq 0$). The conservation law (\ref{BianchiGeneral}) allows for a general situation characterized by a transfer of energy
between matter-radiation  $\rmr$, vacuum energy $\rv$ and at the same time  accompanied by a possible running of $G$.

Four specific  scenarios can be highlighted:
\begin{itemize}

\item  i)  $G=$const. {and} $\rv=$const.
It corresponds to the standard (or `concordance') $\CC$CDM model \cite{peebles:1993,Turner:2022gvw} with a rigid cosmological term $\CC=$const. It leads to the usual local conservation law of matter-radiation:
\begin{equation}\label{standardconserv}
\dot{\rho}_m+3\,H\,(\rmr+P_m)=0.
\end{equation}
Using the chain rule $d/dt=a H d/da$, it can be written in terms of the scale factor $a$. Using prime for $d/da$, we have
\begin{equation}\label{eq:standardconserv2}
\rho'_m(a)+\frac{3}{a}(1+\wm)\,\rmr(a)=0\,,
\end{equation}
where $\wm=P_m/\rmr$ is the EoS of matter ($0$ for non-relativistic and $1/3$ for relativistic matter).
The solution of \eqref{eq:standardconserv2} reads
\begin{equation}\label{solstandardconserv}
\rmr(a)=\rmo\,a^{-3(1+\wm)}=\rmo\,(1+z)^{3(1+\wm)}\,,
\end{equation}
$z=(1-a)/a$ being the cosmological redshift, and $\rmo$ the current matter density (i.e. its value at $a=1$, or equivalently at $z=0$).

\item ii) $G=$const {and} $\drv\neq 0$.
In this case, Eq.(\ref{BianchiGeneral}) leads to a mixed conservation law which is independent of $G$:
\begin{equation}\label{mixed conslaw}
\drv+\dot{\rho}_m+3\,H\,(\rmr+P_m)=0\,.
\end{equation}
In this scenario, an exchange of energy between matter and vacuum is mandatory in order to fulfill the Bianchi identity, so that neither matter nor vacuum energy are conserved in this case.  However, equation\,\eqref{mixed conslaw} cannot be solved unless a specific ansatz  is proposed for $\rmr$ (other than \eqref{solstandardconserv})  or for a variable $\rv$.  As an example, the RVM makes a theoretically well motivated proposal for $\rv=\rv(H)$, see Eq.\,\eqref{eq:RVMcanonical}, which was originally tested in \cite{Espana-Bonet:2003qjh} and in many subsequent analyzes, e.g. \cite{Sola:2015wwa,Sola:2016jky,SolaPeracaula:2016qlq,Sola:2016hn,Sola:2016zeg,SolaPeracaula:2017esw,Sola:2017znb}, including variants of it in recent works \cite{SolaPeracaula:2021gxi,SolaPeracaula:2023swx,deCruzPerez:2025dni}. Alternatively, one can start directly from an anomalous conservation law of matter\cite{Wang:2004cp,Alcaniz:2005dg}, although it is a more phenomenological approach.

\item iii) $\dot{G}\neq 0$ {and} $\rv=$const. Again, a mixed conservation law is involved, but now dependent on $G$ and its time derivative:
\begin{equation}\label{dGneqo}
\dot{G}\,(\rmr+\rv)+G\,\left[\dot{\rho}_m+3H(\rmr+P_m)\right]=0\,.
\end{equation}
Here, matter is not conserved at the expense of a variable $G$. To solve it for $G$ in a nontrivial case we need a non-conservative ansatz for $\rmr$. Specific examples of this scenario are discussed e.g. in \cite{Guberina:2006fy,Fritzsch:2012qc}.

\item iv) $\dot{G}\,\neq 0$ {and} $\drv\neq
    0$.
The simplest implementation of this scenario is by
    assuming covariant conservation of
    matter-radiation, i.e Eq.\,(\ref{standardconserv}). Then Eq.\,(\ref{BianchiGeneral}) reduces to
\begin{equation}\label{Bianchi1}
(\rmr+\rv)\,\dot{G}+G\,\drv=0\,.
\end{equation}
In this case we have a dynamical interplay between $G$ and $\rv$. To solve the model, we need an ansatz or some theoretical proposal for the time variation of $G$ or $\rv$.  Different variants of this kind of scenario are considered e.g. in \cite{Shapiro:2004ch,Sola:2007sv,Fritzsch:2012qc,Fritzsch:2015lua,Fritzsch:2016ewd,Hanimeli:2019wrt,Montani:2024ejp}.

\end{itemize}

Apart from the standard model scenario i), which is always compared to,  herein we shall focus briefly on scenarios ii) and iv) for which the vacuum is `running' in both cases.  The VPC pattern associated with these scenarios is naturally linked with the RVM, although connections with quantum gravity or string theory are also possible\cite{Mavromatos:2020kzj,PhantomVacuum2021}.

\subsection{The micro and macro connection in the laws of Nature} \label{sect:micromacro}

Let us now focus in more detail  on  class ii) in the list of scenarios given in the previous section. It leads to an anomalous matter conservation law along with dynamical vacuum energy.  We assume that  $\rv$ evolves with the Hubble rate according to the canonical RVM expression \,\cite{Sola:2013gha} (see the formal part of our presentation, specifically Eq.\,\eqref{eq:RVMcanonical} of Sec.\ref{eq:RVMPheno}):
\begin{equation}\label{RGlaw2}
 \rv(H)=\rvo+ \frac{3\nu}{8\pi}\,\mpl^2\,(H^{2}-H_0^2)\,,
\end{equation}
where, as always $G_N=1/\mpl^2$, and $\rvo$ is the present value of the VED, i.e., its value at $H=H_0$.
As we have seen in the previous sections, the  above form \eqref{RGlaw2} for the VED evolution can be derived from explicit QFT calculations\,\cite{Moreno-Pulido:2020anb,Moreno-Pulido:2022phq}. These calculations show that in the current universe the VED evolves as the power $H^2$ of the Hubble rate, which is the lowest order power being consistent with general covariance.  The same QFT calculations show that the parameter $\nu$ in \eqref{RGlaw2} is the $\beta$-function coefficient of the running $\rv$. It is a small (dimensionless) parameter whose nonvanishing value receives loop contributions from boson and fermion fields\,\cite{Moreno-Pulido:2023ryo} and makes possible the renormalization group running of the VED as a function of $H^2$ (cf. Sec. \ref{eq:RVMPheno}).

Taking as a starting point  the evolution law of the VED  as in \eqref{RGlaw2}, the cosmological equations can be solved under different assumptions on the conservation or non-conservation of matter, see the four scenarios described in the previous section.  In the particular case of scenario ii), matter is interacting with vacuum and this determines the anomalous conservation law of matter, which in terms of the cosmological redshift reads
\begin{equation}\label{mRG2}
\rM(z;\nu) =\rMo\,(1+z)^{3(1-\nu)}\,.
\end{equation}
Here,  $\rMo$
is the current matter density, which is mostly non-relativistic.
As we can see, there is a tiny departure from the standard conservation law $\rho_m\sim a^{-3}=(1+z)^3$, which is  parameterized by a nonvanishing value of $\nu$.  On the other hand, for the VED we find
\begin{equation}\label{CRG2}
\rv(z)=\rvo+\frac{\nu\,\rM^0}{1-\nu}\,\left[(1+z)^{3(1-\nu)}-1\right]\,.
\end{equation}
The matter density parameter at present is
$\OMo=\rMo/\rco\simeq 0.3$, where $\rco$ is the critical density today.
We also need $\Ovo=\rvo/\rco\simeq 0.7$, i.e. the current normalized vacuum
energy density.
A nonvanishing value of the parameter $\nu$ introduced in Eq.\,\eqref{RGlaw2} is clearly responsible for the time evolution of the vacuum energy density. As we have seen,  $\nu$ also determines the anomalous conservation of matter, since the exchange between matter and vacuum is governed by this parameter, and indeed we recover the standard law \eqref{solstandardconserv} for the present universe ($\wm\simeq0$) only for $\nu=0$. This deviation from the conventional matter conservation law is admissible because it is small and is offset by a corresponding small change of the VED. The two compensating changes duly take care of covariance, since $\nabla^{\mu}\tilde{T}_{\mu\nu}=0$ is satisfied, with  $\tilde{T}_{\mu\nu}$ given by \eqref{tildeEMT}.

Next we define the amount of violation of the standard matter conservation law at a given redshift:
$\delta\rho_m\equiv \rM(z;\nu)-\rM(z)$. This expression must be proportional to $\nu$, as we subtract the usual
 density of matter (corresponding to $\nu=0$). We can easily confirm this, since to order $\nu$, Eq.\eqref{mRG2} gives $\delta\rho_m=-3\,\nu\, \rMo (1+z)^3\ln(1+z)$.
Thus, the relative time variation of the matter density is easily found:
\begin{equation}\label{eq:deltadotrho}
\frac{\delta\dot{\rho}_m}{\rM}=3\nu\,\left(1+3\ln(1+z)\right)\,H+{\cal O}(\nu^2)\,,
\end{equation}
where we used $\dot{z}=(dz/da)\dot{a}=(dz/da)aH=-(1+z)H$. In practice, measurements aimed at detecting VPC are carried out for relatively small values of the redshift, so we can neglect the log term in the above formula and in good approximation we have
\begin{equation}\label{eq:deltadotrho2}
 \frac{\delta\dot{\rho}_m}{\rM}\simeq 3\nu\,\,H\,.
\end{equation}
The corresponding result for the VED is obtained from (\ref{CRG2}):
\begin{equation}\label{eq:deltaLambda}
\frac{\drv}{\rv}\simeq -3\nu\,\frac{\OMo}{\Ovo}\,(1+z)^3\,H+{\cal O}(\nu^2)\,.
\end{equation}
For small redshift,it is of the same order of magnitude as (\ref{eq:deltadotrho2}) but exhibits the opposite sign, as could be expected from the fact that there is energy transfer between  vacuum and matter.

The canonical RVM form \eqref{RGlaw2} for the running vacuum energy has been extensively scrutinized in the literature by fitting it to the main sources of cosmological data. This means  e.g. type Ia supernovae (SNIa), the Cosmic Microwave Background anisotropies (CMB), data on Baryonic Acoustic Oscillations (BAO), data on the large scale  structure (LSS) formation and also the so-called cosmic chronometer values of the Hubble rate (CCH), see e.g.\,\cite{Sola:2016jky,SolaPeracaula:2016qlq,Sola:2016hn,Sola:2016zeg,SolaPeracaula:2017esw,Sola:2017znb,SolaPeracaula:2021gxi,SolaPeracaula:2023swx} and references therein. The range of values pinned down for $\nu$  is the following:
\begin{equation}\label{eq:numodeliicosmology}
|\nu|^{\rm cosm.}\lesssim {\cal O}(10^{-5}-10^{-3})\,,\ \ \ \ \
\end{equation}
depending on particular realizations of the RVM.
In the above range, the fit quality of the RVM is comparable or superior to that of the $\CC$CDM, as shown in the aforementioned references. Moreover, values of $\nu$ of this order are completely in accordance with old theoretical estimates\cite{Sola:2007sv} and modern BBN bounds\cite{Asimakis:2021yct}.

An immediate application within the VPC program now follows. If we use the cosmological limit \eqref{eq:numodeliicosmology} in eq.\eqref{eq:deltadotrho2}, we can predict a relative cosmic drift of the matter density (that is, a gain or loss of matter density per unit time in the universe, depending on the sign of $\nu$)  at the level of
\begin{equation}\label{eq:deltadotrhoexp}
 \left|\frac{\delta\dot{\rho}_m}{\rM}\right|\simeq 2\times \left(10^{-15}-10^{-13}\right) {\rm yr}^{-1}\,,
\end{equation}
where the current value of the Hubble parameter reads  $H_0=1.0227\,h\times
10^{-10}\,{\rm yr}^{-1}\,$  with $h\simeq 0.7$.
Matter leakages as tiny as \eqref{eq:deltadotrhoexp} are not irrelevant. They are correlated with a time evolution of the vacuum energy  and hence  produce a time variation of the cosmological `constant' $\CC=8\pi G\rv$ through \eqref{eq:deltaLambda}.  Furthermore, as the golden rule of the field says,  once a fundamental constant is found to vary, all constants can vary!  The obtained result, therefore, aligns perfectly well with the sought-for connection of the RVM formulation of the expansion history with the VPC search, and in a way  fully consistent with general covariance. Additional hints are given in the next section.

\subsection{Cosmic drift of Newton's $G$ and  Sommerfeld's $\alpha$}

We can also estimate the phenomenological impact of the predicted variation of $G$ within the RVM.  Consider the model type iv) listed in Sec.\ref{eq:GeneralRVM}. In it, matter is conserved in the standard way \eqref{standardconserv}. This is possible thanks to the simultaneous dynamical change of $\rv$ and $G$ encoded in Eq.\,\eqref{Bianchi1}. This equation can be solved using the RVM vacuum running law \eqref{RGlaw2}  and the relation $\rho_m+\rv=3H^2/(8\pi G)$ from  Friedmann's equation (\ref{Friedmann}). It is easy to see that $G$ must satisfy the equation
\begin{equation}\label{eq:diffGH}
\frac{dG}{G^2}=-\frac{\nu}{G_N}\frac{dH^2}{H^2}\,.
\end{equation}
which can be  integrated at once using the boundary condition $G(H_0)=G_N$:
\begin{equation}
\label{Gevolution}
 G(H) = \frac{G_N}{1 + \nu \ln \frac{H^2}{H^2_0} }\,.
\end{equation}
The obtained result implies a very mild running of $G$ with the expansion, since it entails a logarithmic evolution with the Hubble rate $H$, and in addition $\nu$ is  small in absolute value. The relative variation of $G$ with the cosmic time immediately follows from Eq.\,\eqref{Gevolution}. To order $\nu$, it gives
\begin{equation}\label{GdotoverGo}
\frac{\dot{G}}{G}= -2\nu\,\frac{\dot{H}}{H}=2\,(1+q)\,\nu\,H\,,
\end{equation}
where use has been made of $\dot{H}=-(1+q)H^2$, with
$q=-\ddot{a}/(aH^2)$  the standard deceleration parameter. Typically, the existing bounds
on the relative time variation of $G$ yield
$|\dot{G}/G|\lesssim 10^{-12}\,{\rm yr}^{-1}\,$\,\cite{Uzan:2010pm,Chiba:2011bz}. The value in our time of the deceleration parameter
 for a flat universe with $\OMo=0.30$ is $q_0=3\OMo/2-1=-0.55$. Therefore,
\begin{equation}\label{GdotoverGo2}
\left|\frac{\dot{G}}{G}\right|\lesssim \,0.9 |\nu|\,H\,.
\end{equation}
From  the current value $H=H_0$ (given above) of the Hubble parameter, which we mentioned before,  we find $|\nu|\lesssim
10^{-2}$. This bound, however, can be improved one order of magnitude. In fact, using more recent data based on  the double pulsar PSR J0737–3039A/B\cite{Kramer:2021jcw} a more stringent range of variation for $G$ can be extracted:
\begin{equation}\label{eq:lc1}
\frac{\dot{G}}{G}=\left(-0.8\pm 1.4\right)\times 10^{-13} \frac{1}{{\cal F}_{AB}} \,{\rm yr}^{-1}\,,
\end{equation}
in which ${\cal F}_{AB}\simeq 1$ for weakly self-gravitating bodies, see\cite{Kramer:2021jcw}. The above measurement translates immediately into a tighter bound  for the running parameter of the RVM: $|\nu|\lesssim 10^{-3}$. It is rewarding to see that this bound is fully consistent with the BBN bound on this parameter, as shown in \cite{Asimakis:2021yct}\footnote{The attentive reader may have compared equations \eqref{Gevolution} and  \eqref{eq:runGH}. In fact, the most rigorous one is the latter, which was derived from explicit QFT calculations within the RVM. These calculations actually introduce a pressure correction \eqref{eq:fullpressure} which amounts to a modification of the Bianchi identity\eqref{Bianchi1} into the more complete one \eqref{ModifiedConsLaw}. The latter has been discussed in Sec.\ref{sec:RunningG} and from it we have rederived the more correct form \eqref{eq:runGH2}, which reconfirmed  Eq.\,\eqref{eq:runGH}.  Notwithstanding, for the present discussion this does not make much difference, if $\nu$ is considered as a mere phenomenological parameter to be fitted against the cosmological data.}

There are many other observables which are interesting to look at from the point of view of the VPC program. As mentioned previously, much attention has been paid to the proton-to-electron mass ratio $\mu\equiv m_p/m_e$\cite{Uzan:2010pm,Chiba:2011bz,Martins:2017yxk,Safronova:2017xyt}. Here, for lack of space, we shall not address this subject and refer the reader to\cite{Fritzsch:2012qc} and to the aforementioned reviews.

However, we would not like to close this short tasted of the phenomenological implications of the RVM on the VPC program without mentioning  another aspect that involves the possible variation of the proton mass, specifically its possible cosmic drift and correlation with that of Sommerfeld's electromagnetic fine structure constant, $\alpha$, whose possible variation with time and space has also been under extensive surveillance over the years\,\cite{Murphy:2000pz,Webb:2010hc}. A few  observations will suffice to demonstrate that the RVM framework can also have a bearing on this important topic in the pursuit of the variation of the physical constants.

The connection of the RVM and the time variation of $\alpha$, however, is a bit more subtle since gravitation and particle physics are still not part of the same `super-grand GUT' comprising really all of the fundamental interactions of Nature. However, an indirect connection can be established in the context of conventional GUT's\cite{Calmet:2001nu,Fritzsch:2012qc} if we assume that all particle masses are susceptible to evolve with cosmic expansion owing to its possible interaction with the vacuum. To better understand how this may come about in the RVM context, let us consider the
baryonic density in the universe. It essentially consists of the mass density of protons, hence we can write $\rM^B=n_p\,m_p$, where $n_p$ is the number density
of protons and $m_p^0=938.272 013(23)$ MeV is the current proton mass. Now, if
this mass density is not conserved, two possibilities can be conceived:
\begin{itemize}
\item 1) Either $n_p$ does not follow
the normal dilution law with the expansion, i.e. $n_p\sim a^{-3}=(1+z)^3$,
{but} the anomalous law
\begin{equation}\label{eq:nonconservednumber}
n_p(z)=n_p^0\,(1+z)^{3(1-\nu)}\ \ \ \ ({\rm at\ fixed\ proton\ mass}\ m_p=m_p^0)\,,
\end{equation}
and/or
\item  2) the proton mass $m_p$ does not stay constant with time and
redshifts with the cosmic evolution:
\begin{equation}\label{eq:nonconservedmp}
m_p(z)=m_p^0\,(1+z)^{-3\nu}\ \ \ \ ({\rm with\ normal\ dilution}\ n_p(z)=n_p^0\,(1+z)^3)\,.
\end{equation}
\end{itemize}
Within the RVM context,  the vacuum can offset the difference originating from any of these two options since
$\rv=\rv(z)$ ``runs with the expansion'' (see the previous section).  Possibility 1)  implies
that during the expansion a certain number of particles (protons, in this
case) are transferred into the vacuum (if $\nu<0$; or ejected from it, if
$\nu>0$). Possibility 2), in contrast, implies that the number of particles is strictly conserved. Here, the number density follows the normal dilution law with the
expansion, but the mass of each particle changes slightly (decreases for
$\nu<0$, or increases for $\nu>0$) with the cosmic evolution.

For the sake of the present argument, in what follows we adopt Possibility 2), {i.e.}
Eq.\,(\ref{eq:nonconservedmp}).
It is known that the bulk of the proton mass is proportional to the fundamental QCD parameter $\Lambda_{\rm QCD}\simeq 332\pm 17$ MeV\,\cite{Deur:2016tte}, that is,
$m_p\simeq c_{\rm QCD}\,\Lambda_{\rm QCD}$,  up to small corrections from the light quark masses (which we will neglect)\cite{Fritzsch:2012qc}. Thus, if there is a cosmic evolution of the  proton mass through an energy exchange with vacuum, the QCD scale will also evolve with expansion $\LQCD=\LQCD(H)$.
Now the QCD scale  is well-known to be related at 1-loop with the strong fine structure constant $\alpha_s=g_s^2/(4\pi)$ as follows\cite{Deur:2016tte}
\begin{equation}\label{alphasLQCD}
\alpha_s(\mu_R)=\frac{4\pi}{\left(11-2\,n_f/3\right)\,\ln{\left(\mu_R^2/\LQCD^2\right)}}\,,
\end{equation}
where  $\mu_R$ is the renormalization point and  $n_f$ is the number of quark flavors.
Thus, if  $\LQCD=\LQCD(H)$, the strong coupling will also inherit this feature and it will run  not only with the ordinary renormalization scale $\mu_R$ but also with the cosmic scale $H$\,\cite{Fritzsch:2012qc}.  $\alpha_s=\alpha_s(\mu_R,H)$.  If we assume a GUT scenario,  it is now a simple matter to uncover the connection with the cosmic running of Sommerfeld's  constant $\alpha$ and conclude that the latter also runs with the cosmic evolution.  In fact,
in a typical GUT,  the gauge couplings $\alpha_i$ meet at a unification point at very high energy,  $M_X\sim 10^{16}$ GeV.
The scaling of the SM gauge couplings ($g_1, g_2, g_3\equiv g_s$) can be given in terms of the respective fine structure constants $\alpha_i^2=g_i^2/4\pi$. In  the 1-loop approximation, the scaling laws are
\begin{equation}\label{RGevolution}
\alpha_i^{-1}(\mu_R)=\alpha_U^{-1}+\frac{b_i}{2\pi}\,\ln\frac{\mu_R}{M_X}\,.
\end{equation}
Here  $\alpha_U=\alpha_i(M_X)$ is the common unification point of the three couplings at $\mu_R=M_X$, and  $b_i$ are the coefficients of the $\beta$-functions. For $b_i>0$ we have asymptotic freedom (viz. $\alpha_i$ decreases with increasing $\mu_R$). Obviously, these coefficients do not depend on the cosmic expansion.  Thus, from \eqref{RGevolution} we infer that
${d\alpha_i^{-1}/dH}$ is independent of $\mu_R$.  Finally, bearing in mind that in the EW theory the guage couplings can be related to $\alpha$ as  $\alpha_1=5\alpha/(3\cos^2\theta_W)$ and $\alpha_2=\alpha/\sin^2\theta_W$ ($\theta_W$ being the weak mixing angle), one can easily show that the cosmic running of $\alpha$ is connected to that of the QCD scale $\Lambda_{\rm
QCD}$ precisely as follows\,\cite{Calmet:2001nu,Fritzsch:2012qc}:
\begin{equation}\label{eq:timealpha}
\frac{1}{\alpha}\frac{d\alpha(\mu_R;H)}{\,dH}=\frac83\,\frac{\alpha(\mu_R;H)/\alpha_s(\mu_R;H)}{\ln{\left(\mu_R/\LQCD\right)}}\,\left[\frac{1}{\LQCD}\,\frac{d{\Lambda}_{\rm
QCD}(H)}{dH}\right]\,.
\end{equation}
The measured values of the couplings at the  $Z$-pole mass  $\mu_R=M_Z\simeq 91.18$ GeV are well-known, $\alpha(M_Z)\simeq 1/128.9$ and $\alpha_s(M_Z)\simeq 0.118$. Using $\ln \left(M_Z/{\Lambda}_{\rm
QCD}\right)\simeq 5.61$, we finally arrive at the desired relation interconnecting the running of Sommerfeld's fine structure constant with that of the proton mass\cite{Fritzsch:2012qc}:

\begin{equation}\label{eq:timealpha2}
\frac{1}{\alpha}\frac{d\alpha(\mu_R;H)}{\,dH}\simeq
0.031\left[\frac{1}{\LQCD}\,\frac{d{\Lambda}_{\rm
QCD}(H)}{dH}\right]\simeq 0.031\left[\frac{1}{m_p}\,\frac{d m_p(H)}{dH}\right]\,.
\end{equation}
Thus, if there is a GUT supporting the unification of the gauge couplings, the cosmic running of Sommerfeld's  $\alpha$ should be some $30$ times
slower  than that of the proton mass $m_p$.  The upshot is that searching for  the cosmic drift of $\alpha$ is more difficult than searching for the cosmic drift of $m_p$, but there is a precise connection between the two. For other interesting implications of the VPC searches on different sectors of particle and nuclear physics, including dark matter, Higgs bosons, ultralight scalar dark matter models, proton-to-electron mass ratio, strong-field gravity tests etc, see e.g. \cite{Fritzsch:2012qc,Fritzsch:2015lua,Fritzsch:2016ewd,Stadnik:2015kia,Sola:2016our,SolaPeracaula:2018dsw,Calmet:2017czo,Chakrabarti:2021sgs,Kramer:2021jcw,Calmet:2001nu} and the reviews \cite{Safronova:2017xyt,Barontini:2021mvu,Bass:2023hoi,Ginges:2003qt,Uzan:2024ded,SolaPeracaula:2023wqw}, and references therein.

\newpage
\section{Discussion and outlook} \label{sec:conclusions}

In this review, I summarize the Running Vacuum Model (RVM) approach to cosmology, which nowadays can be considered a unification paradigm for Dark Energy (DE) and Inflation formulated within one and the same Quantum Field Theory (QFT) framework in curved spacetime. This means that these two central concepts of modern cosmology are two sides of the same coin, and there is, in principle, no need to introduce \textit{had hoc} fields to explain DE in the current universe and inflation in the very early universe. In the RVM context, Dark Energy and Inflation originate from quantum effects on the dynamical background stemming from the quantized matter fields in Grand Unified Theories (GUT's). This does not exclude, of course,  the possible existence of quantum gravity (QG) effects which should be added up on top of the present semiclassical approach. However, while QG is still under construction, I have focused on the quantized matter fields alone as the prime actors.

The fast and fruitful theoretical developments undergone by the RVM framework over time can easily be traced in the literature by considering the derivation of its basic equations for the vacuum energy density (VED), originally guessed from  old semiqualitative arguments based on the renormalization group (RG) approach to cosmology -- see  \cite{Sola:2013gha,Sola:2011qr,Sola:2014tta,Shapiro:2004ch,Sola:2007sv,Sola:2015rra,Shapiro:1999zt,ShapSol,Shapiro:2009dh,Babic:2001vv,Guberina:2002wt,Babic:2004ev} and references therein -- until reaching  its mature status at present through explicit renormalization calculations within QFT in curved spacetime\cite{SolaPeracaula:2022hpd}. These calculations were first put forward in  \cite{Moreno-Pulido:2020anb,Moreno-Pulido:2022phq,Moreno-Pulido:2022upl,Moreno-Pulido:2023ryo}. The most recent theoretical studies on the subject have, on the one hand, built upon the $H^4$ inflationary mechanism of inflation in the RVM\cite{SolaPeracaula:2025yco}, and on the other on a comparison with the scenario of unstable de Sitter vacuum decaying into FLRW spacetime in the radiation-dominated epoch\cite{SolaPeracaula:2026trz}. This is a very fruitful comparison since both scenarios can produce a period of inflation triggered by the power $H^4$ during a short period of time where the Hubble rate $H$ remains constant, followed by a fast decay of vacuum into the standard radiation epoch of the $\CC$CDM model. As with the aforementioned general running $\CC$ ideas, which are slightly more than a quarter of century old,  the first studies on $\sim H^{2n} (n=2,3...)$ -inflation  trying to extend the RVM to include the inflationary period,  were  essentially of phenomenological nature \cite{Lima:2013dmf,Perico:2013mna,Basilakos:2013xpa,Lima:2014hia,Lima:2015mca ,Sola:2015rra,Sola:2015csa,Sola:2014tta,Yu2020} until they have been integrated into the QFT unification paradigm presented here, which is only a little over five years old\cite{Moreno-Pulido:2020anb,Moreno-Pulido:2022phq,Moreno-Pulido:2022upl,Moreno-Pulido:2023ryo,SolaPeracaula:2025yco,SolaPeracaula:2026trz}.

There is no inflaton field at all in RVM-inflation  (nor in the unstable de Sitter vacuum approach), since inflation is here the pure work of gravity under the vacuum fluctuations of the quantized matter fields. In both of these scenarios, the graceful exit transition is automatically guaranteed due to the subleading power $H^2$ of the VED, which is again predicted to exist as a quantum vacuum effect. This means that there is no such thing as a highly non-adiabatic event represented by the reheating mechanism, which is typical in inflaton-like scenarios\cite{KolbTurner,LiddleLyth,Kallosh:2025ijd}. Furthermore, despite naive resemblances, RVM-inflation is also very different from Starobinsky's inflation\cite{Starobinsky:1980te},  since in the latter there is a reheating period associated with scalaron decay and moreover there  is no period for which $H$ can remain approximately constant, so these are entirely different inflationary mechanisms. The effectiveness of RVM-inflation is similar to that of usual models based on inflaton fields, but without any real inflaton at all. In fact, RVM-inflation predicts an approximate de Sitter period where the VED remains essentially constant, which substitutes the postulated existence of an effective potential for a scalar field triggering inflation while the field remains approximately constant in its ground state.  Finally, we find that after the RVM inflationary epoch, a remnant of vacuum energy remains in the late universe which is mildly dynamical and can provide a natural (and fundamental) explanation for the dark energy (DE) that we observe in the current universe, as well as for its dynamical character recently favored by DESI measurements\cite{DESI:2024mwx,DESI:2024aqx,DESI:2025zgx,DESI:2025fii}. That prediction of dynamical DE is also shared by the unstable de Sitter vacuum scenario\cite{SolaPeracaula:2026trz}.

From the foregoing, it is clear that the RVM approach emphasizes the notion of quantum vacuum in the context of fundamental physics. This applies, in particular, to QFT in curved spacetime and suggests that the VED  should play a relevant role in the description of the cosmological evolution from first principles. Not surprisingly, the quantum dynamics of vacuum inherent to the RVM naturally  predicts the dynamics of the DE in the present universe and a mechanism of inflation in the early universe. As we should reiterate, the QFT approach within the RVM is semiclassical in that only the quantized matter fields determine the quantum vacuum energy. QG considerations can wait for a final quantum theory of the gravitational field, which unfortunately does not exist at present notwithstanding the numerous efforts devoted to it.

Renormalization is, of course, necessary in the RVM approach since QFT is involved. As a distinctive characteristic of the modern RVM formulation, we adopt what we have called the  ``off-shell adiabatic prescription'' to renormalize the energy-momentum tensor, following the proposal of\cite{Moreno-Pulido:2020anb,Moreno-Pulido:2022phq}.  The renormalized VED in that context depends, as in any renormalization calculation, on a floating scale $M$. The existence of such a scale is characteristic  of QFT due to the intrinsic breaking of conformal invariance by quantum effects. The value of $M$ plays the role of renormalization point. The various epochs of the cosmic history (mapped out by the value of the Hubble rate, $H$) are explored by setting the scale $M$ precisely to the value of $H$ at each epoch. This is, after all,  in the very spirit of the RG methods in ordinary gauge theories but applied to cosmology. The setting $M=H$ defines the characteristic energy scale of the FLRW metric in natural units and  $H^2$ defines the degree of $4$-curvature of the geometric background in each cosmic era.   The VED emerges as an expansion in an even number of time derivatives of the scale factor,  as demanded by general covariance of the effective action.

The expansion may conveniently be rephrased in terms of $H$ and its time derivatives $\rho_{\rm vac}=\rho_{\rm vac}(H, \dot{H},\ddot{H},...)$.  The obtained expression for the VED can then be used to explore the entire history of the universe from the very early times where inflation occurs (triggered by the quantum vacuum effects), going through the radiation- and matter-dominated epochs until reaching the current DE epoch. Despite the fact that the RG methods in cosmology are old enough\cite{Nelson:1982kt}, the RVM approach equipped with an off-shell renormalization prescription to explore cosmic evolution is unprecedented in the literature\cite{Moreno-Pulido:2020anb,Moreno-Pulido:2022phq}. Thanks to it, a unified QFT paradigm for DE and inflation becomes possible. In fact, what we call the dark energy density can be thought of as the remnant tail of the huge VED that brought about the exponential inflation at the very early times and is still decaying very slowly at present. In this sense, the RVM predicts the existence of the DE as vacuum energy from QFT origin, and also that the DE we observe today is mildly dynamical.

In a number of phenomenological studies, the RVM has been successfully confronted with a large set of cosmological observations over the years. We have illustrated this fact only very briefly in this review (Sec.\,\ref{sec:RVM today}). It was long established that the RVM can be a serious competitor to the purely phenomenological description of the $\CC$CDM with a rigid (and \textit{ad hoc}) cosmological constant\,
\cite{Sola:2015wwa,Sola:2016jky,Sola:2017znb,SolaPeracaula:2016qlq,Sola:2016hn,Sola:2016zeg,SolaPeracaula:2017esw,Basilakos:2009wi,Basilakos:2010rs,Grande:2011xf,Gomez-Valent:2014rxa,Gomez-Valent:2014fda,Gomez-Valent:2015pia}. This fact has recently been reconfirmed in\,\cite{SolaPeracaula:2021gxi,SolaPeracaula:2023swx,deCruzPerez:2025dni}. The currently preferred fitting values for the parameter $\nu_{\rm eff}$ that controls the running of the VED fall in the ballpark of $\sim 10^{-4}- 10^{-3}$ (depending on particular realizations, see the aforementioned papers). Such  analyses also show that the $H_0$ and growth tensions\cite{CosmoVerseNetwork:2025alb} can both be significantly alleviated in the RVM context.

On top of  these positive phenomenological implications,
perhaps the principal message of the RVM picture of the cosmic expansion is one of profound conceptual nature: viz. neither the VED at the present time, $\rho_{\rm vac}^0 =\rho_{\rm vac}(H_0)$, nor the associated cosmological `constant' $\Lambda=8\pi G\,\rho_{\rm vac}^0$ that we have measured are really constants of nature in a fundamental QFT context. And this applies to the gravitational coupling as well, $G$.  In fact, at any cosmic time, the VED is given by the above mentioned function of $H$ and its time derivatives, and the associated  cosmological term is dynamical too: $\Lambda(H)=8\pi G(H)\,\rho_{\rm vac}(H)$, with $G=G(H)$ a very mild (logarithmic) function of the expansion. The dynamics is smooth enough so as to make them appear as approximately constant and hence preserve the basic properties of the $\CC$CDM.   In this sense  the RVM behavior around the present time (and for that matter during the entire post-inflationary epoch)  is essentially $\Lambda$CDM-like. The VED change between two nearby epochs in recent history is $\delta \rho_{\rm vac}\propto \nu_{\rm eff}\mpl^2H^2$, where the small parameter $|\nu_{\rm eff}| \ll 1$ is calculable in QFT and plays the role of $\beta$-function coefficient of the VED running\,\cite{SolaPeracaula:2022hpd}.

Another no less remarkable fact is that the VED evolution is free of the undesired quartic mass contributions $\sim m^4$  from any quantum matter field with a nonvanishing rest mass\,\cite{Moreno-Pulido:2020anb,Moreno-Pulido:2022phq}. If they were present, these contributions would recreate the need for extreme fine-tuning, and hence would reproduce such an ugly feature of the  cosmological constant problem\,\cite{Weinberg:1988cp}.  Obviously, this theoretical property of the RVM is much welcome and remarkable.


One more aspect on which we have dwelled upon in this review is the possibility that the so-called fundamental constants of Nature are not really constant throughout the cosmic history (cf. Sec.\,\ref{eq:GeneralRVM}).  The pursuit of the possible variation of the physical `constants' (VPC)  program has a long scientific history until now, and it may eventually be able to prove the reality of such a variation. It includes not only the cosmological `constants' such as $\CC$ and $G$, but also those from the subatomic particle physics world, such as the electromagnetic and strong couplings as well as the ratio of the proton mass to the electron mass. It is difficult to believe that the set of physical parameters was fixed once and forever since the beginning of time.  Barring anthropic considerations\cite{Weinberg:1987dv}, the fate of the mankind should have no bearing whatsoever on the ultimate design of the laws of Nature. Tiny changes in the values of the fundamental  `constants'  should be possible, even if fully imperceptible at the human scale.  These minute changes can only be perceived by extremely accurate experiments made in the lab using sophisticated quantum optics techniques, atom interferometry, optomechanical devices, atomic and nuclear clocks using
quantum entanglement strategies\cite{Safronova:2017xyt,Barontini:2021mvu,Bass:2023hoi,Ginges:2003qt}; and at the same time, alternatively, through the  observation of binary pulsars\cite{Kramer:2021jcw} or even by tracking the values of the relevant `constants' over cosmological spans of time, what requires gathering information on their values in the depths of our cosmic past (using e.g. absorption systems in the spectra of distant quasars\,\cite{Murphy:2000pz,Webb:2010hc,Reinhold:2006zn,Ubachs:2015fro}).

 If signs of VPC are eventually collected, we need to be prepared to understand them within the framework of our fundamental theories, or at least be ready to refine our theories appropriately to accommodate the new features. Such a refinement may be more than that, it could be the firsts steps into a pathway leading  to a new physical paradigm.   We should recall once more that the possible change of a single fundamental constant in Nature opens the Pandora box for many other changes. For example, a change in $G$ can be offset by an appropriate change of  the vacuum energy density such that the Bianchi identity is still preserved, and with it the general covariance of the theory. In actual fact, the RVM under study proves to be a specific QFT framework where such a correlated change between $\Lambda$ (or VED, $\rv$) and $G$ really occurs (cf. Sec.\,\ref{sec:RunningG} ).
 The dynamical nature of the VED, and hence of  $\CC$,  calls either for the corresponding dynamics of $G$ and/or the presence of new interactions leading to the non-conservation of the number of matter particles and/or the non-conservation of the particle masses themselves at the expense of the vacuum dynamics.

 Overall, general covariance is \textit{always} preserved, but a rich VPC phenomenology may nonetheless ensue, which gives hope for a deeper understanding of the laws of Nature. This was indeed the essence of the running vacuum approach to the VPC program as formulated in \,\cite{Fritzsch:2012qc}.
 The RVM framework studied here, therefore,  provides not only a unified paradigm for understanding inflation and dark energy, but also a possible solution to the cosmological tensions and  a fundamental approach to understanding scenarios of VPC, which have already given many hints of potentially new exciting physics over the years.

 In an even more ambitious vein, the RVM gives  a  clue to solving the fine tuning trouble inherent to the CCP and it may even shed light on the kernel aspect of the CCP, namely the ``old cosmological constant problem'' mentioned in the Introduction. To start with, the fact that the VED (and the corresponding $\CC$ term) scales with the renormalization point $M$ makes it dynamical through the association $M=H$ and this  leads to a gentle connection between any two points of the cosmological expansion. The connection is smooth due to the cancellation of the quartic mass contributions $\sim m^4$, a nontrivial property that expresses the naturalness of our off-shell renormalization procedure.  Therefore, although we cannot compute the value of the VED at a given scale  $M=H$, since after all this is not the final aim  of the renormalization group program - which (let us refresh our memory) consists of interrelating values of a given quantity at different scales -- we can nevertheless observe from the smoothness of the scaling evolution of the VED that all of its contributions in the current universe should naturally lie in the ballpark of ${\cal O}(\mpl^2H^2)$ rather than taking  inordinately large values of order  $\sim \mpl^4$. This should possibly offer a  viable pathway not only to a solution of the fine tuning problem but  a final clue into the hard core aspect of the cosmological constant problem, to wit: the value itself of $\CC$.

 In a nutshell: the Running Vacuum Model framework discussed in this review furnishes a unification paradigm for Dark Energy and Inflation that can be formulated within first principles of QFT and GR. As a result, it offers important clues for understanding the dynamics of vacuum, gravity, and matter in an overarching formulation of the expansion history of the universe from the very early times to our day. At the same time, it provides intriguing clues as to the micro and macro connection between the laws of physics, i.e. between the subatomic world of the particle interactions and the large scale structure of the universe. It suggests that all `constants' of Nature actually drift with the expansion, but always preserving the general covariance principle inherent to GR.

 \vspace{0.5cm}


\section*{Acknowledgements}



The author thanks Prof. K.K. Phua for the invitation to contribute to this special issue to celebrate the 40th anniversary of IJMPA and MPLA. The special volume is edited by Professors I. Antoniadis, K.K. Phua, and G. Shiu.  I am funded in part by PID2022-136224NB-C21 (MICIU), 2021-SGR-249 (Generalitat de Catalunya) and CEX2024-001451-M (ICCUB). I also acknowledge participation
in Cost Association Actions CA21136 “Addressing observational tensions in cosmology with systematics and fundamental physics (CosmoVerse)”" and CA23130 ”Bridging high and low energies in search of QG (BridgeQG)”. Last but not least, it is my pleasure to thank  J. de Cruz P\'erez,  A. G\'omez-Valent,   A. Gonz\'alez-Fuentes and  C. Moreno-Pulido  for fruitful collaboration in works exploring different phenomenological and theoretical aspects and implications of the RVM paradigm on the hot problems of modern cosmology. The author extends his gratitude to N. E. Mavromatos for discussions and common work on the StRVM version of the the RVM, which, however,  has not been covered here, see the forthcoming\cite{NickJoan_PR} for a comprehensive exposition.







\newpage
\vspace{1.5cm}
{\bf \Large Appendices}
\appendix
\counterwithin{equation}{section} 
\renewcommand{\theequation}{\thesection.\arabic{equation}}

\section{Off-shell adiabatic expansion of the mode functions}\label{sec:appendixA}
In this appendix, we provide some details of the calculation on the adiabatic expansion of the solution of the WKB equation \eqref{WKBIteration} in the context of the off-shell ARP (cf. Sec.\ref{sec:AdiabaticVacuum}) up to adiabatic order fourth. We summarize the derivation of the expansion terms in Eq.\,\eqref{WKBexpansions}. At the same time we illustrate the connection between this expansion and the effective action approach used in Sec\,.\ref{sec:effAction}.

Recall that our starting point is the action \eqref{eq:Sphi} for a quantum field $\phi$ of physical mass $m$ non-minimally coupled to gravity. Since there is no closed-form solution to the  mode functions \eqref{eq:phaseIntegral}  in FLRW spacetime, we use the adiabatic expansion introduced in Sec.\ref{sec:AdiabaticVacuum}:
\begin{equation}\label{WKB2}
W_k=\omega_k^{(0)}+\omega_k^{(2)}+\omega_k^{(4)}+\cdots
\end{equation}
In the first two steps of this expansion, two important quantities will appear and deserve a comment, namely the arbitrary mass scale $M$ and the mass squared difference with respect to the actual particle mass, $\Delta^2=m^2-M^2$.  These quantities then propagate to all higher adiabatic orders owing to the recursive WKB series.  Let us start with the  first (zeroth order, i.e. the leading)  term of the above expansion \eqref{WKB2}.  We take it  as $\omega_k^{(0)}=\omega_k(\tau, M)\equiv \sqrt{k^2+a^2(\tau) M^2}$,  where the physical mass $m$ has been replaced by an arbitrary mass scale $M$. This is licit insofar as it respects the order of adiabaticity since both terms $\omega_k(\tau, M)$ and $\omega_k(\tau, m)$  are of zeroth adiabatic order.  In fact, the crucial point of the initialization process is that the leading term of the WKB expansion need not be defined on mass-shell, but can be taken off-mass shell and in this sense  the adiabatic expansion is not a mere  WKB series.  This sort of  `ambiguity' at the initial step  is the clue to the off-shell adiabatic renormalization prescription (ARP) and reflects the fact that it should be perfectly possible to renormalize the theory at an arbitrary scale or  `renormalization point'  $M$ rather than on mass-shell.  In more formal terms, we can say that the presence of  a floating scale $M$ in the adiabatic prescription  is nothing but  the practical  implementation of the  renormalization group (RG) wisdom in this formulation.  An  alternative justification is obtained at the level of the effective action formalism (cf. Sec.\ref{sec:effAction} and Appendix \ref{sec:appendixB}) where all the renormalized couplings of the theory, including the VED,  $\rv(M)$,  and Newton's gravitational coupling $G(M)$,  appear explicitly as  `running' quantities scaling with $M$. The consistency between the two calculational procedures to renormalize the EMT at the arbitrary scale $M$  is nontrivial and  was first proven in \cite{Moreno-Pulido:2022phq}.  In Appendix \ref{sec:appendixB}, we comment on the connection between the two methods.

But let us come back to our zeroth order adiabatic  `seed'  $W_k=\omega_k(\tau, M)$  defined at the arbitrary mass scale $M$. We can proceed  straightforwardly to compute the next adiabatic order, namely $\omega_k^{(2)}$.   To this end, we substitute $W_k=\omega_k^{(0)}+\omega_k^{(2)}=\omega_k(\tau, M)+\omega_k^{(2)}\equiv \omega_k+\omega_k^{(2)}$ on both sides  of Eq.\,\eqref{WKBIteration}, where we used again the abridged notation $\omega_k$ for $\omega_k(\tau, M)$, as we did in the main text, Eq.\,\eqref{eq:omegaM}.  Now the derivative terms  have to be considered,  and we have
\begin{equation}\label{WKBdetail1}
\left(\omega_k+\omega_k^{(2)}\right)^2=\Omega_k^2-\frac12\,\frac{\omega_k^{\prime\prime}+\omega_k^{(2)\prime\prime}}{\omega_k+
\omega_k^{(2)}}+\frac34\,\left(\frac{\omega_k^{\prime}+\omega_k^{(2)\prime}}{\omega_k+\omega_k^{(2)}}\right)^2\,.
\end{equation}
Using Eq.\,\eqref{eq:ODEmodefunctions} for $\Omega_k^2$ and bearing in mind that now the scalar of curvature $R$ enters as a second order adiabatic contribution, we can expand the above expression on both sides consistently up to this order and we find
\begin{equation}\label{WKBdetail2}
k^2+a^2M^2+2\omega_k\omega_k^{(2)}\simeq\omega_k^2(m)+a^2\,(\xi-1/6)R-\frac12\,\frac{\omega_k^{\prime\prime}}{\omega_k}+\frac34\,\frac{\omega_k^{\prime 2}}{\omega_k^2}\,.
\end{equation}
or
\begin{equation}\label{WKBdetail3}
2\omega_k\omega_k^{(2)}\simeq  a^2(m^2-M^2)+a^2\,(\xi-1/6)R-\frac12\,\frac{\omega_k^{\prime\prime}}{\omega_k}+\frac34\,\frac{\omega_k^{\prime 2}}{\omega_k^2}\,.
\end{equation}
Finally, dividing out through $2\omega_k$ and taking into account that $\Delta^2=m^2-M^2$ we obtain
\begin{equation}
\omega_k^{(2)}= a^2\,\frac{\Delta^2 + R\,(\xi-1/6)}{2\omega_k}-\frac{\omega_k^{\prime \prime}}{4\omega_k^2}+\frac{3\omega_k^{\prime 2}}{8\omega_k^3}\,,\label{WKBexpansions1}
\end{equation}
which is exact up to second adiabatic order. In this way  we have just proven the  result quoted in the second row of  Eq.\,\eqref{WKBexpansions}. To compute the fourth order adiabatic term $\omega_k^{(4)}$, we can just iterate the above procedure using $W_k=\omega_k^{(0)}+\omega_k^{(2)}+\omega_k^{(4)}$ on both sides of Eq.\,\eqref{WKBIteration} and expand it while keeping only terms up to fourth adiabatic order. Rearranging, we find
\begin{equation}\label{WKBdetail43}
\begin{split}
2\omega_k\omega_k^{(4)}&\simeq  a^2(m^2-M^2)+a^2\,(\xi-1/6)R- 2\omega_k\omega_k^{(2)}-\left(\omega_k^{(2)}\right)^2  -\frac{\omega_k''+\omega_k^{(2)\prime\prime}}{2\omega_k} \left(1-\frac{\omega_k^{(2)}}{\omega_k}\right)\\
& +\frac{3\left(\omega_k'^2 + 2\omega_k'\omega_k^{(2)\prime}\right)}{4\omega_k^2} \left(1- 2\frac{\omega_k^{(2)}}{\omega_k}\right)
\end{split}
\end{equation}
Next we insert the expression for the second order result  \eqref{WKBexpansions1} in the third term $- 2\omega_k\omega_k^{(2)}$ on the \textit{r.h.s.} of the above expression and some cancellations take place. We obtain
\begin{equation}\label{WKBdetail44}
\begin{split}
2\omega_k\omega_k^{(4)}&
\simeq \frac{\omega_k^{\prime \prime}}{2\omega_k}-\frac{3\omega_k^{\prime 2}}{4\omega_k^2}-\left(\omega_k^{(2)}\right)^2 -\frac{\omega_k''}{2\omega_k} +\frac{\omega_k^{(2)}\omega_k''}{2\omega_k^2}\\
&-\frac{\omega_k^{(2)\prime\prime}}{2\omega_k}+\frac{3\omega_k'^2}{4\omega_k^2}-3\frac{\omega_k'^2 \omega_k^{(2)}}{2\omega_k^3}+\frac{3\omega_k'\omega_k^{(2)\prime}}{2\omega_k^2}\\
&= -\left(\omega_k^{(2)}\right)^2+\frac{\omega_k^{(2)} \omega_k''}{2\omega_k^2}-\frac{\omega_k^{(2)\prime\prime}}{2\omega_k}-\frac{3\omega_k^{(2)} \omega_k'^2}{2\omega_k^3}+\frac{3\omega_k'\omega_k^{(2)\prime}}{2\omega_k^2}\,.
\end{split}
\end{equation}
Dividing through $2\omega_k$, we finally get the fourth order term of the adiabatic expansion:
\begin{equation}
\omega_k^{(4)}=-\frac{1}{2\omega_k}\left(\omega_k^{(2)}\right)^2+\frac{\omega_k^{(2)}\omega_k^{\prime \prime}}{4\omega_k^3}-\frac{\omega_k^{(2)\prime\prime}}{4\omega_k^2}-\frac{3\omega_k^{(2)}\omega_k^{\prime 2}}{4\omega_k^4}+\frac{3\omega_k^\prime \omega_k^{(2)\prime}}{4\omega_k^3}\,.
\label{WKBexpansionfourthorder}
\end{equation}
which is just the one given by the third row of  Eq.\,\eqref{WKBexpansions} ({\bf q.e.d.}).  The adiabatic  expansion can continue with the sixth-, eighth-order terms, and so on, but calculations become more and more difficult. For example, the sixth adiabatic order $\omega_k^{(6)}$ was explicitly computed in \cite{Moreno-Pulido:2022phq} and it is already a rather cumbersome task, especially because right next one has to use the result to subsequently compute the $h_k$ field modes in Eq.\,\eqref{eq:phaseIntegral} and then, of course,  the  EMT components from which to derive the VED and vacuum pressure.  However, let us note that the adiabatic expansion is an asymptotic expansion, which means that its truncation is mandatory after a few relatively low adiabatic orders, beyond which its convergence gets degraded.


\section{Mode expansion versus effective action method}\label{sec:appendixB}

We can see from Eq.\,\eqref{WKBexpansions1} that the sum of terms  $\Delta^2+ R(\xi-1/6)$ behaves effectively as a block of adiabatic order 2 since $R$ is indeed of order 2. This explains why the quantity $\Delta^2$ must also be treated as being of second adiabatic order, although in what follows we can provide a more formal reason\footnote{The fact that $m^2$ and $M^2$ are of zeroth adiabatic order whereas $\Delta^2\equiv m^2-M^2$ is of adiabatic order 2 can naively and very roughly be compared to the following situation: $1$ and $\cos x$, both are finite quantities for $x\to 0$ (hence of zeroth infinitesimal order), whereas their difference $1-\cos x\sim x^2/2$ is an infinitesimal quantity of order 2.} 

The above result can also be motivated as follows.  We start from  the Green's function $G_F(x,x^\prime)$ of the field $\phi$ associated with the sourced KG equation
\begin{equation}\label{eq:KGsource}
\left(\Box_x-m^2-\xi R(x)\right)\phi(x)=-J(x)\,.
\end{equation}
The Green's function then satisfies
\begin{equation}\label{KGPropagatorOnShell}
\left(\Box_x-m^2-\xi R(x)\right)G_F(x,x^\prime)=-\left(-g(x)\right)^{-1/2}\delta^{(4)}(x-x^\prime)\,.
\end{equation}
It is readily verified that the solution of equation \eqref{eq:KGsource} is
\begin{equation}\label{KGsourced}
\phi(x)=\int d^4 x^\prime\,\sqrt{-g}\  G_F(x,x^\prime)\, J(x^\prime)\,.
\end{equation}
The initialization procedure indicated above by which we introduce an off-shell scale $M$ in the WKB expansion  translates in the Green's function language into rewriting Eq.\,\eqref{KGPropagatorOnShell} as follows:
\begin{equation}\label{KGPropagatorOffShell}
\left(\Box_x-M^2-\Delta^2-\xi R(x)\right)G_F(x,x^\prime)=-\left(-g(x)\right)^{-1/2}\delta^{(4)}(x-x^\prime)\,,
\end{equation}
where again $\Delta^2=m^2-M^2$.  If such quantity is to be used to explore the off-shell regime it must be dealt with as being of  adiabatic order higher than $M$ (which is of order zero). Hence  $\Delta^2$ must be considered as being of adiabatic order $2$, which is the next-to-leading order compatible with general covariance.  Bearing in mind that $\xi R$  in \eqref{KGPropagatorOffShell} is also of adiabatic order $2$, the combination $\Delta^2+\xi R$  can be treated as a block  of adiabatic order $2$.  This adiabaticity assignment for $\Delta^2$  is consistent with the WKB method outlined above.  Taking into account that the adiabaticity order of the terms must be hierarchical respected, the fact  that the mass scale $M$ is of adiabatic order zero whilst the special quantity $\Delta^2$ is of adiabatic order $2$ is just what makes that the solution to the Green's function equation \eqref{KGPropagatorOffShell}  will be different from the solution to the original (on-shell) equation \eqref{KGPropagatorOnShell}.   In fact, the adiabatic expansion of the solution to Eq.\,\eqref{KGPropagatorOffShell} will generate new ($\Delta^2$-dependent) terms which are genuinely distinct as compared to the adiabatic expansion of the solution to the original Eq.\,\eqref{KGPropagatorOnShell}.   The solution of the propagator equation \eqref{KGPropagatorOffShell} using the DeWitt-Schwinger expansion fully confirms the consistency of the two approaches, as shown in Sec. \ref{sec:effAction}.

In practice, obtaining the  renormalized vacuum Lagrangian \eqref{eq:LWrenM} from which to construct the effective action \eqref{eq:effActionLren} and then retrieve the vacuum EMT  \eqref{eq:TrenMm0FLRWMm}, is not difficult, but is quite laborious.
In the on-shell case, the method is described e.g. in \cite{Birrell:1982ix,Fulling89,Parker:2009uva}. But in the off-shell case, which is crucial in our RVM approach, we have to carefully trace the corrections introduced by the term  $\Delta^2\equiv m^2-M^2$.  Details are given in \cite{Moreno-Pulido:2022phq}.
Here we shall only indicate that to obtain the adiabatic solution of the off-shell Green's function equation \eqref{KGPropagatorOffShell}, the calculation can be conveniently performed  in Riemann normal coordinates, up to four derivatives of the metric (hence  up to fourth adiabatic order). In these coordinates, the metric admits the following local expansion up to $4th$ order around flat spacetime\cite{Bunch:1979uk}:
\begin{equation}\label{NormalCoordExp}
\begin{split}
g_{\mu\nu}(y)=&\eta_{\mu \nu}-\frac{1}{3}R_{\mu \alpha \nu \beta}y^\alpha y^\beta-\frac{1}{6}R_{\mu \alpha \nu \beta;\gamma}y^\alpha y^\beta y^\gamma\\
&+\left[-\frac{1}{20}R_{\mu\alpha\nu\beta ; \gamma \delta}
+\frac{2}{45}R_{\alpha\mu\beta\lambda}R^\lambda_{\ \gamma\nu\delta}\right]y^\alpha y^\beta y^\gamma y^\delta+\dots
\end{split}
\end{equation}
Here $y$ stands for the difference between the spacetime coordinate $x$ and the source point $x^\prime$ taken as a reference point in normal coordinates, i.e. $y=x-x^\prime$. The various curvature tensors and their derivatives that appear in the expansion are to be computed at the point  $x=x'$ (i.e. at $y=0$).
A simplification in normal coordinates is that we can raise and lower indices with the Minkowskian metric, as easily shown from the expansion \eqref{NormalCoordExp}.

The solution of the off-shell propagator equation \eqref{KGPropagatorOffShell} in curved spacetime can be obtained -- not without some computation effort\cite{Moreno-Pulido:2022phq} -- in the form of an adiabatic series based on counting the number of time derivatives of the metric. The different adiabatic orders will be denoted $\mathcal{G}_F^{(i)}(k)$, and the series reads as follows:
\begin{equation}
\mathcal{G}_F(k)=\mathcal{G}_F^{(0)}(k)+\mathcal{G}_F^{(1)}(k)+\mathcal{G}_F^{(2)}(k)+\mathcal{G}_F^{(3)}(k)+\mathcal{G}_F^{(4)}(k)+\cdots
\end{equation}
Here, for convenience, we have defined $\mathcal{G}_F(x,x^\prime)=\left(-g(x)\right)^{1/4}G_F(x,x^\prime)$.
Introducing this expansion in the propagator equation \eqref{KGPropagatorOffShell} one can generate a recurrent solution for the different terms. Only the even adiabatic orders survive.  The results, up to $4th$-order, are the following:
\begin{equation}
\begin{split}
&\mathcal{G}_F^{(0)}(k)=\frac{1}{k^2+M^2}, \qquad \mathcal{G}_F^{(1)}(k)=0, \\
&\mathcal{G}_F^{(2)}(k)=-\frac{1}{(k^2+M^2)^2}\left(\left(\xi-\frac{1}{6}\right)R+\Delta^2\right),\\
&\mathcal{G}_F^{(3)}=-\frac{i}{2}\left(\xi-\frac{1}{6}\right)R_{;\alpha}\frac{\partial}{\partial k_\alpha}\left(\frac{1}{(k^2+M^2)^2}\right),\\
&\mathcal{G}_F^{(4)}=\frac{1}{3}Q_{\alpha \beta}\frac{\partial^2}{\partial k_\alpha \partial k_\beta}\left(\frac{1}{(k^2+M^2)^2}\right)\\
&\phantom{XX}+\left[\left(\xi-\frac{1}{6}\right)^2R^2+\Delta^4+2\Delta^2R \left(\xi-\frac{1}{6}\right) -\frac{2}{3}{Q^\lambda}_\lambda\right]\frac{1}{(k^2+M^2)^3}\,,
\end{split}
\end{equation}
where  we have defined
\begin{equation}\label{eq:Qalphabeta}
\begin{split}
Q_{\alpha\beta}\equiv&\frac{1}{2}\left(\xi-\frac{1}{6}\right)R_{;\alpha\beta}+\frac{1}{120}R_{;\alpha\beta}-\frac{1}{40}{R_{\alpha\beta;\lambda}}^\lambda+\frac{1}{30}{R_{\alpha}}^\lambda R_{\lambda \beta}\\
&-\frac{1}{60}{{{R^\kappa}_\alpha}^\lambda}_\beta R_{\kappa \lambda}-\frac{1}{60}{R^{\lambda \mu \kappa}}_\alpha R_{\lambda \mu \kappa \beta}.
\end{split}
\end{equation}
We can easily recognize that the nonvanishing terms $\mathcal{G}_F^{(i)}(k)$ obtained  are of adiabatic orders  $i=0,, 2,4$, respectively. They represent successive corrections to the propagator solution up to $4th$ adiabatic order. We remark that all these terms are evaluated at the off-shell scale $M$.

Using standard  Fourier integral formulas in normal coordinates (with $ky\equiv \eta^{\alpha\beta}y_\alpha k_\beta$), e.g.
\begin{equation}\label{FourierTransform1}
\mathcal{G}_F(x,x^\prime)=\frac{1}{(2\pi)^n}\int d^n k e^{iky}\mathcal{G}_F(k),
\end{equation}
\begin{equation}\label{FourierTransform2}
i\eta^{\alpha \beta}y_\beta\mathcal{G}_F(x,x^\prime)=\frac{1}{(2\pi)^n}\int d^n k e^{iky}\frac{\partial}{\partial k_\alpha}\mathcal{G}_F(k)\,,
\end{equation}
we can convert  the above solution in momentum space into position space. Integrating by parts  and neglecting the boundary terms, we find:
\begin{equation}\label{eq:FourierGF}
\begin{split}
\mathcal{G}_F (x,x^\prime)=\frac{1}{(2\pi)^n}\int d^n k e^{iky}&\left\{ \hat{a}_0 (x,x^\prime)+\hat{a}_1 (x,x^\prime)\left(-\frac{\partial}{\partial M^2}\right)\right.\\
&\left. +\hat{a}_2(x,x^\prime)\left(-\frac{\partial}{\partial M^2}\right)^2+\cdots\right\}\left(\frac{1}{k^2+M^2}\right),
\end{split}
\end{equation}
where the various `bilocal coefficients' $\hat{a}_i$ are given by
\begin{equation}\label{eq:bilocalWScoeff}
\begin{split}
&\hat{a}_0 (x,x^\prime)=1,\\
&\hat{a}_1 (x,x^\prime)=-\left(\xi-\frac{1}{6}\right)R-\Delta^2-\frac{1}{2}\left(\xi-\frac{1}{6}\right)R_{;\alpha}y^\alpha-\frac{1}{3}Q_{\alpha\beta}y^\alpha y^\beta ,\\
&\hat{a}_2 (x,x^\prime)=\frac{1}{2}\left(\xi-\frac{1}{6}\right)^2 R^2+\frac{\Delta^4}{2}+\Delta^2 R \left(\xi-\frac{1}{6}\right)-\frac{1}{3}{Q^\lambda}_\lambda\,.
\end{split}
\end{equation}
As we can see, these coefficients receive $\Delta^2$-dependent corrections in our case, in contrast to the ordinary adiabatic expansion\cite{Birrell:1982ix,Fulling89,Parker:2009uva}.
The quantity  ${Q^\lambda}_\lambda$  in the coefficient $\hat{a}_2$ can be found by taking the trace of \eqref{eq:Qalphabeta}:
\begin{equation}\label{eq:traceQ}
{Q^{\lambda}}_\lambda= -\frac{1}{60} R^{\alpha\beta\gamma\delta}R_{\alpha\beta\gamma\delta}+\frac{1}{60} R^{\alpha\beta}R_{\alpha\beta}+\frac12 \left(\xi-\frac15\right)\Box R\,.
\end{equation}
This expression can then be rephrased, if desired,  in terms of the standar Euler density and the square of the Weyl tensor, see Eq.\,\eqref{eq:traceQ1} in the main text.

The poles in \eqref{eq:FourierGF} in the different terms are understood to be shifted $M^2\rightarrow M^2-i\epsilon$ in order to have a time ordered product. Finally, using the standard Schwinger's proper time representation\cite{Schwinger:1951nm,DeWitt1975,ParkerCargese1978}\,of the zeroth order propagator, through the following identities and corresponding derivatives with respect to the off-shell scale $M$,
\begin{equation}\label{eq:propertime}
\begin{split}
&(k^2+M^2-i\epsilon)^{-1}=i\int_0^\infty ds e^{-is(k^2+M^2-i\epsilon)},\\
&\left(-\frac{\partial}{\partial M^2}\right)^j (k^2+M^2-i\epsilon )^{-1}=i\int_0^\infty (is)^j e^{-is(k^2+M^2-i\epsilon )}ds\,,
\end{split}
\end{equation}
we can interchange the order of integration and perform the following Gaussian integral in momentum space first:
\begin{equation}
\int d^nk e^{iky-isk^2}=i\left(\frac{\pi}{is}\right)^{n/2}e^{-\sigma(x,x^\prime)/(2is)}\,,
\end{equation}
where the characteristic function $\sigma(x,x^\prime)$ (sometimes called the world function\,\cite{Fulling89}) is one-half of the square of the geodesic distance between $x$ and $x^\prime$, that is,  $\sigma(x,x^\prime )= \frac{1}{2}y_\alpha y^\alpha\equiv \frac{1}{2}\,(x-x^\prime)^2$.

At this point we are ready  to finally reach the proper time representation of the Green's function that solves the above propagator equation. The final result is a bilocal expansion in the points $x$ and $x'$:
\begin{equation}\label{eq:HK}
\begin{split}
G_F(x,x^\prime)&=\frac{i{\cal D}^{1/2}(x,x^\prime)}{(4\pi)^{n/2}}\int_0^\infty i ds \frac{e^{-iM^2 s-\sigma (x,x^\prime)/(2is)}}{(is)^{n/2}}\\
&\times\left[\hat{a}_0(x,x^\prime)+is \hat{a}_1 (x,x^\prime)+(is)^2 \hat{a}_2 (x,x^\prime)+\cdots\right],
\end{split}
\end{equation}
where  ${\cal D}(x,x^\prime)=(-g(x))^{-1/2}$ for normal coordinates.  The limit $x\to x'$ of the bilocal coefficients $\hat{a}_i(x,x')$ above is given in Sec.\,\ref{sec:effAction} -- cf.  Eq.\eqref{eq:ModifDWScoeff} -- where we should emphasize once more that in our case they carry the important off-shell corrections, which depend on the term $\Delta^2\equiv m^2-M^2$. These off-shell contributions are then reflected, of course, in the final result for the vacuum effective action and Lagrangian  given in  Eq.\eqref{eq:effLagrangian} of the main text.  We refer the reader once again to reference \cite{Moreno-Pulido:2022phq} for more details.

\newpage
\section{The quantum vacuum in QED: is the zero-point energy `real'?}\label{sec:appendixC}

Quantum Mechanics first emerged in the very short period 1925-26, one hundred years ago, out of a trilogy of founding works  headed by the famous  ``Umdeutung paper'', a truly epoch-making scientific contribution made by Heisenberg alone in 1925\cite{Umdeutung1925}, in which a reinterpretation (`Umdeutung') was made, without recourse anymore to the kind
of ‘inspired guesswork’ that characterized the Old Quantum Theory, into the incipient matrix language of quantum mechanics\footnote{According to S. Weinberg\cite{Weinberg1992}, ``...Heisenberg's 1925 paper was pure magic''. In fact, a paper not easy to follow. And he adds: `If the reader is mystified at what Heisenberg was doing, he or she is not alone'. See \cite{Aitchison:2004cic} for some help in the detailed `exegesis' of Heisenberg' Umdeutung paper, for which he received (alone) the Nobel Prize a few years later.}. It was  followed by an article coauthored by Born and Jordan\cite{Born:1925mph}. The last paper in the trilogy (a rather lengthy paper of $70$ pages) involved the three authors \cite{Born:1926uzf}, and in this landmark article the foundations of the quantum theory of the free electromagnetic field in the absence of any sources were put on firm ground.  Thus, remarkably enough, the genesis of the theory that has subsequently been called Quantum Electrodynamics (QED), entailed already from the very beginning a formal approach to the quantum vacuum. For the first time, the existence of genuine  electromagnetic fluctuations of the quantum vacuum was predicted from first principles. The application of the theory to the emission and absorption of radiation was made by Dirac shortly afterwards\cite{Dirac:1927dy} who at the same time made significant contributions to the developments of the theoretical QED framework.  In a sense, one could say that QED bestowed physical existence on the zero-point energy (ZPE) of the electromagnetic field. Of course, the primary notion of ZPE had already appeared in the old Planck formulation of the quantum theory, specifically in the so-called `second quantum theory' (1912)\cite{Planck1912}, but it was not taken too seriously, namely as something really measurable. The first suggestion that there might be a measurable  ZPE filling out all space was actually made four years later  by Nernst\cite{Nernst1916}, in fact 10 years before the breakthrough of quantum mechanics.

Notwithstanding the support received from the formal theory of QED, the ZPE is still a rather subtle concept whose physical reality has been repeatedly questioned and, as a matter of fact, has not been firmly settled beyond what could be called a heuristic interpretation; hence, it deserves further consideration. After all, there is the theory of GR, which demands a place in this story, and there is the CCP, which is currently alive and kicking (see the Introduction); in fact, after innumerable efforts, we are still striving hard with this mind-boggling problem involving GR and QFT at a time\cite{Weinberg:1988cp,Sahni:1999gb,Carroll:2000fy,PeeblesRatra2003,Padmanabhan2003,Nobbenhuis:2004wn,Copeland2006,Aitchison2009,Sola:2013gha,SolaPeracaula:2022hpd,Bauer:2010wj,Burgess:2013ara}. Here we retake our discussion on zero-point energies in the context of QED, which we initiated in Sec.\ref{sec:finetuningCCP}. Although the issues we now outline are well-known, it may be convenient to bring them forward explicitly as they are closely related to the main theme of this review.

For simplicity, let us place ourselves in a non-ultrarelativistic regime, hence we assume that particle velocities satisfy $v^2/c^2\ll v/c$. In this way,  the electromagnetic field can essentially be discussed in terms of the electric field ${\bf E}(t, {\bf x})$ alone, i.e. we neglect the magnetic field acting on the  moving particle. In Sec.\ref{sec:finetuningCCP} we recalled the notion of normal ordering, which is so frequently used in QED as an artificial procedure to get rid of the quantum vacuum fluctuations. These vacuum-to-vacuum effects produce an infinite, but constant,  contribution which is usually discarded using the normal ordering `trick'.  Now the deletion of the ZPE through normal ordering does not completely remove the vacuum field, i.e.  the free field part of ${\bf E}(t, {\bf x})$, which we will denote ${\bf E}_0(t, {\bf x})$.  The latter is actually necessary for the formal consistency of the theory in flat spacetime. After all, ${\bf E_0}(t, {\bf x})$ is the complementary solution of the wave equation derived from Maxwell theory; or to be more precise in the quantum mechanical context, the homogeneous solution of the operator Maxwell equations in the Heisenberg picture.

Let us summarize the situation. If one considers the electromagnetic field in infinite free space, one starts solving  the wave equation for the vector potential  ${\bf A}(t, {\bf x})$ for a discrete spectrum under periodic boundary conditions in a volume $V$ before setting the continuum limit ($V\to \infty$). Following standard steps, the vector potential can be decomposed into a complete basis of mode functions ${\bf A}_{{\bf k}\lambda}({\bf r})=V^{-1/2} {\bf e}_{k\lambda} e^{i{\bf k}\cdot {\bf r}}$ that satisfy the Helmholtz equation
\begin{equation}\label{Helmholtz}
\nabla ^2{\bf A}({\bf r})+ k^2{\bf A}({\bf r})=0\,,
\end{equation}
where  ${\bf e}_{{\bf k}\lambda}$  ($\lambda=1,2$) is  the polarization vector  for each mode,  with ${\bf e}_{{\bf k}\lambda}\cdot{\bf e}_{{\bf k}\lambda'}=\delta_{\lambda,\lambda'}$ and ${\bf k}\cdot {\bf e}_{{\bf k}\lambda}=0$.  The mode functions are normalized
\begin{equation}\label{eq:normalizedA}
\int_V d^3r  {\bf A}_{{\bf k\lambda}}^*({\bf r}) \cdot {\bf A}_{{\bf k'}\lambda'}({\bf r})=\delta_{{\bf k}{\bf k'}}\delta_{\lambda\lambda'}
\end{equation}
since $(2\pi)^3\delta^3({\bf k}-{\bf k'})=(2\pi)^3\delta^3({\bf 0}) \delta_{{\bf k}{\bf k'}}=V\delta_{{\bf k}{\bf k'}}$.
These functions are, as usual, accompanied by Fourier coefficients, which in the proces of quantization are promoted into photon creation and annihilation operators $a_{\bk,\lambda }^\dagger$ and $a_{\bk,\lambda }$,  in this case also involving the polarization index $\lambda$ compared to the scalar field case \eqref{CommutationRelation}. These ladder operators satisfy standard commutation relations,
\begin{equation}
[a_{\bk,\lambda}, a_{\bk',\lambda' }^\dagger]=\delta^ 3_{{\bf k}{\bf k'}}\delta_{\lambda\lambda'}, \qquad [a_{\bk\lambda},a_{\bk'\lambda'}]=0\,.\label{QEDCommutationRelation}
\end{equation}
After appropriate normalization, one has
\begin{equation}\label{eq:Aexpansion}
{\bf A}({\bf r},t)=\sum_{{\bf k}\lambda} {\bf e}_{{\bf k}\lambda}\left(\frac{2\pi\hbar c^2 }{\omega_kV}\right)^{1/2}\left[a_{{\bf k}\lambda}(t) e^{i{\bf k}\cdot{\bf r}}+a^{\dagger}_{{\bf k}\lambda}(t) e^{-i{\bf k}\cdot{\bf r}}\right]\,.
\end{equation}
The corresponding electric field obtains in the standard manner from the relation  ${\bf E}=-(1/c)\partial{\bf A}/\partial t$, which holds in the radiation gauge without sources, and the magnetic field from ${\bf B}=\nabla\times{\bf A}$ (although we already said that we will neglect the latter in our considerations). Now, as we know from the standard QED wisdom, the free electromagnetic fields resulting from \eqref{eq:Aexpansion} are mathematically equivalent, frequency by frequency, to a harmonic oscillator in quantum mechanics.  Thus, the free electromagnetic field can be described by an infinite collection of quantum mechanical harmonic oscillators. It follows that the field Hamiltonian is given by the sum over all these quantum mechanical oscillators with all possible frequencies and polarizations:
\begin{equation}\label{eq:HF}
H_F=\sum_{{\bf k}\lambda}\hbar\omega_k \left(a_{{\bf k}\lambda}^\dagger a_{{\bf k}\lambda} +\frac12\right)=\sum_{{\bf k}\lambda}\hbar\omega_k \left(N_{{\bf k}\lambda} +\frac12\right)\,.
\end{equation}
In the above field Hamiltonian,  $N_{{\bf k}\lambda}=a_{{\bf k}\lambda}^\dagger a_{{\bf k}\lambda}$ is the number operator counting the number of photons with a given momentum and polarization. The state $|n\rangle$ has $n$ photons: $N|n\rangle=n|n\rangle$. One can proceed similarly with the field momentum, with the result
\begin{equation}\label{eq:PF}
{\bf P}_F=\sum_{{\bf k}\lambda}\hbar {\bf k} \left(a_{{\bf k}\lambda}^\dagger a_{{\bf k}\lambda} +\frac12\right)=\sum_{{\bf k}\lambda}\hbar {\bf k}\ a_{{\bf k}\ \lambda}^\dagger a_{{\bf k}\lambda}=\sum_{{\bf k}\lambda}\hbar {\bf k}\, N_{{\bf k}\lambda}\,.
\end{equation}
The above formulas nicely embody what could be naturally  expected from the quantum field theory of the free electromagnetic field, in which stationary states of the field are described by photons with energies  $\epsilon_k=\hbar\omega_k$ and linear momenta ${\bf p}_k=\hbar{\bf k}$, with $|{\bf k}|\equiv k=\omega_k/c$.
Clearly, the vacuum state $|0\rangle$ has no photon $N_{{\bf k}\lambda}|0\rangle=0$ but has a nonvanishing zero-point energy $\frac12 \sum_{{\bf k}\lambda}\hbar\omega_k$.  These are of course, the results before normal ordering, cf. Sec.\,\ref{sec:finetuningCCP}, as otherwise e.g. the ZPE term would be absent. We want to keep it here, though,  as we wish to dwell on its possible physical meaning. As for the zero-point momentum, $\sum_{{\bf k}\lambda} \frac12\hbar {\bf k}$, it obviously vanishes (even without normal ordering) because free space is isotropic and the average of the wave vector ${\bf k}$ is simply zero.

As mentioned, the electric field for the free theory can be easily obtained from \eqref{eq:Aexpansion} upon using  ${\bf E}=-(1/c)\partial{\bf A}/\partial t$. Noting from the commutation relations that $[N,a]=-a$, $[N,a^\dagger]=a^\dagger$, and then using Heisenberg equations for the ladder operators, $i\hbar\,\dot{ a}=[a, H_F]=\hbar\omega [a,N]$ and its hermitian conjugate, one immediately finds  $\dot{a}=-i\omega a$ and $\dot{a}^\dagger=i\omega a^\dagger$. Thus,
\begin{equation}\label{eq:Efree}
{\bf E}({\bf r},t)=i\sum_{{\bf k}\lambda} {\bf e}_{{\bf k}\lambda}\left(\frac{2\pi\hbar \omega_k}{V}\right)^{1/2}\left[a_{{\bf k}\lambda}(t) e^{i{\bf k}\cdot{\bf r}}-a^{\dagger}_{{\bf k}\lambda}(t) e^{-i{\bf k}\cdot{\bf r}}\right]\,.
\end{equation}
To further proceed, but now in the context of a nontrivial situation with sources (charges), let us consider a practical computational example,  e.g. in  models of thermal equilibrium between oscillating dipoles and electromagnetic radiation\cite{Milonni1994}, and similarly for the nonrelativistic equation of motion of the free electron\cite{Milonni:1981mv}. Following this approach, consider the equation of motion of a particle of charge $e$,  mass $m$ and natural frequency $\omega_0$  oscillating in vacuum.  The total Hamiltonian of this system is given by the sum of the kinetic energy, elastic potential energy and electromagnetic energy:
\begin{equation}\label{eq:FullH}
H=\frac{1}{2m}\,\left({\bf p}-\frac{e}{c}\,{\bf A}\right)^2+\frac12 m\omega_0^2{\bf x}^2+H_F\,,
\end{equation}
where ${\bf p}=m\dot{{\bf x}}+\frac{e}{m}{\bf A}$ is the canonical momentum and $H_F$ is the  electromagnetic field contribution \eqref{eq:HF}.

As stated, in the non-ultrarelativistic regime we can just concentrate on the electric field contribution and neglect the magnetic field effects. The corresponding Heisenberg equations of motion $i\hbar\,\dot{\bf O}=[{\bf O}, H]$  for each quantum mechanical operator ${\bf O}={\bf x}, {\bf p}$ can be combined to give
\begin{equation}\label{eq:eqmotione}
m\ddot{\bf x}+m\omega_0^2{\bf x}=e {\bf E}(t, {\bf x})\,,
\end{equation}
where we have neglected the Lorenz force $(e/c) \dot{\bf x}\times {\bf B}$ exerted by the magnetic field.
However, what is  the electric field  ${\bf E}(t, {\bf x})$ on the \textit{r.h.s} of \eqref{eq:eqmotione}? A reasonable ansatz can be the following:
\begin{equation}\label{eq:Einteraction}
{\bf E}(t)=i\sum_{{\bf k}\lambda} {\bf e}_{{\bf k}\lambda}\left(\frac{2\pi\hbar \omega_k}{V}\right)^{1/2}\left[a_{{\bf k}\lambda}(t)-a^{\dagger}_{{\bf k}\lambda}(t)\right]\,,
\end{equation}
which is inspired by the free field form of Eq.\eqref{eq:Efree}, but  it should be noted that the spatial dependence of the field has been neglected. This approximation is valid if e.g. the equation of motion corresponds, as previously indicated,  to a linear dipole oscillator, in which  one neglects spatial variations of the field over the distance separating the particles (it is called the electric dipole approximation). This approximation notwithstanding, the above expression for the electric field is nontrivial since now the ladder operators are to be determined from the Heisenberg equation of motion  $i\hbar\,\dot{a}_{{\bf k}\lambda}=[{a}_{{\bf k}\lambda}, H]$ with the full Hamiltonian \eqref{eq:FullH}.
Because the solution must also involve the free case (i.e. the vacuum component) considered before, it will be possible to decompose the electric field in \eqref{eq:eqmotione} into the vacuum part and the source part, i.e. ${\bf E}(t, {\bf x})={\bf E}_0(t, {\bf x})+{\bf E}_s(t, {\bf x})$.  The result indeed reads as follows:
\begin{equation}\label{eq:solfora}
a_{{\bf k}\lambda}(t)=a_{{\bf k}\lambda}(0) e^{-i\omega_kt}+ie \left(\frac{2\pi}{\hbar \omega_kV}\right)^{1/2} \int _0^t dt'  {\bf e}_{{\bf k}\lambda}\cdot \dot{\bf x}(t') e^{i\omega_{\bf k} (t'-t)}\,,
\end{equation}
where the first term on the \textit{r.h.s.} corresponds to the free theory. Substituting this expression in \eqref{eq:Einteraction}, we can finally proceed to the mentioned decomposition of ${\bf E}(t, {\bf x})$. On the one hand, we have the free field part or vacuum component, which stems from the first term on the \textit{r.h.s} of \eqref{eq:solfora}, and hence has the same form as \eqref{eq:Efree} but with the ladder operators at $t=0$ (and ${\bf r}=0$):
\begin{equation}\label{eq:E0}
{\bf E}_0(t)=i\sum_{{\bf k}\lambda} {\bf e}_{{\bf k}\lambda}\left(\frac{2\pi\hbar \omega_k}{V}\right)^{1/2}\left[a_{{\bf k}\lambda}(0) e^{-i\omega_kt}-a^{\dagger}_{{\bf k}\lambda}(0) e^{i\omega_kt}\right]\,.
\end{equation}
As for the source part, let us note that, in this case,  the source electric field is caused by the charge acting on itself, so we expect that   ${\bf E_s}(t, {\bf x})$ should coincide, in the non-ultrarelativistic limit under consideration,  with the  `radiation reaction field' of classical electrodynamics in the same limit \cite{Jackson1999}, viz. ${\bf E}_{\rm RR}(t, {\bf x})=\frac{2 e}{3 c^3} \dddot{\bf x}$. In fact, it is so.  This result can be obtained from  the second term on the \textit{r.h.s} of \eqref{eq:solfora}.  Relabeling ${\bf E}_{s}\equiv {\bf E}_{\rm RR}$, we find
\begin{equation}\label{eq:ERR}
{\bf E}_{\rm RR}(t)=-\frac{4\pi e}{V}\sum_{{\bf k}\lambda}  {\bf e}_{{\bf k}\lambda}\int _0^t dt'  {\bf e}_{{\bf k}\lambda}\cdot \dot{\bf x}(t') \cos \omega_{\bf k} (t'-t)\,.
\end{equation}
To check explicitly by direct calculation that this expression gives indeed the mentioned radiation reaction part from classical electrodynamics, we recall that the sums over modes in all these expressions are to be understood in the mode continuum limit:
\begin{equation}\label{eq:continuum}
\sum_{{\bf k}\lambda}\rightarrow  \frac{V}{(2\pi)^3} \sum_{\lambda} \int d^3k\,.
\end{equation}
For example, the ZPE density of the free theory in the continuum limit becomes
\begin{equation}\label{eq:ZPEcontinuum}
\frac{1}{V}\sum_{{\bf k}\lambda} \frac12\hbar \omega_{\bf k}\rightarrow \frac{2}{(2\pi)^3} \int d^3k \frac12\hbar \omega_{\bf k}=\frac{\hbar  c}{2\pi^2} \int dk k^3\,.
\end{equation}
after trivially integrating the solid angle. This result for the ZPE of the photons
coincides, as could be expected,  with Eq.\,\eqref{eq:Minkoski} for the scalar field up to an additional factor of $2$ (and $c=1$), as we have two polarizations for the photon, and $m=0$. We can verify the consistency of this result by recomputing it directly from the free field solution \eqref{eq:E0}, whose associated energy density (the ZPE) is
\begin{equation}\label{eq:ZPEE0}
\begin{split}
\frac{1}{4\pi}\langle{\bf E}_0^2(t)\rangle&=\sum_{{\bf k}\lambda} \sum_{{\bf k'}\lambda}\left(\frac{2\pi\hbar \omega_k}{V}\right)^{1/2} \left(\frac{2\pi\hbar \omega_{k'}}{V}\right)^{1/2} \langle a_{{\bf k}\lambda}(0)\rangle a^\dagger_{{\bf k'}\lambda'}(0)\rangle\\
&=\frac{1}{V}\sum_{{\bf k}\lambda} \frac12\hbar \omega_{\bf k}\,,
\end{split}
\end{equation}
where use has been made of the commutation relations \eqref{QEDCommutationRelation}. So, we have indeed obtained the ZPE associated with the expression of $H_F$ in Eq.\,\eqref{eq:HF}. Notice that while the VEV of the square of the vacuum field is nonvanishing and leads to the ZPE, the VEV of the field itself is zero, of course,  $\langle 0|{\bf E}_{0}(t)|0\rangle=0$, as it is obvious from the fact that $a_{\bk,\lambda}|0\rangle=0 \ (\forall \bk,\lambda)$.  This is once more analogous to the situation with the scalar field studied in Sec.\,\ref{sec:AdiabaticVacuum}.

Next, let us apply the continuum limit to  the calculation of the radiation reaction part given by Eq.\eqref{eq:ERR}:
\begin{equation}\label{eq:ERR2}
{\bf E}_{\rm RR}(t)=-\frac{4\pi e}{V}\frac{V}{(2\pi)^3}\int d^3 k \sum_\lambda  {\bf e}_{{\bf k}\lambda}\int _0^t dt'  {\bf e}_{{\bf k}\lambda}\cdot \dot{\bf x}(t') \cos \omega_{\bf k} (t'-t)\,.
\end{equation}
Decomposing $\dot{\bf x}$ on the basis $\left\{\hat{\bf k},{\bf e}_{{\bf k}\lambda}\right\}$, where $\hat{\bf k}={\bf k}/k$ is the unit propagation vector, we can write the discrete sum in the integrand as  $\sum_\lambda(\dot{\bf x}\cdot {\bf e}_{{\bf k}\lambda}){\bf e}_{{\bf k}\lambda}= \dot{\bf x} - (\hat{\bf k}\cdot \dot{\bf x}) \hat{\bf k}$,  and we find
\begin{equation}\label{eq:ERR3}
{\bf E}_{\rm RR}(t)=-\frac{e}{2\pi^2}\int _0^t dt' \int_0^\infty dk k^2 \cos \omega_{\bf k} (t'-t)  \int d\Omega_{\bf k} \left[\dot{\bf x}(t')- (\hat{\bf k}\cdot \dot{\bf x}(t')) \hat{\bf k}\right]\,.
\end{equation}
We can easily integrate over the solid angles using standard formulae, and the result does not depend on ${\bf k}$ but only on the modulus $k=\omega/c$:
\begin{equation}\label{eq:ERR4}
{\bf E}_{\rm RR}(t)=-\frac{4e}{3\pi c^3}\int _0^t dt'\, \dot{\bf x}(t') \int_0^\infty d\omega\, \omega^2 \cos \omega (t'-t)\,.
\end{equation}
The integral involved in the above result is UV-divergent, but this can be handled without problem since it just amounts to mass renormalization. In fact, by explicitly separating the infinite piece, the final result can be cast as follows, after some calculations\footnote{Using e.g. $\int_0^\infty d\omega\, \omega^2 \cos \omega (t'-t)=\pi \frac{\partial^2}{\partial t'^2} \delta(t'-t)$ and then integrating $\int dt'...$ twice by parts.}:
\begin{equation}\label{eq:ERR4}
{\bf E}_{\rm RR}(t)=\frac{2e}{3c^3}\dddot{\bf x}-\frac{\delta m}{e}\,\ddot{{\bf x}}\,,
\end{equation}
where the last term is the infinite part, given by
\begin{equation}\label{eq:deltaintegral}
\delta m=\frac{4e^2}{3\pi c^3}\int_0^\infty d\omega= \frac{4e^2}{3 c^3} \delta(0)\,.
\end{equation}
The UV-divergent term $\delta m$ acts as a mass counterterm and can be absorbed into the initial coefficient of $\ddot{\bf x}$ in Eq.\,\eqref{eq:eqmotione}, which is to be treated as a bare mass. Then the sum  $m+\delta m$  becomes the new coefficient and  constitutes the  renormalized (observed) mass, which we call again $m$, i.e. we set $m+\delta m\to m$. In this way, the UV-divergent part of the self reaction radiation field can be removed by means of the renormalization of the electron mass. As noted, this mass renormalization procedure is also invoked in the classical radiation theory\,\cite{Jackson1999,Lechner2018}, although it gains plenty more significance in the QED context, of course. As promised, the final common result is indeed  ${\bf E}_{\rm RR}(t, {\bf x})\rightarrow\frac{2 e}{3 c^3} \dddot{\bf x}$ after removing the divergent term, and with the tacit understanding that in the QED case  ${\bf x}(t)$ is the Heisenberg position operator, of course. Hence, the net outcome upon mass renormalization  is ${\bf E}(t, {\bf x})={\bf E}_0(t, {\bf x})+{\bf E}_s(t, {\bf x})\to {\bf E}_0(t, {\bf x})+\frac{2 e}{3 c^3} \dddot{\bf x}$ and the  above equation of motion \eqref{eq:eqmotione} can be restated as follows:
\begin{equation}\label{eq:eqmotione2}
\ddot{\bf x}+\omega_0^2{\bf x}-\tau \dddot{\bf x}=\frac{e}{m}{\bf E}_0(t, {\bf x})\,,
\end{equation}
where
\begin{equation}\label{eq:deftau}
\tau\equiv\frac{2e^2}{3m c^3}=\frac{2r_0}{3c}
\end{equation}
is a well-known fundamental time scale in classical radiation theory (of order $\tau\sim 10^{-23}$ sec), which we have also expressed in terms of the classical radius of the electron ($r_0=e^2/m c^2$, in Gaussian units, which we are using here, or $r_0=e^2/(4\pi mc^2)$ in Heaviside-Lorentz units). As we shall see in a moment, $\tau$ plays a crucial role in what follows.
Apart from the oscillatory term $\omega_0^2{\bf x}$, equation \eqref{eq:eqmotione2} is the so-called Abraham-Lorentz equation of motion of the electron\cite{Jackson1999}. Again, we remark that in QED this equation must be considered as an operator equation in the Heisenberg representation. The key point for our discussion is to have the nonvanishing vacuum field component on its \textit{r.h.s.}, which, as mentioned, is the homogeneous solution  of the field PDE's, and hence a fully legitimate contribution.

The importance of keeping ${\bf E}_0(t, {\bf x})$ alive in the equation of motion  of the electron is almost evident.
For simplicity, let us consider just the $z$-component of Eq.\,\eqref{eq:eqmotione2} in our discussion.  Let  $E_{0z}(t)$  be the component of the vacuum field in this direction.  As advanced in  Sec.\ref{sec:finetuningCCP}, failing to include $E_{0z}(t)$  would be fatal for the internal consistency of the theory, since the radiation reaction term, which was transferred to the \textit{l.h.s.} of the equation of motion, would exponentially damp the oscillation. This is obvious from the fact that the equation of motion without the vacuum field reads
$\ddot{z}+\omega_0^2 z-\tau \dddot{z}=0$\,.
That the oscillations are damped can be seen by considering  the small-damping approximation $\tau\omega_0\ll 1$, for which
$\ddot{z}\simeq -\omega_0^2 z$, thus $\dddot{z}\simeq -\omega_0^2\dot{z}$ and as a result the previous equation becomes $\ddot{z}+\tau \omega_0^2\dot{z}+\omega_0^2 z=0$, which is the standard equation of the damped oscillator. Its solutions carry a damping factor $e^{-\frac12\tau\omega_0^2 t}$. Therefore, in the absence of the vacuum field, the oscillations become quickly suppressed after a time $t\gg (\tau\omega_0^2)^{-1}$.

This feature, if present in the QED description of the particle motion,   would kill the quantum commutator $[z(t), p(t)]=[z(t), m\dot{z}(t)]$ in the same short period of time. Here we used the canonical momentum $p_z(t)= m \dot{z}(t)+\frac{e}{m} A_z(t)$ and the equal-time commutativity of particle and field operators, $[z(t), A_z(t)]=0$. If such a breaking of the commutation relations were to happen, it would  result  in a blunt violation of the unitarity of quantum theory. In contrast, in the presence of the vacuum field,  the commutation relations are preserved at all times\cite{Milonni1994,Milonni:1981mv}.  We can verify this nontrivial property by formally solving the equation of motion \eqref{eq:eqmotione2} by means of standard Fourier techniques, using the explicit form of the vacuum field \eqref{eq:E0} in the continuum limit and with the help of the commutation relations \eqref{QEDCommutationRelation}.  The following exact result emerges after some calculations, valid $\forall t$:
\newline
\begin{equation}\label{eq:calculation1}
[z(t), p(t)]=\left(\frac{i\hbar e^2}{2\pi^2 mc^3}\right)\left(\frac{8\pi}{3}\right)\,\int_0^\infty\frac{d\omega\, \omega^4}{(\omega^2-\omega_0^2)^2+\tau^2\omega^6}= i\hbar\,,
\end{equation}
where the integral can be worked out exactly using the residue theorem and recalling that $\tau$ is given by Eq.\,\eqref{eq:deftau}. However, one can better grasp the physical meaning of the result  by using again the small-damping approximation ($\tau\omega_0\ll1$), which turns out to give the same result. In that approximation, the integrand of \eqref{eq:calculation1} is sharply peaked at $\omega=\omega_0$. After a simple change of variable ($\zeta=\omega^2-\omega_0^2$, with $d\zeta\simeq 2\omega_0 d\omega$) and using $\int_0^\infty=\frac12\int_{-\infty}^\infty$ since the integrand is an even function, we get
\newline
\begin{equation}\label{eq:calculation2}
[z(t), p(t)]\simeq\left(\frac{2i\hbar e^2 \omega_0^3}{3\pi mc^3}\right)\,\int_{-\infty}^\infty\frac{d\zeta}{\zeta^2+\tau^2\omega_0^6}= \left(\frac{2i\hbar e^2 \omega_0^3}{3\pi mc^3}\right) \left(\frac{\pi}{\tau \omega_0^3}\right)=i\hbar\,,
\end{equation}
where in this case the integration is elementary. As we can see, the same result \eqref{eq:calculation1} is obtained with a clearer physical meaning. Thus, thanks to the vacuum field, unitarity is preserved in QED, this being true even if the ZPE of QED is retained and takes an infinite value without normal ordering.

This implies that a charged particle in the Minkowski vacuum must always `feel' the zero-point electromagnetic field to avoid the breakdown of unitarity in QED. However, if we have to be consistent also with GR (which is mandatory, of course), the total contribution  of QED to the vacuum energy (density) must be kept to zero so as  not to induce any cosmological term in Einstein's equations. Otherwise, QED would bluntly contradict the assumed flatness of spacetime.

However, the vanishing of the vacuum energy may pose a serious difficulty in our understanding of basic facts of  the early history of quantum theory. Indeed, let us recall that the primeval notion of the ZPE emerges from Planck's `second quantum theory' for the black-body radiation\cite{Planck1912}.  According to Planck, the average energy of an
oscillator of frequency $\omega$ in equilibrium with radiation at
temperature $T$, is the following:
\begin{equation}\label{eq:PlanckBlackbody}
E_{\omega}=\frac{\hbar\omega}{e^{\hbar\omega/k_BT}-1}+\frac12\,\hbar\omega\,.
\end{equation}
For $T\to 0$, $E_{\omega}\to (1/2)\,\,\hbar\omega$. This is the zero-point energy of the oscillator, which Planck put forward. However, the most significant limit is actually the opposite, the classical limit $k_BT\gg\hbar\omega$ (in which the Boltzmann constant $k_B$ matters). Indeed, upon expansion of the denominator up to second order, it yields  $E_{\omega}\simeq
k_BT-(1/2)\,\,\hbar\omega+(1/2)\,\,\hbar\omega=k_BT$, as can be easily checked.  The obtained result is the correct classical limit (mathematically it is equivalent to take the  limit $\hbar\to 0$ of the above formula). Therefore, the isolated ZPE term on the \textit{r.h.s.} of
(\ref{eq:PlanckBlackbody}), which is a pure quantum effect, seems crucial to  recover the expected classical result
for the average energy of an oscillator in thermal equilibrium at
temperature $T$. This notable observation was made not by Planck (who was already quite skeptical even about the first limit $T\to 0$) but by Einstein and Stern\cite{EinsteinStern1913}, who with this insight provided additional support to the possibility that the ZPE could have some physical meaning. As mentioned above, a few years later Nernst\cite{Nernst1916} further exploited the idea of ZPE by proposing that the `empty' space is actually a reservoir of zero-point electromagnetic radiation, that he quantitatively estimated at a level of an energy density of order $10^{23}$erg/cm$^3$. In natural units, such a VED lies in the ballpark of $10^{-15}$GeV$^4$, which is not small at all if we compare it with the current value of the VED  ($\rv\sim 10^{-47}$GeV$^4$).

From different perspectives, one would like the ZPE of the quantized electromagnetic field to be there, at least within the right order of magnitude.
Whether these contributions are a reality or not is not a settled issue. Let us note that it is not just a matter of QED considerations, as gravity is also there and imposes a toll fee to be paid, which is, of course, to require consistency with the spacetime geometry that is used  and Einstein's equations. For example, as previously noted, the ZPE is frequently adduced as an explanation for  the Casimir effect\cite{Bordag:2001qi}. Recall that for two uncharged, perfectly conducting  parallel plates, there exists a force per unit area ($F/A$) on the plates which is a pure quantum effect (proportional to
$\hbar$) that goes as the inverse of the quartic power of the separation $a$ of the plates. The (nowadays well-measured) force between the plates is attractive and is given by
\begin{equation}\label{eq:CasimirForce}
\frac{F}{A}=-\frac{1}{A}\frac{d E}{da}=-\hbar\,c\,\frac{\pi^2}{240\,a^4}\,,
\end{equation}
where
\begin{equation}\label{eq:CasimirEnergy}
\epsilon(a)=\frac{E(a)}{V}=-\hbar\,c\,\frac{\pi^2}{720\,a^4}
\end{equation}
is the finite `binding' energy  per unit volume ($V=Aa$) that causes the mutual attraction of the plates. It results from subtracting the potential energy density $U(a)$ in the presence of the plates and the energy density in the absence of them, $U(\infty)$. The energy  difference  $\epsilon(a)=U(a)-U(\infty)$ is the  work needed to bring the plates from a very large ($a\to\infty$) separation to a short separation $a$. The two contributions are actually divergent, but the difference turns out to be a finite quantity given by the above formulas, the final result being always the same no matter what regularization you use to perform the calculation; see the standard literature for details, e.g.\cite{Bordag:2001qi,Milonni1994,Bordag12009} and references therein.  The  distinctive vibrational vacuum modes left from the difference of $U$ values, can be associated to the presence of the boundaries and should be responsible for the Casimir force. In Sec.\,\ref{sec:CasimirAnalogy}, we traced an analogy with the vacuum energy in the universe, except that in the universe we do not have plates. We may still have them somehow, though, through $3$-dimensional boundary effects if $K\neq0$; but even if $K=0$ the $4$-dimensional curvature is there since the universe is expanding and that spacetime curvature is associated with a (time-evolving) VED.

The quantum vacuum interpretation of the standard Casimir effect with parallel plates was originally put forward by H. B. G. Casimir himself in 1948\cite{Casimir:1948dh}. It  seems to work pretty well and provides the simplest calculational approach to the above formulas. Nevertheless, it has been repeatedly disputed that the observed Casimir force is not  direct evidence of quantum vacuum energy since other interpretations are perfectly possible, for example in terms of van der Waals molecular forces between dielectrics, which lead exactly to the same quantitative prediction for the Casimir formula\,\cite{Milonni1994,Bordag12009}. Such is the case of Lifshitz theory of the macroscopic molecular forces between solids\cite{Lifshitz:1956zz}, or in Schwinger's approach from source field theory\cite{SchwingerSourceTherory}, ``where the vacuum is regarded as truly a state with all physical properties equal to zero''\cite{Schwinger:1977pa}.

Does this severe dispute mean that the ZPE notion must be abandoned in QED?  It is hard to say.  The derivation of the Casimir force is so straightforward  with the help of the quantum vacuum fluctuations as compared to the other approaches that makes the ZPE almost `real'.
However, despite it being the most economic computational method, the existence of alternative interpretations make the quantum vacuum approach ultimately questionable, as it could just be a heuristic explanation. In point of fact, we should honestly recognize that there is no experimental evidence for the reality of zero point energies in QFT  without gravity \cite{Jaffe:2005vp}. This fact alone should give us a strong hint that in order to endow the VED of physical significance  we should really go beyond flat spacetime (cf. Sec.\,\ref{sec:CasimirAnalogy}). For example, in the context of our non-minimally coupled scalar field calculation within the RVM,  we have shown that the VED is zero in Minkowski space (cf. Sec.\,\ref{sec:VEDMinkowski}). In fact, otherwise the VED calculations \eqref{RenVDEexplicit} and \eqref{RenVDEexplicitMinkow} for curved and flat spacetimes, respectively, would not match; the latter must indeed be zero, as it must coincide with the limit of the former when $H\to 0$, and such a limit must also be zero: $\rv\to 0$.

Now a vanishing VED in Minkowski space is not synonymous of vanishing  renormalized ZPE. If we write $\rv=\rho_\Lambda+$ZPE we can still have $\rv=0$ at the expense of having the ZPE value and the renormalized $\rL(M)$ value equal and with opposite signs, see Sec.\,\ref{sec:VEDMinkowski}. In such a scenario, the classical limit of Planck's equation \eqref{eq:PlanckBlackbody} and  the explanation of the Casimir effect could both still rely on nonvanishing  zero-point energies of the electromagnetic field. However, this is not necessarily true in the case of the Casimir effect, because of the alternative explanations mentioned.

That the full VED is exactly zero in QFT in Minkowski spacetime is out of question if GR is to prevail as the correct theory of the gravitational field.  Whether this still leaves room for a nonvanishing ZPE that can help explain the aforementioned physical effects cannot be presently  asserted with certainty, unfortunately.
Therefore, the physical role of the quantum vacuum in QED is inconclusive. While it is necessary for the formal internal consistency of the theory, as we have seen in some detail above, we cannot identify specific physical effects that are unambiguously attributable to it at present.

In truth, the situation with standard QED and the vacuum energy is rather peculiar. For, if we consider QED in infinite space without boundaries, the electromagnetic  ZPE is constant and infinite, but it is rendered zero after renormalization or normal ordering. If we instead set boundaries, then finite differences of ZPE can be computed, which could be responsible for the Casimir effect. But, as we know, not even this fact can be considered  as its ultimate explanation. On the other hand, the ZPE in the standard  QED context does not have any obvious relation with the vacuum energy in cosmology.  Gravity cannot be just sensitive to the ZPE part of the VED if the latter must be zero in flat spacetime. We may imagine ourselves doing standard QED in a local inertial frame where we have made the artificial adjustment $\rv=\rho_\Lambda+$ZPE$=0$ (cf. Sec.\,\ref{sec:CasimirAnalogy}). This is no fine tuning at all, it is nothing but a formal mathematical declaration that we are imposing Minkowsi space in that neighborhood, and only in that neighborhood.  However, at no time will we be able to detect a single trace of vacuum energy that is responsible for any physical effect related to cosmology since we cannot compare Minkowskian neighborhoods (tangent spaces) in global cosmological spacetime.

As a matter of fact, standard QED cannot produce any significant contribution to the cosmological term,  even if the electromagnetic field is formulated in curved spacetime, because the photon is massless, so it is impossible to obtain a soft vacuum electromagnetic contribution of the type $\sim m^2 H^2$ (cf. Sec.\,\ref{eq:RVMPheno}).
Under normal conditions, one can expect at most  $\sim H^4$ effects from QED in cosmological spacetime, which are completely negligible in the current universe, except for possible $\sim H^2$ terms in chiral models with KR axions in string theory\cite{BasMavSol,basilakos2, basilakos3}.

From the foregoing discussion, the only safe thing we can assert at this point  about the role played by the vacuum field in QED is that it must remain to ensure the internal consistency of the theory. However, this is not tantamount to saying that the vacuum energy fluctuations of the electromagnetic field are there. If we have to be scientifically honest, we have to admit that there is no empirical evidence that they are ultimately responsible for the Casimir effect, and much less that they have a definite impact on the value of $\CC$ in cosmology -- if the basic principles of QED are to be preserved;  in particular, the fundamental principle of gauge invariance. If this principle were to be violated, nothing can prevent the existence of  an effective mass term $m_A^2 A_\mu A^\mu$ for the photon field and a corresponding soft contribution $\sim m_A^2H^2$ to the vacuum energy, which would anyway be extremely tiny; and there is no replacement for that in pure QED.  It is hard to obtain sizable soft effects $\sim H^2$ for the current universe from  standard QED alone in curved spacetime without compromising fundamental aspects of the theory. They are not impossible\cite{BasMavSol,basilakos2, basilakos3}, but would lead us too far astray from the standard lore of QED\footnote{Another valuable attempt, not exempt however of the mentioned difficulties\cite{BeltranJimenez:2008enx}, points to VED contributions of order $\sim A_0^2 H^2$, which turn out to behave as an effective cosmological constant during the cosmic evolution. Here, $A_0$ is the time component of the $4$-vector potential $A_\mu$. However, this mechanism obviously relies on the hypothesis that gauge invariance is broken at cosmological scales. It also assumes that the primordial QED quantum fluctuations were produced during inflation at the electroweak scale, as then one has  $(A_0^2 H^2)_{\rm EW}\sim H_{\rm EW}^4\sim \left(M_{\rm EW}^2/\mpl\right)^4\sim 10^{-47}$ GeV$^4 \sim \rvo$, if taking a ballpark estimate\cite{Arkani-Hamed:2000ifx} of the EW scale around $M_{\rm EW}\sim 4\times 10^3$ GeV. }.

Fortunately, in contexts much wider than pure QED, such as in generic GUT's with a large number of massive boson and fermion fields, we have seen that the ZPE can appear as a full-fledged function of the cosmological expansion. Such is the case in  our RVM calculation for a non-minimally coupled massive scalar field in cosmological spacetime  (cf. Sec.\,\ref{sec:RVM-QFT}, particularly \ref{sec:RenormEMT}), which leads to physical consequences that can be cosmologically tested\cite{SolaPeracaula:2021gxi,SolaPeracaula:2023swx,deCruzPerez:2025dni}. The RVM affords a better fit to the cosmological data than the concordance $\CC$CDM model -- as first hinted  a decade ago in \cite{Sola:2015wwa,Sola:2016jky,SolaPeracaula:2016qlq,Sola:2016hn,Sola:2016zeg,Sola:2017znb,SolaPeracaula:2017esw} using large sets of observational data.

In our RVM framework, a nonvanishing renormalized VED emerges as a function of the Hubble rate and its time derivatives, which include  contributions at two hierarchical levels: on the one hand, vacuum effects of order $H^4$,  which can be the physical triggers of a period of inflation in the early universe; and, on the other hand, soft contributions  $m^2H^2$ ((for $m$ of order of a typical GUT scale), which may provide an explanation of the observed dark energy in the current universe.  Such a  result acquires full  physical  significance only in curved spacetime, as only in a curved spacetime  context the VED can be unambiguously associated with the existence of a nonvanishing  $\CC$-term in full consistency with Einstein's equations.

At the end of the day, we have to admit that the possible existence of physical quantum vacuum effects  can only be meaningfully attributed to the presence of the gravitational field. In its absence, there is no nontrivial geometry, and hence there is  no consistent support from Einstein's equations. Intriguingly enough, the fact that the VED, and hence the $\CC$-term, become dynamical quantities in the RVM context, and that any two nearby cosmic epochs display VED differences of order  $\delta\rv\propto \mpl^2H^2$, might be the clue for an explanation of the observed Dynamical Dark Energy\cite{DESI:2024mwx,DESI:2024aqx,DESI:2025zgx,DESI:2025fii} on fundamental QFT grounds.


\end{document}